\documentclass[leqno,a4paper,12pt,oneside]{article}
\usepackage[latin1]{inputenc}
\usepackage[english]{babel}
\usepackage[T1]{fontenc}
\usepackage{amsfonts,amssymb,amsmath,amsthm,latexsym,amscd,mathtools,exscale}
\usepackage{makeidx}
\title{The early historical roots of Lee-Yang theorem}
\author{\small Giuseppe Iurato\\\footnotesize \it University of Palermo, IT}
\date{}
\begin{document}\setcounter{tocdepth}{22}\maketitle\begin{abstract}A deep and detailed historiographical
analysis of a particular case study concerning the so-called \it Lee-Yang theorem \rm of theoretical statistical
mechanics of phase transitions, has emphasized what real historical roots underlie such a case study. To be
precise, it turned out that some well-determined aspects of entire function theory have been at the primeval
origins of this important formal result of statistical physics.
\end{abstract}\titlepage\tableofcontents\newpage

\section*{Preamble}\addcontentsline{toc}{section}{Preamble}In history of exact and natural sciences (and not only in these),
it is oftentimes possible to identify certain recurrent or persistent ideas, methods or techniques which go
through a whole sector of a certain discipline, or take part in different lands or grounds, like, for instance,
those concerning mathematics and physics, and without which it would have been very difficult to carry on in
building up a certain theory, in attaining a given research's program or in pursuing a possible line of thought.
In this paper, we wish to claim the attention upon one of these historiographical cases just regarding the
crucial and inextricable relationships between mathematics and physics, and that has nevertheless been quite
neglected, that is to say, the one related to the so-called \it Lee-Yang theorem \rm of statistical mechanics.
This formal result of the theoretical framework of statistical physics, was successfully achieved by T.D. Lee
and C.N. Yang in the early 1950s, and marked the rise of a new and fruitful avenue of pure and applied
mathematics. However, as far as we know hitherto, a little attention has been devoted to the history of this
important outcome, except a few notes given by one of the authors of this result, namely C.N. Yang, who has
highlighted some related historical aspects in his comment to the works achieved by him, together T.D. Lee, and
inserted into the collectanea (Yang 2005). Starting from these brief although precious historical remarks, we
have gone much beyond, until up reach the truly early origins of this result, that is to say, some notable
results of entire function theory. To be precise, we have identified into the K. Weierstrass and J. Hadamard
works on the factorization of entire functions, the essential formal tools and techniques without which it is
much likely that a result achieved by Lee and Yang has been very hard to obtain. Indeed, the main basic formal
outcome thanks to which these two physicists were able to gather their conclusions towards this theorem, was
provided by a 1926 work of G. P\'{o}lya about some certain Fourier analysis integral representations of the
celebrated Riemann $\xi$ function which, in turn, refers to very central and pivotal techniques of complex
analysis, that is to say, the so-called \it factorization theorems \rm of entire functions, which revolve around
the pioneering works of Weierstrass and Hadamard achieved after the mid-1800s. These outcomes have really put up
the central pillars around which then build up new avenues or casted new bridges between different contexts or
disciplines, either in pure and applied mathematics. Since the latest historiographical trends even more lead to
put attention toward the probabilistic aspects for an historical event have to occur, it is clear that, when a
given idea, method or tool, is present or happens with a certain frequency in such a way to become central, then
we assign a certain historical relevance to it, bringing to think that, without its attendance, very likely the
historical course would have been quite different, also at the expense to make appeal to counterfactuals
reasonings, which, nevertheless that, have their unquestionable historiographical contextual relevancy. Well,
following the line of this historiographical argument, it will be possible to outline a reliable historical
route which might be assumed as one of the most probable among those which have could lead to that historical
event or case study under examination, that is, the Lee-Yang theorem. In our case, we have tried to descry such
a historical path concerning Lee-Yang theorem, starting from Weierstrass-Hadamard factorization theorems, hence
going on with P\'{o}lya work, till to reach such a result achieved by Lee and Yang in the 1950s. As we will see
at the end of this historical recognition, it will turn out to be evident that the Riemann $\xi$ function has
played the role of a final, mysteriously charming and never attainable specter, a sort of opera ghost kindly
provided to us by the candid Riemann soul, which has gently blown over the scene! And it still continue brazenly
to do it, for a good peace to all men and women. Irony aside, first of all, we have paid a lot of the paper to
describe a very detailed and deep historical analysis of factorization theorems of entire function theory,
moreover little considered, due to the fact that such formal tools of complex analysis have taken a central and
prominent role in either pure and applied mathematics. Then, we have focussed on those historical moments of
this story which have concerned with, and from which started, some works of George P\'{o}lya regarding that meet
ground between entire function theory and Riemann zeta function theory which, in turn, played, sometimes
implicitly, a key role in these works of T.D. Lee and C.N. Yang. We are aware of the heavy (maybe also
excessive) attention put to the history of entire function factorization theorems but, due to the fact that it
is still lacking notwithstanding the pivotal importance of these formal tools, we have caught the occasion to
explain it from an historical standpoint, having had as main target the formal scenario involving Lee-Yang
theorem. Therefore, the sketch of the historical pathway which we have wished to follow, see involved the
following main figures$$\mbox{\rm Riemman,\ Weierstrass\ \&\ Hadamard}\longrightarrow \mbox{\rm
P\'{o}lya}\longrightarrow\mbox{\rm Lee\ \&\ Yang}.$$Lastly, a final remark. With this historical recognition, we
would want to highlight a possible useful role that history of scientific ideas may play for the science's
development itself: indeed, from the historical outlook we have delineated in this paper, it is possible to have
a clearer and wider sight of the question under examination, its origins and related evolution, hence its
achievements as well as its limits and failures, in such a manner to may subsequently infer possible further
insights about other related issues concerning the subject with which we have dealt. For instance, in our
historical case, a central formal problematic aspect that will emerge from the analysis we have done, concerns
the mathematical properties of the partition function of a thermodynamical system as well as its determination.
Now, such a problem may receive more light just from its historical recognition which puts the question under
examination into its own general cultural context in which it relies, so emphasizing its multiple and various
facets from whose knowledge such a problem may get, on its turn, further clarifications as well as possible
insights for approaching its unsolved issues.

\newpage\section*{1. Outlines of history of entire function factorization theorems}
To sum up following (Remmert 1998, Chapter 1), infinite products first appeared in 1579 in the works of F.
Vieta, whilst in 1655 J. Wallis gave the famous product for $\pi/2$ in his \it Arithmetica Infinitorum\rm. But
L. Euler was the first to systematically work with infinite products and to formulate important product
expansions in his 1748 \it Introductio in Analysin Infinitorum. \rm The first convergence criterion is due to
Cauchy in his 1821 \it Cours d'analyse\footnote{See its English annotated translation (Bradley \& Sandifer
2009).}, \rm whilst the first comprehensive treatment of the convergence theory of infinite products was given
by A. Pringsheim in 1889 (see (Pringsheim 1889)). Infinite products found then their permanent place in analysis
by 1854 at the latest, through the works of Weierstrass and others. In this section, we wish to deeply outline
some historical aspects and moments regarding the dawning of infinite product techniques in complex analysis,
with a view towards some of their main applications.\\\\\addcontentsline{toc}{section}{1. Outlines of history of
entire function factorization theorems}\bf 1.1 On the Weierstrass' contribution. \rm Roughly, the history of
entire function theory starts with the theorems of factorization of a certain class of complex functions, later
called \it entire transcendental functions \rm by Weierstrass (see (Loria 1950, Chapter XLIV, Section 741) and
(Klein 1979, Chapter VI)), which made their explicit appearance around the mid-1800s, within the wider realm of
complex function theory which had its paroxysmal moment just in the 19th-century. But, if one wished to
identify, with a more precision, that chapter of complex function theory which was the crucible of such a
theory, then the history would lead to elliptic function theory and related factorization theorems for doubly
periodic elliptic functions, these latter being meant as a generalization of trigonometric functions. Following
(Vesentini 1984, Chapter VII), the theory of elliptic integrals was the first main historical motif from which
elliptic function theory sprung out, whilst the polydromy of such integrals finds its natural environment of
development in the geometry of algebraic curves (see (Enriques \& Chisini 1985, Volume 1) and (Dieudonné \&
Grothendieck 1971)). Again, following (Stillwell 2002, Chapter 12, Section 12.6), the early idea which was as at
the basis of the origin of elliptic functions as obtained by inversion of elliptic integrals, is mainly due to
Legendre, Gauss, Abel and Jacobi (together two pupils of this last, namely G.A. G\"{o}pel (1812-1847) and J.A.
Rosenhain (1816-1887) (see (Hermite 1873, pp. 296-297)). Following (Fasano \& Marmi 2002, Appendix 2), the
elliptic functions was introduced for the first time by J. Wallis in 1655 in computing the arc length of an
ellipse whose infinitesimal element is not equal to the differential of an already known elementary function. In
their most general form, they are given by $\displaystyle\int R(x,y)dx$ where $R(x,y)$ is a rational function of
its arguments and $y=\sqrt{P(x)}$ with $P(x)$ a fourth degree polynomial. Legendre, in 1793, proved that a
general elliptic integral is given by a linear combination of elementary functions and three basic elliptic
integrals said to be integrals of the first, second and third kind. Gauss had the idea of inversion of elliptic
integrals in the late 1790s but didn't publish it; Abel had the same idea in 1823 and published it in 1827,
independently of Gauss. Jacobi's independence instead is not quite so clear. He seems to have been approaching
the idea of inversion in 1827, but he was only spurred by the appearance of Abel's paper. At any rate, his ideas
subsequently developed at an explosive rate, up until he published the first and major book on elliptic
functions, the celebrated \it Fundamenta nova theori\ae\ functionum ellipticarum \rm in 1829. Following
(Enriques 1982, Book III, Chapter I, Section 6), on the legacy left, amongst others, by J.L. Lagrange, N.H.
Abel, C.G.J. Jacobi, A.L. Cauchy and L. Euler, Riemann and Weierstrass quickly became the outstanding figures of
the 19th-century mathematics. Agreeing with Poincaré in his 1908 \it Science et méthode, \rm Riemann was an
extremely brilliant intuitive mathematician, whereas Weierstrass was primarily a logician, both personifying,
therefore, those two complementary and opposite typical aspects characterizing the mathematical work. Beyond
what had been made by Cauchy, they created the main body of the new complex function theory in the period from
about 1850 to 1880 (see (Klein 1979, Chapter VI)). Both received a strong impulse from Jacobi's work. The first
elements of the theory of functions according to Weierstrass date back to a period which roughly goes on from
1842-43 to 1854; in the meanwhile, Riemann published, in the early 1850s, his first works on the foundations of
complex analysis, followed by the celebrated works on Abelian functions (which are elliptic functions so named
by Jacobi) of the years 1856-57, which dismayed Weierstrass himself, influencing his next research program. This
last point should be taken with a certain consideration. Instead, following (Klein 1979, Chapter VI), in the
period from 1830s to the early 1840s, Weierstrass began to self-taughtly study Jacobi's \it Fundamenta nova
theori\ae\ functionum ellipticarum, \rm hence attended Christoph Gudermann (1798-1852) lectures on elliptic
functions (see also (Manning 1975)). He wrote his first paper in 1841 on modular functions, followed by some
other papers wrote between 1842 and 1849 on general function theory and differential equations. His first
relevant papers were written in the years 1854-56 on hyperelliptic or Abelian functions, which engaged him very
much. Afterwards, in the wake of his previous work on analytic, elliptic and Abelian functions, Weierstrass was
led to consider the so-called \it natural boundaries \rm (that is to say, curves or points - later called \it
essential singularities \rm - in which the function is not regularly defined) of an analytic function to which
Riemann put little attention. The first and rigorous treatment of these questions was given by Weierstrass in
his masterful 1876 paper entitled \it Zur Theorie der eindeutigen anatytischen Funktionen, \rm where many new
results were achieved, amongst which is the well-known \it Casorati-Weierstrass theorem \rm (as we today know
it) and the product factorization theorem. Klein (1979, Chapter VI) states that the content of this seminal
paper surely dated back to an earlier period, and was chiefly motivated by his research interests in elliptic
functions. As pointed out in (Hancock 1910, Introduction), nevertheless, it is quite difficult to discern the
right contribution to the elliptic function theory due to Weierstrass from other previous mathematicians,
because of the objective fact that Weierstrass started  to publish his lessons and researches only after the
mid-1860s.

Weierstrass' theory of entire functions and their product decompositions, according to Klein, has found its most
brilliant application in the (Weierstrass) theory of elliptic functions, to be precise, in the construction of
the basic Weierstrassian $\sigma$-function $\sigma(u)$; perhaps - Klein says - Weierstrass' theory of entire
functions even originated from his theory of elliptic functions (see also (Bottazzini \& Gray 2013, Chapter 6,
Section 6.6.3)). Nevertheless, already Gauss and Abel were gone very close to this $\sigma$-function and its
properties. Again Klein says that he wishes to conclude his discussion of Weierstrass' theory of complex
functions, adding only some remarks on the history, referring to R. Fricke distinguished review on elliptic
functions for more information (see (Burkhardt et al. 1899-1927, Zweiter Teil, B.3, pp. 177-348)). If we now ask
- again Klein says - from where Weierstrass got the impulse to represent his functions by infinite products, we
find his principal forerunner in G. Eisenstein (1823-1852), a student of Gauss, who was also a friend of Riemann
with whom often talked about mathematical questions and who very likely stimulated, according to André Weil (see
(Narkiewicz 2000, Chapter 4, Section 4.1, Number 2) and references therein), the interest towards prime number
theory in Riemann. Following textually (Weil 1975, Second Lecture, p. 21),\\

<<\rm [...] \it the case of Riemann is more curious. Of all the great mathematicians of the last century, he is
outstanding for many things, but also, strangely enough, for his complete lack of interest for number theory and
algebra. This is really striking, when one reflects how close he was, as a student, to Dirichlet and Eisenstein,
and, at a later period, also to Gauss and to Dedekind who became his most intimate friend. During Riemann's
student days in Berlin, Eisenstein tried (not without some success, he fancied) to attract him to number-theory.
ln 1855, Dedekind was lecturing in Gottingen on Galois theory, and one might think that Riemann, interested as
he was in algebraic functions, might have paid some attention. But there is not the slightest indication that he
ever gave any serious thought to such matters. It is clearly as an analyst that he took up the zeta-function.
Perhaps his attention had been drawn to the papers of Schl\"{o}milch and Malmquist in 1849, and of Clausen in
1858. Anyway, to him the analytic continuation of the zeta-function and its functional equation may weIl have
seemed a matter of routine; what really interested him wa8 the connection with the prime number theorem, and
those aspects which we now classify as ''analytic number-theory'', which to me, as 1 have told you, is not
number theory at all. Nevertheless, there are two aspects of his famous 1859 paper on the zeta-function which
are of vital importance to us here \rm [i.e., his functional equation for $\zeta$ function and the famous
Riemann hypothesis]\it>>.\\\\ \rm In his long-paper (see (Eisenstein 1847)), Eisenstein did not attain the fully
symmetric normal form, because he still lacked the exponential factors attached to the individual prime factors
that will be then introduced by Weierstrass for inducing the product to converge in an absolutely manner. As he
himself declared, Weierstrass got this idea from Gauss, who had proceeded in a similar way with his product
expansion of the gamma function in 1812 (see the paper on the hypergeometric series in Weierstrass' \it
Mathematische Werke, \rm Band II; see also Weierstrass' works on elliptic and other special functions included
in Band V). It therefore turns out clear that elliptic and hyperelliptic function theory exerted a notable role
in preparing the \it humus \rm in which grew up the Weierstrass work on factorization theorem, and not only
this: in general, it exerted a great influence on Riemann and Weierstrass work (see (Bottazzini \& Gray 2013,
Chapter 4, Section 4.5)). Following (Burkhardt et al. 1899-1927, Zweiter Teil, B.3, Nr. 15-17, 25, 45, 55),
amongst others, Abel\footnote{In (Greenhill 1892, Chapter IX, Section 258), the author states that the
well-known expressions for the circular and hyperbolic functions in the form of finite and infinite products,
have their analogues for the elliptic functions as laid down by Abel in some his researches on elliptic
functions published in the celebrated \it Crelle's Journal\rm, Issues 2 and 3, years 1827 and 1828. Following
(Hancock 1910, Chapter V, Article 89), Abel showed, in the 1820s, that elliptic functions, considered as the
inverse of the elliptic integrals, could be expressed as the quotient of infinite products, then systematically
reconsidered in a deeper manner by Jacobi.}, Euler, Jacobi, Cayley and Gauss (see (Bottazzini \& Gray 2013,
Chapter 1, Section 1.5.1.1; Chapter 4, Section 4.2.3.1-2)) had already provided product expansions of certain
elliptic functions, but it was Eisenstein (see (Eisenstein 1847)), with his infinite product expansion \it
ansatz, \rm the closest forerunner of the Weierstrass work on his $\sigma$ function, in turn based on the
previous work made by Jacobi and Gauss (see (Weil 1976)). Following (Remmert 1998, Chapter 1), in his 1847
long-forgotten paper, had already systematically used infinite products, also using conditionally convergent
products and series as well as carefully discuss the problems, then barely recognized, of conditional and
absolute convergence, but he does not deal with questions of compact convergence. Thus logarithms of infinite
products are taken without hesitation, and series are casually differentiated term by term, and this
carelessness may perhaps explain why Weierstrass nowhere cites Eisenstein's work. Furthermore, already
Cauchy\footnote{In this regard, see (Bellacchi 1894, Chapter X), where an interesting discussion of the Jacobi
series is made, amongst other things highlighting that already Abel, on the wake of what was done by Johann
Bernoulli, had introduced infinite product expansions of certain elliptic functions (Jacobi's triple product)
that later Jacobi, in turn, converted into infinite series by means of trigonometric arguments, so giving rise
to new elliptic functions (see (Greenhill 1892, Chapter IX, Section 258) and (Remmert 1998, Chapter 1, Section
5)).}, ever since 1843, gave some useful formulas involving infinite products and infinite series with related
convergence properties which maybe could have played a certain role in the 1859 Riemann paper in deducing some
properties of that functional equation related to his $\xi$ function (whose an earlier form was also known to
Euler over a hundred years before Riemann, and to which Euler had arrived in the real domain by use of divergent
series methods; see(Kolmogorov \& Yushkevich 1996, Volume II, Chapter 2) and (Bateman \& Diamond 2004, Chapter
8, Section 8.11)), also because of the simple fact that Riemann himself known very well Cauchy's work.
Therefore, Weierstrass himself acknowledges, in different places, his debit both to Gauss and Cauchy, in
achieving his celebrated results on entire function factorization theorem. Furthermore, following (Fou\"{e}t
1904-07, Tome II, Chapter IX, Section 272), many mathematicians have acknowledged in Abel one of the most
influential scholars who have contributed to the intellectual development of Weierstrass. On the other hand,
following (P\'{o}lya 1930), also J-B. Fourier, in (Fourier 1830, Exposé synoptique, Nos. (15) and (16) III$^e$
and IV$^e$, pp. 65-66), as early as the late 1820s, considered infinite products in algebraic questions inherent
transcendental equations of the type $\phi(x)=0$, with applications to the case $\sin x=0$, whose outcomes could
be therefore known to Riemann. To be precise, in the words of P\'{o}lya, he states a more general theorem
which it is worth quoting verbally as follows\\

\it<<III$^e$. Une fonction transcendante ou algébrique $\phi x$ étant proposée, si l'on fait l'énumération de
toutes les valeurs réelles ou imaginaires de $x$, savoir $\alpha, \beta, \gamma, \delta$, etc. qui rendent nulle
la fonction $f(x)$, et si l'on désigne par $f(x)$ le produit $\displaystyle\Big(1-\frac{x}{\alpha}\Big),
\Big(1-\frac{x}{\beta}\Big), \Big(1-\frac{x}{\gamma}\Big), \Big(1-\frac{x}{\delta}\Big),...$ de tous les
facteurs simples qui correspondent aux racines de l'equation $\phi x=0$, ce produit pourra différer de la
fonction $\phi x$, en sorte que cette fonction, au lieu d'étre équivalente $f x$, sera le produit d'un premier
facteur $f(x)$ par un second $F(x)$. Cela pourra arriver si le second facteur $F(x)$ ne cesse point d'étre une
grandeur finie, quelque valeur réelle ou imaginaire que l'on donne à $x$ , ou si ce second facteur $F(x)$ ne
devient nul que par la substitution de valeurs de $x$ qui rendent infini le premier facteur $f(x)$. Et
réciproquement si l'équation $F(x)$ des racines, et si elles ne rendent point infini le facteur $f(x)$, on est
assuré que le produit de tous les facteurs du premier degré correspondant aux racines de $\phi x=0$ équivaut à
cette fonction $\phi x$. En effet: $1^0$ s'il existait un facteur $fx$ qui ne pût devenir nul pour aucune valeur
réelle ou imaginaire de $x$, par exemple si $Fx$ était une constante $A$ et si $fx$ était $\sin x$, toutes les
racines de $A\sin x=0$ seraient celles de $\sin x=0$, et le produit de tous les facteurs simples correspondant
aux racines de $A\sin x=0$ serait seulement $\sin x$, et non $A\sin x$. Il en serait de méme si le facteur $Fx$
n'était pas une constante $A$. Mais s'il pouvait exister un facteur $Fx$ qui ne cesserait point d'avoir une
valeur finie, quelque valeur réelle ou imaginaire que l'on attribuât à $x$, toutes les racines de l'équation
$\sin xFx=0$ seraient celles de $\sin x=0$, puisqu'on ne pourrait rendre nul le produit $\sin x$. $Fx$ qu'en
rendant $\sin x$ nul. Donc le produit de tous les facteurs correspondants aux racines de $\phi x=0$ serait $\sin
x$, et non $\sin xFx$. On voit donc que dans ce second cas il serait possible que le produit de tous les
facteurs simples ne donnât pas $\phi x$.  $2^0$ Si l'équation $Fx=0$ a des racines, ou réelles ou imaginaires,
ce qui exclut le cas où $Fx$: serait une constante $A$, ou serait un facteur dont la valeur est toujours finie,
et si les racines de $Fx=0$ rendeut $fx$ infini, le produit $fxFx$ devient $0/0$, et peut avoir une valeur
très-différente de $fx$. Mais si les racines de $Fx=0$ donnent pour $fx$ une valeur finie, le produit $fxFx$
deviendrait nul lorsque $Fx=0$: donc l'enumération complète des racines de l'équation $\phi x=0$, ou $fxFx=0$,
comprendrait les racines de $Fx=0$. Or nous avons représenté par $fx$ le produit de tous les facteurs simples
qui répondent aux racines de $\phi x=0$: il serait donc contraire à l'hypothèse d'admettre qu'il y a un autre
facteur $Fx$, tel que les racines de $Fx=0$ sont aussi des facteurs de $\phi x=0$. Cela supposerait que l'on n a
pas fait une enumération complète des racines de l'équation $\phi x=0$, puisqu'on a exprimé seulement par $fx$
le produit des facteurs simples qui correspondent aux racines de cette équation.

IV$^e$. Étant proposée une équation algébrique ou transcendante $\phi x=0$ formée d'un nombre fini ou infini de
facteurs réels ou imaginaires $\displaystyle \Big(1-\frac{x}{\alpha}\Big), \Big(1-\frac{x}{\beta}\Big),
\Big(1-\frac{x}{\gamma}\Big), \Big(1-\frac{x}{\delta}\Big)$, etc. on trouve le nombre des racines imaginaires,
les limites des racines réelles, les valeurs de ces racines, par la méthode de résolution qui a été exposée dans
les premiers livres et qui sera la même soit que la différentiation répétée réduise $\phi x$ une valeur
constante, soit que la différentiation puisse être indéfiniment continuée. L'équation $\phi x=0$ a précisément
autant de racines imaginaires qu'il y a de valeurs réelles de $x$ qui , substituées dans une fonction dérivée
intermédiaire d'un ordre quelconque, rendent cette fonction nulle, et donnent deux résultats de même signe pour
la fonction dérivée qui la précède et pour celle qui la suit. Par conséquent si l'on parvient à prouver qu'il
n'y a aucune valeur réelle de $x$ qui, en faisant évanouir une fonction dérivée intermédiaire, donne le même
signe à celle qui la précède, et à celle qui la suit, on est assuré que la proposée ne peut avoir aucune racine
imaginaire. Par exemple en examinant l'origine de l'équation transcendante (1) $\displaystyle
0=1-\frac{x}{1}+\frac{x^2}{(1\cdot 2)^2}-\frac{x^3}{(1\cdot 2\cdot 3)^2}+\frac{x^4}{(1\cdot 2\cdot 3\cdot
4)^2}-$etc. nous avons prouvé qu'elle est formé du produit d'un nombre infini de facteurs; et en considérant une
certaine relation récurrente qui subsiste entre les coefficients des fonctions dérivées des divers ordres, on
reconnaît qu'il est impossible qu'une valeur réelle de $x$ substituée, dans trois fonctions dérivées
consécutives, réduise la fonction intermédiaire à zéro, et donne deux résultats de même signe pour la fonction
précédente et pour la fonction suivante. On en conclut avec certitude que l'équation (1) ne peut point avoir de
racines imaginaires>>.\\\\\rm P\'{o}lya says that no proof of this theorem, by Fourier or another mathematician,
is known. In 1841, M.A. Stern gave an invalid proof, and repeated in greater detail some affirmations of
Fourier. Since then, the question seems to have been neglected. However, to further emphasize the Weierstrass
work on entire function factorization, we report textual words of Picard (see (Picard 1897) and (Dugac 1973,
Section 5.1)), who is one of the founders of the theory of entire functions, as we will see later\\

\it <<L'illustre analyste a publié en 1876 un mémoire sur la Théorie des Fonctions uniformes; ce mémoire, en
faisant connaître à un public plus étendu les résultats développés depuis longtemps déjà dans l'enseignement du
maître, a été le point de départ d'un très grand nombre de travaux sur la Théorie des Fonctions. Cauchy et ses
disciples fran\c{c}ais, en étudiant les fonctions analytiques uniformes, n'avaient pas pénétré bien profondément
dans l'étude de ces points singuliers appelés ''points singuliers essentiels'', dont le point $z=0$ pour la
fonction $\exp(1/z)$ donne l'exemple le plus simple. Weierstrass, en approfondissant cette étude, a été conduit
à un résultat qui est un des plus admirables théorèmes de l'Analyse moderne, je veux parler de la décomposition
des fonctions entières en facteurs primaires. D'après le théorème fondamental de l'Algèbre, un polyn\^{o}me peut
\^{e}tre décomposé en un produit de facteurs linéaires; pour une fonction entière, c'est-à-dire pour une
fonction uniforme continue dans tout le plan telle que sins),ne peut-on chercher à obtenir aussi une
décomposition en facteurs? Cauchy avait obtenu sur ce sujet des résultats importants, mais sans le traiter dans
toute sa généralité. Il était réservé à Weierstrass de montrer qu'une fonction entière peut être décomposée en
un produit d'un nombre généralement infini de facteurs primaires, chacun de ceux-ci étant le produit d'un
facteur linéaire par une exponentielle de la forme $\exp(P(z))$, où $P(z)$ est un polyn\^{o}me. C'est sans doute
en étudiant l'intégrale Eulérienne de seconde espèce que Weierstrass a été mis sur la voie de ce beau théorème,
et nous rappellerons à ce sujet cet important résultat que l'inverse de cette intégrale est une fonction
entière>>.\\\\\rm Rolf Nevanlinna points out the main role played by Weierstrass in realizing a class of
elementary analytic functions, amongst which Abelian and elliptic functions, whose construction has led
Weierstrass to the creation of a general theory of entire and meromorphic functions of which one of the founding
pillars is just the theorem on the decomposition into primary factors (see (Dugac 1973, Section 5.1)). On the
other hand, Weierstrass himself was fully aware of the importance played by this result within the general
context of complex function theory. Furthermore, as (Kudryavtseva et al. 2005, Section 7) point out, the two
decades following the publication of the celebrated 1859 Riemann's paper, were largely uneventful. Weierstrass,
who was eleven years older than Riemann, but whose rise to fame - from an obscure schoolteacher to a professor
at Berlin - happened in a way very different from Riemann's one, began working and lecturing on complex numbers
and the general theory of entire functions already during the 1860s, but it wasn't until 1876, when Weierstrass
finally published his famous memoir, that mathematicians became aware of some of his revolutionary ideas and
results. Weierstrass' factorization theorem, together with Riemann's memoir, set the stage for the great work of
Hadamard and de la Valée-Poussin in the 1890s. In particular, Hadamard made more explicit and applicable
Weierstrass' factorization theorem.\\\\\bf 1.2 On the Riemann's contribution. \rm In 1858, Riemann wrote his
unique paper on number theory, which marked a revolution in mathematics. According to Laugwitz (1999,
Introduction, Sections 4.1 and 4.2), real and complex analysis has always influenced Riemann work: algebraic
geometry appears, in his works, as a part of complex analysis; he treats number theory with methods of complex
function theory; he subsumes physical applications into partial differential equations; he replaces the usual
axiomatic conception of geometry by his novel (Riemannian) geometry, which is a part or real analysis of several
variables; and he develops the topology of manifolds as a new discipline derived from analysis. Riemann knew the
elements of algebraic analysis according to J-L. Lagrange and L. Euler, through the lessons of his teacher, M.A.
Stern (1807-1894), who was one of the last schoolmasters of the subject. Riemann handled the gamma function in a
secure and self-confident way and has dealt with differential equations and recursions in the Euler's manner.
The Stern lessons were of very fundamental importance to achieve many Riemann's results, even if the celebrated
1748 Euler's \it Introductio in Analysin Infinitorum \rm was one of the most influential textbooks of the time.
Nevertheless, Klein (1979, Chapter VI) states that Riemann began already to study elliptic and Abelian functions
since the late 1840s, because this subject, in the meantime, has become of a certain vogue in Germany. In the
1855-56 winter term, following the Dirichlet's research lines, Riemann lectured on functions of a complex
quantity, in particular elliptic and Abelian functions, while in the 1856-57 winter term he lectured on the same
subject, but now with special regard to hypergeometric series and related transcendentals. These lectures, from
which he drew publications on Abelian and hypergeometric functions, were partially repeated in the following
semesters. Klein (1979, Chapter VI) points out that the years 1857-62 marked the high-point of Riemann's
creativity. Moreover, Klein states that before to characterize the specific Riemannian function theory work, he
wishes to put forward a remark that may cause some surprise: Riemann did much important work in the theory of
functions that does not fit into the framework of his typical theory. Klein refers to the notable 1859 paper on
the number of primes lower than a given magnitude, where it is introduced \it <<the Riemann zeta-function
$\zeta(\sigma+it)$ given by an analytic expression, namely an infinite product. This product is converted into a
definite integral, which can then be evaluated by shifting the path of integration. The whole procedure is
function theory à la Cauchy>>. \rm Therefore, according to what Felix Klein states, the mathematical background
that was at the basis of the Riemannian analytic treatment of his $\zeta$ function, essentially lies on the
Cauchy's theory of complex functions. This is also confirmed by (Bottazzini \& Gray 2013, Chapter 5, Section
5.1), coherently with what has just been said above in regards to the importance played by Cauchy's work on the
Riemann's one.

According to (Laugwitz 1999, Chapter 1), notwithstanding the era of ferment that concerned the 19th century
mathematics, an autonomous and systematic account of the foundations of complex analysis is findable, for the
first time, in the Riemann's works and lecture notes through the winter term 1855-56 to the winter term 1861-62,
the latter having been published by the physicist Carl Ernst Abbe (1840-1905) in the summer term 1861 (see
(Ullrich 2003) and references therein) when he was a student of W. Weber and Riemann in G\"{o}ttingen. The only
systematic and congruous historical attempt to organically recognize the various Riemann's lessons has been
pursued by E. Neuenschwander in (Neuenschwander 1996). In any case, Riemann was fully imbedded into the real and
complex analysis scenery of the first middle of the 19th century, which seen involved the outstanding figures of
Cauchy, Weierstrass and Riemann himself, whose researches were intertwined amongst them more times. According to
(Laugwitz 1999, Chapter 1, Section 1.1.5), just in connection with the drawing up of his paper on the same
subject, Riemann was aware of the Weierstrass' papers on Abelian functions wrote between 1853 and 1856-57, for
which it is evident that a certain influence of the latter on the Riemann's one - at least, as concerns such a
period - there was, even if Weierstrass will publish these his works only later. Again following (Neuenschwander
1996), one of the key themes of the last \it Sommersemestern \rm 1861 Riemann lectures on analytic functions
(see sections 11-13), was the determination of a complex function from its singularities (section 13) mainly
following Cauchy's treatment\footnote{Amongst the first lecture notes on complex functions very close to the
Riemann ideas and approach, there are those exposed in (Durège, 1896).}. Then, he clarifies that this problem
regards only single-valued functions defined on $\mathbb{C}\cup\{0\}$ whose unique singularities are poles (the
names \it pole \rm and \it essential singularity \rm are respectively due to C.A. Briot and J.C. Bouquet and to
Weierstrass). In turn, the resolution of this problem requires the previous knowledge of the zeros of the
function which has to be determined. At first, Riemann considered the case of a function having a finite number
of zeros and poles, then he went over the the next question, namely to determine a function with infinitely many
zeros whose unique point of accumulation if $\infty$ (which, inter alia, concerns too the Riemann zeta function
theory). But, again following (Laugwitz 1999, Chapter 1, Section 1.1.6), in doing so Riemann went over very
close to the next Weierstrass' work on the infinite product representation of an entire function, using special
cases to explain the general procedure. Detlef Laugwitz points out that Riemann has pursued this latter task in
such a way that, by following his directions, one could immediately give a proof of the known Weierstrass'
product theorem, even if Riemann ultimately failed in reaching the general case; nevertheless, for what follows,
this last claim has a certain importance from our historical standpoint.

Indeed, in his renowned paper on the distribution of prime numbers\footnote{This paper was presented by Riemann,
after his nomination as full professor in July 1859, to the Berlin Academy for his consequent election as a
corresponding member of this latter. To be precise, following (Bottazzini 2003), just due to this election,
Riemann and Dedekind visited Berlin, where they met E.E. Kummer, L. Kronecker and K. Weierstrass. According to
(Dedekind 1876), very likely, it was just from this meeting that sprung out of the celebrated 1859 Riemann
number theory paper that was, then, sent to Weierstrass himself, to be published in the November issue of the
\it Monatsberichte der Berliner Akademie. \rm Furthermore, and this is an important witness for our ends,
according to the mathematician Paul B. Garrett (see his Number Theory lessons at
http://www.math.umn.edu/~garrett/), general factorizations of entire functions in terms of their zeros are due
to K. Weierstrass, but sharper conclusions from growth estimates are due to J. Hadamard. In relation to his 1859
memoir, Riemann's presumed existence of a factorization for $\xi$ function to see the connection between prime
numbers and complex zeros of zeta function, was a significant impetus to Weierstrass' and Hadamard's study of
products in succeeding decades.}, Riemann stated the following function
\begin{equation}\xi(t)\doteq\Big(\frac{1}{2}\Gamma\Big(\frac{s}{2}\Big)s(s-1)\pi^{-s/2}\zeta(s)
\Big)_{s=1/2+it}\end{equation} later called \it Riemann $\xi$-function. \rm It is an entire function (see
(Titchmarsh 1986, Chapter II, Section 2.12)). Riemann conjectured that $(\xi(t)=0)\Rightarrow(\Im(t)=0)$, that
is to say, the famous \it Riemann hypothesis \rm (RH), as it will be called later. Whereupon, he stated that\\

\it <<This function $\xi(t)$ is finite for all finite values of $t$, and allows itself to be developed in powers
of $tt$ as a very rapidly converging series. Since, for a value of $s$ whose real part is greater than 1,
$\log\zeta(s)=-\sum\log(1-p^{-s})$ remains finite, and since the same holds for the logarithms of the other
factors of $\xi(t)$, it follows that the function $\xi(t)$ can only vanish if the imaginary part of $t$ lies
between $i/2$ and $-i/2$. The number of roots of $\xi(t)=0$, whose real parts lie between 0 and $T$ is
approximately $=(T/2)\log (T/2)\pi-T/2\pi$; because the integral $\int d\log\xi(t)$, taken in a positive sense
around the region consisting of the values of $t$ whose imaginary parts lie between $i/2$ and $-i/2$ and whose
real parts lie between 0 and $T$, is (up to a fraction of the order of magnitude of the quantity $1/T$) equal to
$(T\log(T/2)\pi-T/2\pi)i$; this integral however is equal to the number of roots of $\xi(t)=0$ lying within this
region, multiplied by $2\pi i$. One now finds indeed approximately this number of real roots within these
limits, and it is very probable that all roots are real. Certainly one would wish for a stricter proof here; I
have meanwhile temporarily put aside the search for this after some fleeting futile attempts, as it appears
unnecessary for the next objective of my investigation. If one denotes by $\alpha$ all the roots of the equation
$\xi(t)=0$, one can express $\log\xi(t)$ as
\begin{equation}\sum\log\Big(1-\frac{tt}{\alpha\alpha}\Big)+\log\xi(0)\end{equation}for, since the density of the roots
of the quantity $t$ grows with $t$ only as $\log t/2\pi$, it follows that this expression converges and becomes
for an infinite $t$ only infinite as $t\log\xi(t)$; thus it differs from $\log\xi(t)$ by a function of $tt$,
that for a finite $t$ remains continuous and finite and, when divided by $tt$, becomes infinitely small for
infinite $t$. This difference is consequently a constant, whose value can be determined through setting $t = 0$.
With the assistance of these methods, the number of prime numbers that are smaller than $x$ can now be
determined>>.\\\\\rm So, in his celebrated 1859 paper, Riemann himself had already reached an infinite product
factorization of $\xi(t)$, namely the (2), which can be equivalently written as follows
\begin{equation}\log\xi(t)=\sum_{\alpha}\log\Big(1-\frac{tt}{\alpha\alpha}\Big)+\log\xi(0)=\log\xi(0)\prod_{\alpha}
\Big(1-\frac{t^2}{\alpha^2}\Big)\end{equation}from which it follows that
$\xi(t)=\xi(0)\prod_{\alpha}\big(1-t^2/\alpha^2)$. Thus, questions related to entire function factorizations had
already been foreshadowed in this Riemann work. Therefore, we now wish to outline the main points concerning the
very early history of entire function factorization theorems, having taken the 1859 Riemann paper as an
occasional starting point of this historical question, in which, inter alia, a particular entire function
factorization - i.e. the (3) - had been used. In short, this 1859 Riemann paper has been a valuable
$\kappa\alpha\iota\rho\acute{o}\varsigma$ to begin to undertake one of the many study's branch which may depart
from this milestone of the history of mathematics, to be precise that branch concerning the entire function
theory which runs parallel to certain aspects of the theory of Riemann zeta function, with interesting meet
points with physics. One of the very few references which allude to these Riemann paper aspects is the article
by W.F. Osgood in (Burkhardt et al. 1899-1927, Zweiter Teil, B.1.III, pp. 79-80), where, discussing of the genus
of an entire function, an infinite product expansion of the function $\sin\pi s/\pi s$ is considered; to be
precise, since Johann Bernoulli to Euler, the following form\footnote{As $n\rightarrow\infty$ and $t\neq 0$,
Weierstrass proved to be $\sin\pi t/\pi t=\prod_{n=-\infty}^{\infty}(1-{t}/{n})e^{t/n}$ - see (Bellacchi 1894,
Chapter XI). But, according to (Bellacchi 1894, Chapter XI) and (Hancock 1910, Chapter I, Arts. 13, 14), Cauchy
was the first to have treated (in the \it Exercises de Mathématiques, IV\rm) the subject of decomposition into
prime factors of circular functions and related convergence questions, from a more general standpoint. Although
Cauchy did not complete the theory, he however recognized that, if $a$ is a root of an integral (or entire)
transcendental function $f(s)$, then it is necessary, in many cases, to join to the product of the infinite
number of factors such as $(1-s/a)$, a certain exponential factor $e^{P(s)}$, where $P(s)$ is a power series in
positive powers of $s$. Weierstrass gave then a complete treatment of this subject. On the other hand, besides
what has already been said above, also in (Greenhill 1892, Chapter IX, Section 258)) it is pointed out that,
since Abel's work, the infinite product expansions of trigonometric functions have been formal models from which
to draw inspiration, by analogy, for further generalizations or extensions. Analogously, following (Fou\"{e}t
1904-07, Tome II, Chapter IX, Section II, Number 279), \it <<Cauchy \rm[in the \it Anciens Exercices de
Mathématiques, 1829-1830\rm] \it avait vu que, pour obtenir certaines transcendantes, il fallait multiplier le
produit des binomes du premier degré du type $X-a_n$ par une exponentielle $e^{g(x)}$, $g(x)$ désignant une
fonction entière. Mème l'introduction de cette exponentielle ne suffit à donner l'expression générale des
fonctions admettant les zéros $a_1,..., a_n,...$ que dans le cas où la série $\sum_n|a_n|^{-1}$ converge.
converge. L'étude du développement des (Weierstrass) fonctions $\mathcal{P}$ et $\sigma$ en produits infinis
amena Weierstrass à s'occuper de cette question et fut ainsi l'occasion d'une de ses plus belles découvertes
\rm[see, for example, (Lang 1987, Chapter 1, Section 2)]. \it Ces recherches, exposées en 1874 par Weierstrass
dans son cours à l'Université de Berlin, ont été publiées dans un Mémoire fondamental ''Zur Théorie der
eindeutigen analytischen Functionen'' de 1876. Quinze ans auparavant, Betti, dans ses \rm Le\c{c}ons \it à
l'Université de Pise (1859-1860), avait traité un problème analogue à celui résolu par Weierstrass, mais sans
apercevoir toutes les conséquences de sa découverte; il en fit l'application au développement des fonctions
eulériennes, trigonométriqucs et elliptiques, puis, laissant son Mémoire dans 1860 inachevé, il n'y pensa
plus>>.} had already been deduced (see (Bellacchi 1894, Chapter XI))$$\frac{\sin\pi s}{\pi
s}=\prod_{n=1}^{\infty}\Big(1-\frac{s^2}{n^2}\Big)$$ which has genus 0. This is often said to be the \it Euler's
product formula \rm (see (Borel 1900, Chapter IV), (Tricomi 1968, Chapter IV, Section 8), (Remmert 1998, Chapter
1, Section 3) and (Maz'ya \& Shaposhnikova 1998, Chapter 1, Section 1.10)), and might be considered as one of
the first meaningful instances of infinite product expansion of an entire function, given by Euler in his 1748
\it Introductio in Analysin Infinitorum, \rm through elementary analytic methods (see (Sansone 1972, Chapter IV,
Section 1)). Furthermore, following (Fou\"{e}t 1904-07, Chapter IX, Section III, Number 286), there have always
been a close analogical comparison between the trigonometric functions and the Eulerian integrals (amongst which
the one involved in the Gamma function) together their properties, a way followed, for instance, by E. Heine.
Also looking at the Riemann's lectures on function theory through the 1855-56, 1856-57, 1857-58 \it
Wintersemestern \rm to the 1858-59 and 1861 \it Sommersemestern \rm lessons - see (Neuenschwander 1996, Section
13) as regards the last ones - it would be possible to descry as well some Riemann's attempts to consider
factorization product expansions whose forms seem to suggest, by analogy, a formula similar to (3). Moreover,
also in the 1847 Eisenstein paper, surely known to Riemann, there is also a certain lot of work devoted to the
study of the Euler's sine product formulas (see (Ebbinghaus et al. 1991, Chapter 5)) which perhaps have could
contribute to stimulate the Riemann insight in finding some formulas used in his 1859 celebrated paper, first of
all the (3). To be precise, because of the close friendship between Eisenstein and Riemann, seen too what is
said in (Weil 1989) (see also (Weil 1976)) about the sure influence of Eisenstein work on Riemann one, reaching
to suppose that 1859 Riemann paper was just originated by Eisenstein influence, we would be inclined to put
forwards the historical hypothesis that the deep and complete analysis and critical discussion of infinite
products pursued in the long and rich paper\footnote{Where, amongst other things, already a wide use of
exponential factors was made for convergence reasons related to infinite products.} (Eisenstein 1847), surely
played a decisive role in the dawning of $\xi$ product expansion of Riemann paper, also thanks to the great
mathematical insight of Riemann in extending and generalizing previous mathematical contexts in others.
Furthermore, following (Genocchi 1860, N. 2), an infinite product factorization of the $\xi$ function could be
easily deduced from what is said in (Briot \& Bouquet 1859, Book IV, Chapter II) about infinite product
factorizations\footnote{It is noteworthy to highlight some historical aspects of Charles Briot (1817-1882)
mathematical works which started in the mathematical physics context in studying the mathematical properties of
light propagation in a crystallin medium, like the Ether (as it was supposed to be at that time, until the
advent of Einstein's relativity), undertaking those symmetry conditions (chirality) soon discovered by Louis
Pasteur about certain chemical crystalline substances. In this regard, Briot published, with Claude Bouquet, a
series of three research memoirs on the theory of complex functions first published in the Journal de
l'\'{E}cole Polytechniquein then collected into a unique monograph published by Bachelier in Paris in 1856 (see
also (Bottazzini \& Gray 2013) as well as the e-archive http://gallica.bnf.fr for a complete view of all the
related bibliographical items), to which more enlarged and complete treatises will follow later either in pure
and applied mathematics (see also (Briot \& Bouquet 1859)) as well as in physics.}. Following (Stopple 2003,
Chapter 6, Section 6.1), Euler's idea is to write the function $\sin\pi x/\pi x$ as a product over its zeros,
analogous to factoring a polynomial in terms of its roots. For example, if a quadratic polynomial
$f(x)=ax^2+bx+c$ has roots $\alpha,\beta$ different from $0$, then we can write $f(x)=c(1-x/\alpha)(1-x/\beta)$.
On the other hand, $\sin\pi x=0$ when $x=0, \pm 1, \pm 2, ...$ and since $\sin\pi x/x=1-\pi^2x^2/6+O(x^4)$,
$\sin\pi x/\pi x\underset{x\rightarrow 0}{\rightarrow}1$ and $\sin\pi x/\pi x=0$ when $x=\pm 1, \pm 2, ...$,
Euler guessed that $\sin\pi x/\pi x$ could have a factorization as an infinite product of the type (see also
(Ebbinghaus et al. 1991, Chapter 5))
$$\frac{\sin\pi x}{\pi x}=\Big(1-\frac{x}{1}\Big)\Big(1+\frac{x}{1}
\Big)\Big(1-\frac{x}{2}\Big)\Big(1+\frac{x}{2}\Big)...=$$
$$=\Big(1-\frac{x^2}{1}\Big)\Big(1-\frac{x^2}{4}\Big)\Big(1-\frac{x^2}{9}\Big)\Big(1-\frac{x^2}{16}\Big)...$$which
will lead later to a valid proof of this factorization. Then, even in the context of the history of entire
function factorization theorems, W.F. Osgood points out that already Riemann, just in his famous 1859 paper, had
considered an entire function, the $\xi(s)$, as a function of $s^2$ with genus 0, taking into account the above
mentioned Euler product formula for the sine function but without giving any rigorous prove of this fact, thing
that will be done later by J. Hadamard in 1893 as a by-product of his previous 1892 outcomes on entire function
theory. In the next sections, when we will deepen the works of Hadamard and P\'{o}lya on the entire function
theory related to Riemann zeta function, we also will try to clarify, as far as possible, these latter aspects
of the 1859 Riemann paper which mainly constitute one of the central cores of the present work. Following
(Cartier 1993, I.1.d),\\

\it<<Concernant les zéros de la fonction $\zeta$, on doit à Riemann deux résultats fondamentaux dans son mémoire
de 1859. Tout d'abord, Riemann ajoute un facteur $s(s-1)$ dans la fonction $\zeta(s)$; cela ne détruit pas
l'équation fonctionnelle, mais lui permet d'obtenir une fonction entière
$$\xi(s)=\pi^{-s/2}\Gamma(s/2)\zeta(s)s(s-1)$$car les deux pôles sont compensés (Aujourd'hui, on préfère garder
la fonction méromorphe). Grâce à l'équation fonctionnelle, on montre facilement que les zéros de la fonction
$\zeta$ sont situés dans la bande critique $0 <\Re s < 1$. Il est de tradition, depuis Riemann, d'appeler $\rho$
ces zéros. La fonction $\xi$ est désormais une fonction entière; si l'on connaît l'ensemble de ses zéros, on
doit pouvoir la reconstituer. Riemann affirme alors que $\xi(s)$ s'écrit sous forme du produit d'une constante
par un produit infini qui parcourt tous les zéros de la fonction $\xi$, chaque facteur s'annulant pour le zéro
$s=\rho$ correspondant de $\zeta(s)$. Bien entendu, ce produit infini diverge mais - et c'est un point important
- il converge si on le rend symétrique, i.e. si l'on regroupe judicieusement les facteurs. L'équation
fonctionnelle montre en effet qu'on peut associer $\`{a}$ tout nombre $\rho$ le nombre $1-\rho$ qui en est le
symétrique, géométriquement, par rapport $\`{a}$ $1/2$. De ce fait, si l'on regroupe dans ce produit infini le
facteur correspondant \`{a} $\rho$ et le facteur correspondant à $1-\rho$, on obtient un produit infini
absolument convergent (le prime signifie que l'on ne prend qu'une fois chaque paire $\rho,1-\rho$)
$$\xi(s)\stackrel{\footnote{The product should be extended to every root with its conjugate.}}{=}
c\prod_{\rho}\Big(1-\frac{s}{\rho}\Big)\Big(1-\frac{s}{1-\rho}\Big).$$Le premier problème majeur, dans le
mémoire de Riemann de 1859, était de démontrer cette formule; il l'énonce, mais les justifications qu'il en
donne sont très insuffisantes. L'objectif de Riemann est d'utiliser cette formule du produit pour en déduire des
estimations très précises sur la répartition des nombres premiers. Si l'on note, suivant la tradition, $\pi(x)$
le nombre (Anzahl) de nombres (Zahlen) premiers $p<x$, Legendre (1788) et Gauss (en 1792, mais jamais publié)
avaient conjecturé qu'on avait $\pi(x)\sim(x/\ln x)$. Riemann donne des formules encore plus précises au moyen
de sommations portant sur les zéros et les $\pi(x)$. En fait, il a fallu près de quarante ans pour que Hadamard
(1896) et, indépendamment, de la Vallée-Poussin (1896) démontrent rigoureusement cette formule de développement
en produit infini au moyen d'une théorie générale de la factorisation des fonctions entières - par des arguments
qui étaient essentiellement connus d'Euler et de Riemann, en tout cas certainement de Riemann - et justifient
ainsi rigoureusement la loi de répartition des nombres premiers. Hadamard donne la forme
$\lim_{x\rightarrow\infty}(x/\ln x)$, et de la Vallée-Poussin donne la forme plus forte
$\displaystyle\pi(x)=\int_2^x\frac{dt}{\ln t} + O\big(x e^{-c\sqrt{\ln x}}\big)$ pour une constante
$c>0$>>.\\\\\rm Following (Stopple 2003, Chapter 10, Section 10.1), it was Riemann to realize that a product
formula for $\xi(s)$ would have had a great significance for the study of prime numbers. The first rigorous
proof of this product formula was due to Hadamard but, as himself remember, it took almost three decades before
he reached to a satisfactory proof of it. Likewise, also H.M. Edwards (1974, Chapter 1, Sections 1.8-1.19)
affirms that the parts concerned with (2) are the most difficult portion of the 1859 Riemann's paper (see also
(Bottazzini \& Gray 2013, Chapter 5, Section 5.10)). Their goal is to prove essentially that $\xi(s)$ can be
expressed as an infinite product, stating that\\

\it <<\rm[...] \it any polynomial $p(t)$ can be expanded as a finite product $p(t)=p(0)\prod_{\rho}(1-t/\rho)$
where $\rho$ ranges over the roots of the equation $p(t)=0$ \rm[\it except that the product formula for $p(t)$
is slightly different if $p(0) = 0$\rm]; \it hence the product formula (2) states that $\xi(t)$ is like a
polynomial of infinite degree. Similarly, Euler thought of $\sin x$ as a ''polynomial of infinite
degree\footnote{Following (Bottazzini \& Gray 2013, Chapter 8, Section 8.5.1), amongst the functions that behave
very like a polynomial, there is the Riemann $\xi$ function. In this regard, see also what will be said in the
next sections about Lee-Yang theorems and, in general, the theory concerning the location of the zeros of
polynomials.}'' when he conjectured, and finally proved, the formula $\sin x=\pi
x\prod_{n\in\mathbb{N}}\big(1-(x/n)^2\big)$. On other hand, \rm[...] \it $\xi(t)$ is like a polynomial of
infinite degree, of which a finite number of its terms gives a very good approximation in any finite part of the
plane. \rm [...] \it Hadamard (in 1893) proved necessary and sufficient conditions for the validity of the
product formula $\xi(t)=\xi(0)\prod_{\rho}(1-t/\rho)$ but the steps of the argument by which Riemann went from
the one to the other are obscure, to say the very least>>.\\\\\rm The last sentence of this Edwards' quotation
is historically quite interesting and would deserve further attention and investigation. Furthermore, H.M.
Edwards states too that\\

\it <<\rm [...] \it a recurrent theme in Riemann's work is the global characterization of analytic functions by
their singularities. See, for example, the Inauguraldissertation, especially Article 20 of Riemann's Werke (pp.
37-39) or Part 3 of the introduction to the Riemann article ''Theorie der Abel'schen Functionen'', which is
entitled ''Determination of a function of a complex variable by boundary values and singularities''. See also
Riemann's introduction to Paper XI of the his collected works, where he writes about '' \rm [...] \it our
method, which is based on the determination of functions by means of their singularities (Umtetigkeiten und
Unendlichwerden) \rm[...]''. \it Finally, see the textbook (Ahlfors 1979), namely the section 4.5 of Chapter 8,
entitled ''Riemann's Point of View''>>,\\\\\rm according to which Riemann was therefore a strong proponent of
the idea that an analytic function can be defined by its singularities and general properties, just as well as
or perhaps better than through an explicit expression, in this regard showing, with Riemann, that the solutions
of a hypergeometric differential equation can be characterized by properties of this type. In short, all this
strongly suggests us the need for a deeper re-analysis of Riemann \it \oe uvre \rm concerning these latter
arguments, as well as a historical seek for the mathematical background which was at the origins of his
celebrated 1859 number theory paper. From what has just been said, it turns out clear that a look at the history
of entire function theory, within the general and wider complex function theory framework, is needed to clarify
some of the historical aspects of this influential seminal paper which, as Riemann himself recognized, presented
some obscure points. In this regard, also Gabriele Torelli (see (Torelli 1901, Chapter VIII, Sections 60-64))
claimed this last aspect, pointing out, in particular, the Riemann's ansätz according to which the entire
function $\xi(t)$ is equal, via (3), to the Weierstrass' infinite product of primary factors without any
exponential factor. As is well-known, this basic question will be brilliantly solved by J. Hadamard in his
famous 1893 paper that, inter alia, will mark a crucial moment in the history of entire function theory (see
(Maz'ya \& Shaposhnikova 1998, Chapter 9, Section 9.2) and next sections).\\\\\bf 1.3 An historical account of
entire factorization theorems from Weierstrass onward. \rm To begin, we wish to preliminarily follow the basic
textbook on complex analysis of Giulio Vivanti (1859-1949), an Italian mathematician whose main research field
was into complex analysis, becoming an expert of the entire function theory. He wrote some notable treatises on
entire, modular and polyhedral analytic functions: a first edition of a prominent treatise on analytic functions
appeared in 1901, under the title \it Teoria delle funzioni analitiche, \rm published by Ulrico Hoepli in Milan,
where the first elements of the theory of analytic functions, worked out in the late 19th-century quarter, are
masterfully exposed into three main parts, giving a certain load to the Weierstrass' approach respect to the
Cauchy's and Riemann's ones. The importance of this work immediately arose, so that a German edition was carried
out, in collaboration with A. Gutzmer, and published in 1906 by B.G. Teubner in Leipzig, under the title \it
Theorie der eindeutigen analytischen funktionen. Umarbeitung unter mitwirkung des verfassers deutsch
herausgegeben von A. Gutzmer \rm (see (Vivanti 1906)), which had to be considered as a kind of second enlarged
and revised edition of the 1901 first Italian edition according to what Vivanti himself said in the preface to
the 1928 second Italian edition, entitled \it Elementi della teoria delle funzioni analitiche e delle
trascendenti intere, \rm again published by Ulrico Hoepli in Milan, and wrote following the above German edition
in which many new and further arguments and results were added, also as regards entire functions. Almost all the
Vivanti's treatises are characterized by the presence of a detailed and complete bibliographical account of the
related literature, this showing the great historical attention that he always put in drawing up his works.
Therefore, he also was a valid historian of mathematics besides to be an able researcher (see (Janovitz \&
Mercanti 2008, Chapter 1) and references therein), so that his works are precious sources for historical
studies, in our case as concerns entire functions. The above mentioned Vivanti's textbook on complex analysis
has been one of the most influential Italian treatises on the subject. It has also had wide international fame
thanks to its German edition.

Roughly speaking, the transcendental entire functions may be formally considered as a generalization, in the
complex field, of polynomial functions (see (Montel 1932, Introduction) and (Levin 1980, Chapter I, Section 3)).
Following (Vivanti 1928, Sections 134-135), (Marku\v{s}evi\v{c} 1988, Chapter VII) and (Pierpont 1914, Chapter
VIII, Sections 127 and 140), the great analogy subsisting between these two last function classes suggested the
search for an equal formal analogy between the corresponding chief properties. To be precise, the main
properties of polynomials concerned either with the existence of zeros (Gauss' theorem) and the linear factor
decomposition of a polynomial, so that it was quite obvious trying to see whether these could be, in a certain
way, extended to entire functions. As regards the Gauss' theorem, it was immediately realized that it couldn't
subsist because of the simple counterexample given by the fundamental elementary transcendental function $e^x$
which does not have any zero in the whole of complex plane. On the other hand, just this last function will
provide the basis for building up the most general entire function which is never zero, which has the general
form $e^{G(x)}$, where $G(x)$ is an arbitrary entire function, and is said to be an exponential factor. Then,
the next problem consisted in finding those entire functions having zeros and hence how it is possible to build
up them from their zero set. In this regard, it is well-known that, if $P(z)$ is an arbitrary non-zero
polynomial with zeros $z_1,...,z_n\in\mathbb{C}\setminus{0}$, having $z=0$ as a zero with multiplicity $\lambda$
(supposing $\lambda=0$ if $P(0)=0$), then we have the following well-known finite product
factorization\footnote{It is noteworthy the historical fact pointed out by Giuseppe Bagnera (1927, Chapter III,
Section 12, Number 73), in agreement to what has been likewise said above, according to which already Cauchy
himself had considered first forms of infinite product developments, after the Euler's work. Also Bagnera then,
in this his work, quotes Betti's work on elliptic functions and related factorization theorems. Instead, it is
quite strange that the Italian mathematician Giacomo Bellacchi (1838-1924) does not cite Betti, in his notable
historical work on elliptic functions (Bellacchi 1894) in regards to entire function factorization theorems
which are treated in the last chapter of this his work; this is also even more strange because Chapter XI of his
book is centered around the 1851 Riemann dissertation on complex function theory, without quoting the already
existed Italian translation just due to Betti. Furthermore, Bellacchi studied at the \it Scuola Normale
Superiore \rm of Pisa in the 1860s, for which it is impossible that he had not known Betti (see (Maroni 1924)).
On the other hand, also (Loria 1950, Chapter XLIV, Section 741) refers that Weierstrass found inspiration for
his factorization theorem, a result of uncommon importance according to Gino Loria, generalizing a previous
Cauchy's formula: indeed, both Cauchy and Gauss are quoted at p. 120 of the 1879 French translation of the
original 1876 Weierstrass paper. This, to further confirmation of what has been said above.}
\begin{equation}P(z)=Cz^{\lambda}\prod_{j=1}^n\Big(1-\frac{z}{z_j}\Big)\end{equation}where $C\in\mathbb{C}
\setminus{0}$ is a constant, so that a polynomial, except a constant factor, may be determined by its zeros. For
transcendental entire functions, this last property is much more articulated respect to the polynomial case:
indeed, whilst the indeterminacy for polynomials is given by a constant $C$, for transcendental entire functions
it is larger and related to the presence of an exponential factor which is need to be added to warrant the
convergence of infinite product development. A great part of history of the approach and resolution of this last
problem is the history of entire function factorization. Nevertheless, we also wish to report what says Giacomo
Bellacchi (1894, Chapter XI, Section 98) about this last problem. To be precise, he states that\\

\it <<Se $a_1, a_2, a_3,..., a_n,....$ simboleggino le radici semplici di una funzione olomorfa $f(z)$, ed il
quoziente $f(z):\prod (z-a_n)$ non si annulli per alcuna di esse, la sua derivata logaritmica
$\psi'(z)=f'(z)/f(z)-\sum (1/(z-a_n))$ è olomorfa in tutto il piano; moltiplicando i due membri per $dz$ ed
integrando, Cauchy giunse alla formula $f(z)=Ce^{\psi(z)}\prod (1-z/a_n)$, dove $C$ è una costante>>\\

\rm[\it <<If $a_1, a_2, a_3,..., a_n,....$ represent the simple roots of a holomorphic function $f(z)$, and the
ratio $f(z):\prod (z-a_n)$ is not zero for each root, then its logarithmic derivative $\psi'(z)=f'(z)/f(z)-\sum
(1/(z-a_n))$ is holomorphic in the whole of plane; multiplying both sides by $dz$ and integrating, Cauchy
reached the formula $f(z)=Ce^{\psi(z)}\prod (1-z/a_n)$, where $C$ is a constant>>\rm],\\\\so that it seems,
according to Bellacchi, that already Cauchy had descried the utility of exponentials as convergence-producing
factors, in a series of his papers published in the Tome XVII of the \it Comptes Rendus de l'Acad\'{e}mie des
Sciences (France)\rm; this supposition is also confirmed by Hancock (1910, Chapter I, Art. 14). Nevertheless,
following (Vivanti 1928, Sections 135-141), the rise of the first explicit formulation of the entire function
factorization theorem was given by Weierstrass in 1876 (see (Weierstrass 1876)) and was mainly motivated by the
purpose to give a solution to the latter formal problem, concerning the convergence of the infinite product
development of a transcendental entire function $f(z)$ having an infinite number of zeros, namely $z=0$, with
multiplicity $\lambda$, and $z_1,...,z_n,...$ such that $0<|z_j|\leq|z_{j+1}|, z_j\neq z_{j+1}\ \ j=1,2,...,$
trying to extend the case related to a finite number of zeros $z_1,...,z_n$, in which such a factorization is
given by
\begin{equation}f(z)=e^{g(z)}z^{\lambda}\prod_{j=1}^n\Big(1-\frac{z}{z_j}\Big),\end{equation}
to the case of infinite zeros, reasoning, by analogy, as follows. The set of infinite zeros $z_j$ is a countable
set having only one accumulation point, that at infinite. Therefore, for every infinite increasing natural
number sequence $\{\rho_i\}_{i\in\mathbb{N}}$, it will be always possible to arrange the zeros $z_j$ according
to their modulus in such a manner to have the following non-decreasing sequence $|z_1|\leq|z_2|\leq ...$ with
$\lim_{n\rightarrow\infty}|z_n|=\infty$. In such a case, if one wants, by analogy, to extend (5) as follows
\begin{equation}f(z)=e^{g(z)}z^{\lambda}\prod_{j=1}^\infty\Big(1-\frac{z}{z_j}\Big),\end{equation}then it will not be
possible to fully avoid divergence's problems inherent to the related infinite product. The first hint towards a
possible overcoming of these difficulties, was suggested to Weierstrass (see (Weierstrass 1856a)) by looking at
the form of the inverse of the Euler integral of the second kind\footnote{Following (Amerio 1982-2000, Volume 3,
Part I), the first historical prototype of the Euler integral of the first kind was provided by the so-called
\it Beta function, \rm whilst the first historical prototype of the Euler integral of the second type was
provided by Gamma function.} - that is to say, the \it gamma function \rm - and given by
\begin{equation}\frac{1}{\Gamma(z)}=z\prod_{j=1}^{\infty}\Big(1+\frac{z}{j}\Big)\Big(\frac{j}{1+j}\Big)^z=
z\prod_{j=1}^{\infty}\Big(1+\frac{z}{j}\Big)e^{-z\log\frac{j+1}{j}},\end{equation}from which he descried the
possible utility of the exponential factors there involved to, as the saying goes, force the convergence of the
infinite product of the last equality; these his ideas concretized only in 1876 with the explicit formulation of
his celebrated theorem on the entire function factorization.

As we have said above, Weierstrass (1856a) attributes, however, the infinite product expansion (7) to Gauss, but
some next historical studies attribute to Euler this formula, that he gave in the famous 1748 \it Introductio in
Analysin Infinitorum. \rm Indeed, as has been said above, from the 1879 French translation of the original 1876
Weierstrass paper, it turns out that both Cauchy and Gauss are quoted (at page 120), before to introduce the
primary factors. Nevertheless, P. Ullrich (1989, Section 3.5) says that the real motivation to these
Weierstrass' results about entire function factorization were mainly due to attempts to characterize the
factorization of quotients of meromorphic functions on the basis of their zero sets, rather than to solve the
above problem related to the factorization of a polynomial in dependence on its zeros. Furthermore, Ullrich
(1989, Section 3.5) observes too that other mathematicians dealt with questions concerning entire function
factorization methods, amongst whom are just Enrico Betti and Bernard Riemann, the latter, in his important 1861
\it sommersemestern \rm lectures on analytic functions, arguing, as has already been said, upon the construction
of particular complex functions with simple zeros, even if, all things considered, he didn't give, according to
Ullrich (1989, Section 3.5), nothing more what Euler done about gamma function through 1729 to his celebrated
1748 treatise on infinitesimal analysis\footnote{Following (Lunts 1950), (Marku\v{s}eci\v{c} 1988, Chapter VII)
and references therein, also Loba\v{c}evskij, since 1830s, made some notable studies on gamma function which
preempted times.}. Instead, as we have seen above, D. Laugwitz (1999, Chapter 1, Section 1.1.6) states that
Riemann's work on meromorphic functions was ahead of the Weierstrass' one, having been carried out with
originality and simplicity. To this point, for our purposes, it would be of a certain importance to deepen the
possible relationships between Riemann and Weierstrass, besides to what has been said above: for instance, in
this regard, Laugwitz (1999, Chapter 1, Section 1.1.5) says that Riemann was aware of the Weierstrass' works
until 1856-57, in connection with the composition of his paper on Abelian functions, in agreement with what has
been said in the previous sections. Again according to (Laugwitz 1999, Chapter 1, Section 1.1.6), one of the key
themes of Riemann's work on complex function theory was the determination of a function from its singularities
which, in turn, implies the approach of another problem, the one concerning the determination of a function from
its zeros. In this regard, Riemann limited himself to consider the question to determine a function with
infinitely many zeros whose only point of accumulation is $\infty$. What he is after is the product
representation later named after Weierstrass. Riemann uses a special case to explain the general procedure. He
does it in such a way that by following his direction one could immediately give a proof of the Weierstrass
product theorem. Therefore, it would be hoped a deeper study of these 1857-61 Riemann's lectures on complex
function theory to historically clarify this last question which is inside the wider historical framework
concerning the work of Riemann in complex function theory.

Furthermore, to this point, there seems not irrelevant to further highlight, although in a very sketchily
manner, some of the main aspects of the history of gamma function. To this end, we follow the as many notable
work of Reinhold Remmert (see (Remmert 1998)) which, besides to mainly be an important textbook on some advanced
complex analysis topics, it is also a very valuable historical source on the subject, which seems to remember
the style of the above mentioned Vivanti's textbook whose German edition, on the other hand, has always been a
constant reference point in drawing up the Remmert's textbook itself\footnote{The usefulness of historical notes
are recognized by Remmert making him what was said by Weierstrass, according to whom \it <<one can render young
students no greater service than by suitably directing them to familiarize themselves with the advances of
science through study of the sources>> \rm (from a letter of Weierstrass to Casorati of the 21st of December
1868). Anyway, see (Davis 1959) for a complete history of gamma function.}. Following (Davis 1959), (Remmert
1998, Chapter 2), (Edwards 1974, Chapter 1, Section 1.3), (Bourbaki 1963, Chapter XVIII), (Bourguet 1881),
(Montgomery \& Vaughan 2006, Appendix C, Section C.1) and (Pradisi 2012, Chapter 3), amongst the many merging
mathematical streams from which it arose, the early origins of gamma function should be above all searched into
the attempts to extend the function $n!$ to real arguments starting from previous attempts made by John Wallis
in his 1655 \rm Arithmetica Infinitorum, \rm to interpolate the values of a discrete sequence, say
$\{u_n\}_{n\in\mathbb{N}}$, with an integral depending on a real parameter, say $\lambda$, such that it is equal
to $u_n$ for $\lambda=n$. In 1730, J. Stirling investigated the formula $\log(n!)=\log\Gamma(n+1)$ in his
celebrated \it Methodus differentialis, sive tractatus de summatione et interpolatione serierum infinitarum. \rm
In 1727, Euler was called by Daniel Bernoulli to join San Petersburg Academy of Science, becoming close
co-workers. In the same period, also Christian Goldbach was professor in the same Academy, and it seems have
been just him to suggest to Euler, on the wake of Wallis' work, to extend factorial function to non-integer
values. So, from then onwards, Euler was the first to approach this last Wallis' problem since 1729, giving a
first expression of this function, in a celebrated 13th of October 1729 letter to Goldbach (see also (Whittaker
\& Watson 1927, Chapter XII, Section 12.1) and (Sansone 1972, Chapter IV, Section 5)), providing a first
infinite product expression of this new function, but only for real values. Gauss, who did not know Euler
work\footnote{This explaining why Weierstrass, as late as 1876, gave Gauss credit for the discovery of the Gamma
function.}, also taking into account Newton's work on interpolation (see (Schering 1881, Sections XI and XII)),
around the early 1810s, considered as well complex values during his studies on the hypergeometric function (of
which the $\Gamma$ function is a particular case of it), denoting such a new function with $\Pi$, while it was
Legendre, in 1814 (but (Jensen 1891) reports the date of 1809), to introduce a unified notation both for Euler
and Gauss functions, denoting these latter with $\Gamma(z)$ and speaking, since then, of \it gamma function. \rm
Other studies on gamma function properties were pursued, amongst others, by Cauchy, Hermite, A.T. Vandermonde,
A. Binet and C. Krampt around the late 1700s. Afterwards, in 1854, Weierstrass began to consider an Euler
infinite product expansion of the function $1/\Gamma(z)$, that he denoted with $Fc(z)$ and is given by
$1/\Gamma(z)=ze^{\gamma z}\prod_{j=1}^\infty(1+z/j)e^{-z/j}$, where $\gamma$ is the well-known \it
Euler-Mascheroni constant\footnote{Following (Sansone 1972, Chapter IV, Section 5), the $\gamma$ constant was
discovered by Euler in 1769, then computed by L. Mascheroni in 1790, hence by Gauss in 1813 and by J.C. Adams in
1878. See (Pepe 2012) for a contextual brief history of the Euler-Mascheroni constant, as well as (Sansone 1972,
Chapter IV, Section 5).}\rm, from which he maybe recognized, for the first time, the importance of the use of
exponential factors as infinite product convergence-producing elements. Following (Remmert 1998, Chapter 2) and
references therein, Weierstrass considered the Euler product for $Fc(z)$ the starting point for the theory,
being it, in contrast to $\Gamma(z)$, holomorphic everywhere in $\mathbb{C}$. Weierstrass said that to be
pleased\\

\it <<to propose the name ''factorielle of $u$'' and the notation $Fc(u)$ for it, since the application of this
function in the theory of factorials is surely preferable to the use of the $\Gamma$-function because it suffers
no break in continuity for any value of $u$ and, overall \rm [...], \it essentially has the character of a
rational entire function>>. Moreover, Weierstrass almost apologized for his interest in the function $Fc(u)$,
writing \it <<that the theory of analytic factorials, in my opinion, does not by means have the importance that
many mathematicians used attributed to it>>.\\\\\rm Weierstrass' factorielle $Fc$ is now usually written in the
form $ze^{\gamma z}\prod_{\nu\in\mathbb{N}}(1+z/\nu)e^{-z/}$, where $\gamma$ is the Euler's constant.
Furthermore, Weierstrass observed, in 1854, that the $\Gamma$-function is the only solution of the differential
equation $F(z+1)=zF(z)$ with the normalization condition $F(1)=1$ that also satisfies the limit condition
$\lim_{n\rightarrow\infty}(F(z+n)/n^zF(z))=1$.

However, according to (Whittaker \& Watson 1927, Part II, Chapter XII, Section 12.1), the formula (7) had
already been obtained either by F.W. Newman (see (Newman 1848)), starting from Euler's expression of gamma
function given by (7). Moreover, following (Davis 1959), the factorization formula given by Newman for the
reciprocal to gamma function was the starting point of the early Weierstrass' interest in studying gamma
function, which will lead him then to approach the problem how functions, other than polynomials, may be
factorized, starting from the few examples then available, among which sine function factorization and Newman
formula, which however required a general theory of infinite products. But, following (Jensen 1891) and
references therein, it turns out already Euler was reached the following expression for the Gamma function
\begin{equation}\Gamma(s)=\frac{1}{s}\prod_{\nu=1}^{\infty}\frac{\displaystyle\Big(1+\frac{1}{\nu}\Big)^s}
{\displaystyle\Big(1+\frac{s}{\nu}\Big)},\end{equation}who unfortunately replaced this excellent definition by
definite integrals by which, in consequences, several of the formal properties of the Gamma function escaped his
attention\footnote{Following (Remmert 1998, Chapter 2, Section 3), Euler observed, as early as 1729, in his work
on the Gamma function, that the sequence of factorials $1,2,6,24,...$ is given by the integral
$$n!=\int_0^1(-\ln\tau)^nd\tau,\qquad n\in\mathbb{N}.$$
In general$$\Gamma(z+1)=\int_0^1(-\ln\tau)^zd\tau$$whenever $\Re z>1$. With $z$ instead of $z+1$ and
$t=-\ln\tau$, this yields the well-known equation$$\Gamma(z)=\int_0^{\infty}t^{s-1}e^{-t}dt$$for
$z\in\mathbb{T}\doteq\{z;z\in\mathbb{C},\Re z>0\}$. This last improper integral was called \it Euler's integral
of the second kind \rm by Legendre in 1811, and it was a cornerstone of the rising theory of Gamma function,
becoming the matter-subject of other scholars like R. Dedekind and, above all, H. Hankel, a Riemann's student
who will give important contributions to the theory of Gamma function. In 1766, Euler systematically studied the
integral
$$\int_0^1x^{p-1}(1-x^n)^{\frac{q}{n}-1}dx=\int_0^1\frac{x^{p-1}}{\sqrt[n]{(1-x^n)^{n-q}}}dx,$$from which he derived
the following improper integral$$B(w,z)=\int_0^1t^{w-1}(1-t)^{z-1}dt,$$which is convergent in
$\mathbb{T}\times\mathbb{T}$ and, after Legendre (still in 1811), called \it Euler's integral of the first kind.
\rm Later, in 1839, this integral will be called \it beta function \rm by J.Ph. Binet who introduced too the
notation $B(w,z)$ (see (Sansone 1972, Chapter IV, Section 5)). Euler knew as well, by 1771 at the latest, that
the beta function could be reduced to the gamma function.}. This 1729 Euler's formula is equivalent to the
following one
\begin{equation}\Gamma(s)=\lim_{n\rightarrow\infty}\frac{(n+1)!(n+a)^s}{s(s+1)...(s+n-1)}\end{equation}
which was provided by Gauss in 1813 who undoubtedly was not familiar with Euler's expression (8). Later on, the
expression
\begin{equation}\Gamma(s)=e^{Cs}s\prod_{\nu=1}^{\infty}\displaystyle\Big(1+\frac{1}{\nu}\Big)^s\end{equation}
was due either to O.X. Schl\"{o}milch in 1843 (see (Schl\"{o}milch 1844; 1848) as well as to F.W. Newmann (see
above), besides to have been rediscovered by Weierstrass in his famous 1856 memoir on analytical factorials (see
(Weierstrass 1856a); see also (Burkhardt et al. 1899-1927, Band II, Erster Teil, Erste H\"{a}lfte, A.3, Nr. 12e)
and (Remmert 1998, Chapter 2)).

Following (Vivanti 1928, Section 135-141), (Remmert 1998, Chapter 3) and, above all, (Bottazzini \& Gray 2013,
Section 6.7), Weierstrass extended the product (5) in such a manner to try to avoid divergence problems with the
\it ad hoc \rm introduction, into the product expansion, of certain forcing convergence factors. This attempt
was successfully attended, since 1874, as a solution to a particular question - the one which may be roughly
summarized as the attempt to build up an entire transcendental function with prescribed zeros - which arose
within the general Weierstrass' intent to solve the wider problem to find a representation for a single-valued
function as a quotient of two convergent power series. To be precise, he reached, amongst other things, the
following main result\\

\it <<Given a countable set of non-zero complex points $z_1,z_2,...,$ such that $0<|z_1|\leq|z_2 |\leq ...$ with
$\lim_{n\rightarrow\infty}|z_n|=\infty$, then it is possible to find, in infinite manners, a non-decreasing
sequence of natural numbers $p_1,p_2,...$ such that the series $\sum_{j=1}^{\infty}|z/z_j |^{p_j+1}$ be
convergent for every finite value of $z$, in such a manner that the most general entire function which is zero,
with their own multiplicity, in the points $z_1,z_2,...,$ and has a zero of order $\lambda$ in the origin, is
given by\footnote{Historically, in relation to (8), the function $f(z)$ was usually denoted, d'après
Weierstrass, by $G(z)$, whilst $z^{\lambda}\prod_{j=1}^{\infty}(1-z/z_j)E_j(z)$ was named \it canonical \rm(or
\it primitive\rm) \it function \rm - see (Sansone 1972, Chapter IV, Section 3), where there are too many
interesting historical notes.}
\begin{equation}f(z)=e^{g(z)}z^{\lambda}\prod_{j=1}^\infty\Big(1-\frac{z}{z_j}\Big)E_j(z)\end{equation}where
$E_j(z)=(1-z)(\sum_{i=1}^jz^i/i)$ for $j\geq1$ and $E_0(z)=1-z$, $g(z)$ being an arbitrary entire function, and
the infinite product is absolutely convergent for each finite value of $|z|$>>.\\\\\rm The factors $E_j(z)$ will
be later called \it Weierstrass' factors, \rm whilst the numbers $p_j$ will be called \it convergence
exponents\rm; finally, $e^{g(z)}$ is also called \it Weierstrass' exponential factor. \rm The sequence
${E_j(z)}_{j\in\mathbb{N}_0}$ plays a very fundamental role in the Weierstrass' theorem: from the
equation\begin{equation}1-z=\exp(\log(1-z))=\exp(-\sum_{i\geq 1}z^i/i),\end{equation}Weierstrass obtained the
formula $E_j (z)=\exp(-\sum_{i>j}z^j/j)$ in proving convergence properties which, on the other hand, would have
been easier obtained by means of the following estimates
\begin{equation}|E_j(z)-1|\leq|z|^{j+1},\forall j\in\mathbb{N}_0,\forall z\in\mathbb{C},|z|\leq 1\end{equation}
that have been proved only later. Amongst the first ones to have made this, seems there having been L. Fejér
(see (Hille 1959, Section 8.7)), but the argument appears as early as 1903 in a paper of Luciano
Orlando\footnote{Luciano Orlando (1887-1915) was an Italian mathematician prematurely died in the First World
War - see the very brief obituary (Marcolongo 1918) as well as (Rouse Ball 1937, Appendix II, pp. 430-431). His
supervisors were G. Bagnera and R. Marcolongo who led him to make researches in algebraic integral equation
theory and mathematical physics.} (1903) which starts from Weierstrass' theorem as treated by Borel's monograph
on entire functions. As has already been said above, Weierstrass was led to develop his theory by the chief
objective to establish the general expression for all analytic functions meromorphic in $\mathbb{C}$ except in
finitely many points, reaching the scope after a series of previous futile attempts only in 1876, with notable
results, spelt out in (Weierstrass 1876), concerning the class of transcendental entire functions. But what was
new and sensational in the Weierstrass' construction was just the introduction and the application of the
so-called \it convergence-producing factors \rm (or \it primary factors \rm or \it Weierstrass' factors\rm)
which strangely have no influence on the behavior and distribution of the zeros.\\\\ \bf 1.4 Towards the theory
of entire functions, and other. \rm In the necrology of Weierstrass, Poincaré (1899, Section 6) said that
Weierstrass' major contribution to the development of function theory was just the discovery of primary factors.
Also Hermite was, in a certain sense, astonished and intrigued from the introduction of this new Weierstrass'
notion of prime factor, which he considered of capital importance in analysis and making later notable studies
in this direction; he also suggested to Èmile Picard to do a French translation of the original 1876
Weierstrass' work, so opening a French research trend on this area. En passant, we also point out the fact that,
from the notion of prime factor and from the convergence of the infinite product
$\prod_{j\in\mathbb{N}}E_j(z/a_j)$, representing an entire transcendental function vanishing, in a prescribed
way, in each $a_j$, Hilbert drew inspiration to formulate his valuable algebraic notion of prime
ideal\footnote{Usually, the notion of prime ideal of the commutative algebra, with related operations, would
want to be stemmed from the factorization of natural numbers.}. Following (Pincherle 1922, Chapter IX, Section
137), (Vivanti 1928, Section 136), (Burckel 1979, Chapter XI), (Remmert 1998, Chapters 3 and 6), (Ullrich 1989,
Section 3.5) and (Bottazzini \& Gray 2013, Sections 5.11.5 and 6.7), since the late 1850s, Enrico
Betti\footnote{Following (Bottazzini 2003), the influence of Riemann's ideas on 19th-century Italian
mathematical school had a great impulse thanks to the Betti's interest since 1850s. In 1858, as is well known,
Betti, Brioschi and Casorati went in G\"{o}ttingen to personally know Riemann and his ideas, translating many
works of Riemann. Betti and Casorati were immediately aware of the innovative power of the new Riemann ideas in
complex analysis, introducing in Italy, for the first time, such a theory with appreciated works and treatises.}
had already reached notable results, about convergence properties of infinite products of the type (6), very
near to the Weierstrass' ones related to the resolution of a fundamental problem of entire function theory, the
so-called \it Weierstrass' problem\footnote{Following (Forsyth 1918, Chapter V, Section 50) and (Bottazzini \&
Gray 2013, Section 4.2.3.2), in relation to the infinite product expression of an entire transcendental function
prior to 1876 Weierstrass' paper, attention should be also paid to a previous 1845 work of A. Cayley on doubly
periodic functions. Furthermore, following (Tannery \& Molk 1893, Section 85), into some previous 1847 works of
G. Eisenstein on elliptic functions, some notable problems having to do with the construction of analytic
functions with prescribed zeros as a quotient of entire functions with the involvement of certain transcendental
entire functions of exponential type (similar to the Weierstrass problem as historically related to meromorphic
functions), had already been considered. See also certain function's quotients stemmed from the developments of
certain determinants given in (Gordan 1874). In any case, all these historical considerations confirm, once
again, that the prolegomena of entire function factorization theorems should be searched in the general history
of elliptic functions.} \rm (see (Pincherle 1922, Chapter IX, Section 137)). Betti exposed these outcomes in his
celebrated 1859-60 Pisa lectures on advanced analysis entitled \it La teorica delle funzioni ellittiche \rm (see
(Betti 1903-1913, Tomo I, XXII)), published in the Tomes III and IV of the \it Annali di matematica pura ed
applicata, \rm Series I, after having published, in the Tome II of these \it Annali\rm, an Italian translation
of the celebrated 1851 Riemann's inaugural dissertation on complex function theory, which can be considered as
an introduction to his next lectures on elliptic functions. Indeed, in these latter, Betti, before all, places
an Introduction on the general principles on complex functions, essentially based on these 1851 Riemann
lectures. From the point 3. onward of this Introduction, Betti starts to deal with entire functions, their
finite and infinite zeros (there called \it roots\rm), as well as on possible quotients between them. In
particular, taking into account what is said in (Briot \& Bouquet 1859), he considered infinite products of the
type $\prod_{\rho}(1-z/\rho)$, where $\rho$ are the zeros of an entire function, with the introduction of a
factor of the type $e^w$, where $w$ is an arbitrary entire function, to make convergent this infinite product.
Furthermore, Betti dealt with this type of infinite products starting to consider infinite product
representations of the following particular function $es(z)=z\prod_{m=1}^{\infty}(m/(m+1))^z(1+z/m)$, which
satisfies some functional equations and verifies the relation $\Gamma(z)=1/es(z)$. Therefore, as Weierstrass too
will do later, Betti started from the consideration of the infinite product expansion of the inverse of the
gamma function for studying the factorization of entire functions. Therefore, Betti guessed the utility of the
convergence factors having exponential form, looking at the infinite product expansion of Gamma function,
similarly to what Weierstrass will do. Afterwards, Betti proved some theorems which can be considered particular
cases of the next Weierstrass' results, concluding affirming that\\

\it <<Da questi teoremi si deduce che le funzioni intere potranno decomporsi in un numero infinito di fattori di
primo grado ed esponenziali, e qui comparisce una prima divisione delle funzioni intere. Quelle che hanno
gl'indici delle radici in linea retta, e quelle che le hanno disposte comunque nel piano; le prime, che sono
espresse da un prodotto semplicemente infinito, le chiameremo di prima classe, le seconde, che sono espresse da
un prodotto doppiamente infinito, le diremo di seconda classe. Le funzioni di prima classe si dividono anch'esse
in due specie, la prima, che comprende quelle che hanno gl'indici delle radici disposti simmetricamente rispetto
a un punto, e che possono esprimersi per un prodotto infinito di fattori di primo grado, le altre, che hanno
gl'indici delle radici disposti comunque sopra la retta, le quali si decomporranno in fattori di primo grado ed
esponenziali. Ogni funzione intera di prima classe della prima specie potrà decomporsi nel prodotto di più
funzioni intere della stessa classe di seconda specie, e data una funzione della seconda specie se ne potrà
sempre trovare un'altra che moltiplicata per la medesima dia per prodotto una funzione della prima specie. Le
funzioni di seconda classe si dividono anch'esse in due specie; la prima comprenderà quelle che hanno gl'indici
delle radici disposti egualmente nei quattro angoli di due assi ortogonali, in modo che facendo una rotazione
intorno all'origine di un quarto di circolo, gl'indici di tutte le radici vengano a sovrapporsi, le quali
funzioni possono esprimersi per un prodotto doppiamente infinito di fattori di primo grado; la seconda
comprenderà quelle che hanno gl'indici disposti comunque, e si decompongono in un prodotto doppiamente infinito
di fattori di primo grado e di fattori esponenziali. Data una funzione della seconda specie se ne potrà sempre
trovare un'altra che moltiplicata per quella dia una funzione della prima specie>>.\\

\rm[\it <<From these theorems, we deduce that entire functions might be decomposed into an infinite number of
first degree factors and exponential factors, so that here there is a first classification of entire functions
according to that their root's indexes lie along a line or are arbitrarily placed in the plane; the former are
said to be of first class and are expressed by a simply infinite product, while the latter are said to be of
second class and are expressed by a doubly infinite product. The functions of the first class are, in turn,
classified into two kinds: the first one comprises those functions having the root's indexes symmetrically
placed respect to a point and that can be expressed by an infinite product of first degree factors; the second
one comprises those functions having root's indexes arbitrarily placed along a line and that can be expressed by
an infinite product both of first degree factors and of exponential factors. Each entire function of first class
and of first kind might be decomposed into the product of other entire functions of the same class and of the
second kind; furthermore, given a function of the second kind, it is always possible to find another function
that multiplied by the former, the product gives rise to another function of the first kind. Likewise, the
functions of the second class are divided into two kinds: the first one comprises those functions having the
root's indexes equally placed into the four angles of the two orthogonal cartesian axes in such a manner that
all these are overlapped through a $\pi/2$ radian rotation around the origin, and are decomposable into a doubly
infinite product of first degree factors; the second one includes those functions having the root's indexes
arbitrarily placed and that are decomposable into a doubly infinite product of first degree factors and
exponential factors. Furthermore, given a function of the second kind, it is always possible to find another
function that multiplied by the former, the product gives rise to a function of first kind>>\rm].\\\\\rm Then,
Betti carries on treating entire functions in the first part of his lessons on elliptic functions, followed by a
second part devoted to quotients of functions, mentioning either the paper (Weierstrass 1856a) and the paper
(Weierstrass 1856b). Therefore, Betti's work on entire function factorization, made in the period 1860-63, was
very forerunner of the Weierstrass' one: this is confirmed either by (Rouse Ball 1937, Appendix II, pp.
376-384)) and by (Federigo Enriques 1982, Book III, Chapter I, Section 6), in which it is pointed out that the
fundamental Weierstrass' theorem on the factorization of entire transcendental functions from their zeros, had
already been discovered by Betti, highlighting however as the Pisa's mathematician, with singular personal
disinterestedness, wanted not claim it as due to him. Indeed, following Francesco Cecioni's comments about some
works of Ulisse Dini (see (Dini 1953-59, Volume II)), it turns out that Betti's work could easily reach, only
with very slight modifications, the same generality and abstraction of the Weierstrass' one, as Dini explicitly
proved in (Dini 1881); furthermore, Dini proved too that Betti's work could be able to give a particular case,
given in the years 1876-77, of the general Gösta Mittag-Leffler theorem - see (Mittag-Leffler 1884), (Vivanti
1928, Section 145), (Loria 1950, Chapter XLIV, Section 752) and (Bottazzini \& Gray 2013, Section 6.7.6) -
independently by what Weierstrass himself was doing in the same period, in regards to these latter arguments.
Cecioni says that this Dini's work had already been worked out since 1880, whilst the Weierstrass' theorem was
published in 1876 - see (Weierstrass 1876). Thus, much before, namely in 1860, Betti had proved, as we have
already said, a particular but important case of this theorem, albeit he didn't go beyond, because the results
achieved by him were enough to his pragmatic scopes concerning Abelian and elliptic functions\footnote{In this
regard, also Salvatore Pincherle (1899, Chapter IX, Section 175) reports that Betti solved the Weierstrass'
problem in a quite general case.}, and, as also Pincherle (1922, Chapter IX, Section 135) has claimed, the
Weierstrass' method was essentially the same of the Betti's one with slight modifications. In the years 1876-77,
also G. Mittag-Leffler proved a particular case of a more general theorem that he will give later, to be precise
in 1884, after a long series of previous works in which he gradually, through particular cases, reached the
general form of this his theorem as nowadays we know it. In the meanwhile, Weierstrass reconsidered
Mittag-Leffler's works, since the early 1880s, in relation to what himself have done on the same subject. Also
F. Casorati (1880-82) had some interesting ideas similar to the Mittag-Leffler's ones, giving further
contributions to the subject (see (Loria 1950, Chapter XLIV, Section 750)). Almost at the same time, amongst
others, Ernst Schering (1881), Charles Hermite (1881), Émile Picard (1881), Felice Casorati (1882), Ulisse Dini
(1881), Paolo Gazzaniga\footnote{Some historical sources refer of Paolo Cazzaniga, whereas others refer of Paolo
Gazzaniga, but, very likely, they are the same person, that is to say, Paolo Gazzaniga (1853-1930), an Italian
mathematician graduated from Pavia University in 1878 under the supervision of Felice Casorati. In the years
1878-1883, he was interim assistant professor at Pavia, then he spent a period of study in Germany under the
Weierstrass and Kronecker supervision. Afterwards, from 1888, he became professor at the high school Tito Livio
in Padua, teaching too in the local University. He was also one of the most influential teachers of Tullio
Levi-Civita during his high school studied. Gazzaniga's researches mainly concerned with applied algebra and
number theory. Furthermore, Paolo Gazzaniga has to be distinguished from Tito Camillo Cazzaniga (1872-1900), a
prematurely died Italian mathematician, graduated from Pavia University in 1896, whose researches concerned with
matrix theory and analytic functions according to the research trend of Ernesto Pascal (1865-1940) during his
teaching in Pavia. Both Tito Cazzaniga (see (Rouse Ball 1937, Appendix II, pp. 412-413)) and Paolo Gazzaniga are
quoted in (Vivanti 1901) but not in (Vivanti 1928).} (1882), Claude Guichard (1884) and Paul Painlèvé (1898a,b),
achieved notable results about the general problem to build up a complex function with prescribed singularities,
although related to a generality degree less than that of the Mittag-Leffler results. Thus, the history of the
Mittag-Leffler theorem makes too its awesome appearance within the general history of meromorphic functions, a
part of which may be retraced in the same Mittag-Leffler 1884 paper in which, amongst other things, also the
1881 work of Ulisse Dini is quoted. However, both Schering (1881, Section XVI) and Casorati (1880-82, p. 269,
footnote (***)), in discussing the above mentioned Mittag-Leffler results, quote Betti's work on Weierstrass'
theorem; in particular, the former speaks of Betti's convergence factors and the latter states that\\

\it <<Anche il sig. Dini, nella sua Nota sopra citata, dimostra questo teorema, riducendo lo studio del prodotto
infinito a quello della serie dei logaritmi dei fattori; riduzione di cui s'era già valso felicemente, per il
caso di distribuzione degli zeri a distanze non mai minori di una quantità fissa $d$, il sig. Betti nella
Introduzione della sua Monografia delle funzioni ellittiche (Annali di Matematica, Tomo III, Roma, 1860), dove
precede assai più oltre di Gauss nella via che mena al teorema del sig. Weierstrass>>.\\

\rm[\it<<Also Mr. Dini, in his Note of above, proves this theorem, reducing the study of the infinite product to
the study of the series of the logarithms of the factors; reduction, this, that had already been used by Mr.
Betti in the Introduction to his monograph on elliptic functions (Annali di Matematica, Tome III, Rome, 1860)
for the case of a distribution of zeros having reciprocal distances not less than a fixed quantity $d$; in doing
so, he much foregoes Gauss in a fashion which leads to the theorem of Mr. Weierstrass>>.\rm]\\\\Therefore, from
the Mittag-Leffler's works onwards, together to all those works made by other mathematicians amongst whom are
Dini, Schering, Casorati, Hermite, Picard, Cazzaniga, Guichard, Von Schaper\footnote{Hans Von Schaper, a
doctoral student of Hilbert (see (Borel 1900, Chapitre II, p. 26)), whose doctoral dissertation thesis, entitled
\it Über die Theorie der Hadamardschen Funktionen und ihre Anwendung auf das Problem der Primz\"{a}hlen, \rm and
defended at G\"{o}ttingen in 1898, was just centered around the applications of 1893 Hadamard factorization
theorem of entire function; in it, some further interesting properties on the order of an entire function, like
the distinction between \it real \rm and \it apparent order, \rm were discussed as well.}, Painlèvé and
Weierstrass himself, it starts the theory of entire transcendental functions whose early historical lines have
been traced in the previous sections. In any case, with Mittag-Leffler, we have the most general theorems for
the construction, by infinite products, of a meromorphic function with prescribed singularities (see (Bottazzini
\& Gray, Chapter 6, Section 6.7) for a deeper historical analysis of these representation theorems). On the
other hand, following (Gonchar et al. 1997, Part I, Introduction) and (Vivanti 1901, Section 215), the above
mentioned works by Weierstrass, Mittag-Leffler and Picard, dating back to the 1870s, marked the beginning of the
systematic studies of the theory of entire and meromorphic functions. The Weierstrass and Mittag-Leffler
theorems gave a general description of the structure of entire and meromorphic functions, while the
representation of entire functions as an infinite product \it à la \rm Weierstrass, served as basis for studying
properties of entire and meromorphic functions. Following (Remmert 1998, Chapter 3, Section 1), Weierstrass
developed his 1876 paper with the main objective to establish the general expression for all functions
meromorphic in $\mathbb{C}$ except at finitely many points but, as said above, the really importance of
Weierstrass' construction was the application of the convergence-producing factors which have no influence on
the behavior of the zeros. The awareness that there exist entire functions with arbitrarily prescribed zeros
revolutionized the thinking of function theorists. Suddenly, one could construct holomorphic functions that were
not even hinted at in the classical framework. Nevertheless, this sort of freedom does not contradict the
so-called solidarity of value behavior of holomorphic functions required by the identity theorem because, with
the words of Remmert himself, the 'analytic cement' turns out to be pliable enough to globally bind locally
prescribed data in an analytic way. Weierstrass left it to other the extension of his product theorem to regions
in $\mathbb{C}$. So, as early as 1881, E. Picard considered, for the first time, Weierstrass' products in
regions different from $\mathbb{C}$, albeit he nothing said about convergence questions. In 1884, Mittag-Leffler
proved existence theorems for more general regions but without quoting Picard's work, even if Edmund Landau (see
(Landau 1918)) later will speak of the ''well-known Picard/Mittag-Leffler product construction''. Further
generalization of Weierstrass' theorem was then given too by A. Pringsheim in 1915 (see (Burkhardt et al.
1899-1927, II.C.4, Nr. 26)). Following (Gol'dberg \& Ostrovski\v{i} 2008, Preface), the classical 1868 theorem
of J. Sokhotski and F. Casorati, the above mentioned 1876 theorem of Weierstrass and the 1879 Picard theorem
opened the theory of value distribution of meromorphic functions, while the works of J.L.W. Jensen and J.
Petersen in the late 1890s, had great importance for the further developments of the theory of entire and
meromorphic functions (see (Remmert 1998, Chapter 4, Section 3)) which started, in the same period, to gradually
become a separate and autonomous mathematical discipline after the pioneering investigations mainly pursued by
the French school of Laguerre, Hadamard, Poincaré, Lindelöff, Picard, Valiron and Borel, up until the Rolf
Nevanlinna work of the early 1900s, which gave an almost definitive setting to the theory. All that will be
in-depth studied in the next section, where we shall deal with the main lines of the history of entire and
meromorphic functions whose theory basically starts just from the entire function factorization theorems.
Following (Zhang 1993, Preface), in 1925, Nevanlinna established two main theorems that constituted the basis
upon which build up the theory of value distribution of meromorphic functions, whilst, in 1929, by examining
some examples, he recognized as well that there is an intrinsic relationship between the problem of exceptional
values (deficient values are exceptional values under a certain kind of implication) and the asymptotic value
theory. Moreover, Nevanlinna anticipated that the study of their relationship might help to clarify some of the
profound problems of the theory of entire and meromorphic functions. From his product theorem, Weierstrass
immediately deduced the theorem on quotient representation of meromorphic functions, attracting attention by
this alone. From this work of the ''celebrated geometer of Berlin'', Poincaré worked out his 1883 famous theorem
on the representability of every meromorphic function in $\mathbb{C}^2$ as the quotient $f(w,z)/g(w,z)$ of two
entire functions in $\mathbb{C}^2$ and locally relatively prime everywhere, so giving rise to a new theory that,
through the works of P. Cousin, T.H. Gronwall, H. Cartan, H. Behnke, K. Stein, K. Oka, J-P. Serre, H. R\"{o}hrl
and H. Grauert, is still alive and rich today. With his product theorem, Weierstrass opened the door to a
development that led to new insights in higher-dimensional function theory as well. In particular, the
Weierstrass' product theorem was for the first time generalized to the case of several complex variables as
early as 1894 by Pierre Cousin (1867-1933), a student of Poincaré, in (Cousin 1895) centered around his doctoral
thesis whose main aim was that to generalize the above mentioned 1883 Poincaré theorem to higher dimensions and
more general domains, so giving rise to the celebrated \it I \rm and \it II problem of Cousin, \rm solved by him
for product domains of the type $X=B_1\times...\times B_n\subset\mathbb{C}^n$ (see (Maurin 1997, Part V, Chapter
6) and (Della Sala et al. 2006, Chapter 11, Section 6)). As Cousin himself says, the 1883 Poincaré theorem was
the first successful attempt to extend Weierstrass results to analytic functions several complex variables:
following (Dieudonné 1982, A VIII), that branch of mathematics known as ''analytic geometry'' is nothing but the
modern form of the theory of analytic functions of several complex variable. Then, Cousin recalls too the
attempts made by P. Appell and by S. Dautheville in the 1880s to extend, along the same line, the 1884
Mittag-Leffler work to the $n$ complex variable case. En passant, then, we also note that the Weierstrass'
entire function factorization theorem has had further remarkable applications in many other pure and applied
mathematical contexts. In this place, we wish to point out another possible interesting historical connection.
To be precise, following (Marku\v{s}evi\v{c} 1967, Volume II, Chapters 8 and 9), (Burckel 1979, Chapter VII) and
(Remmert 1998, Chapter 4), a very similar problem to that considered by Weierstrass was the one considered in
(Marku\v{s}evi\v{c} 1967, Volume II, Chapter 8, Theorem 8.5) where, roughly, a bounded analytic function with
prescribed zeros is constructed by means of certain infinite products introduced by Wilhelm Blaschke (see
(Blaschke 1915)), called \it Blaschke products, \rm in relation to questions related to the well-known Giuseppe
Vitali convergence theorem for sequences of holomorphic functions, and defined upon those complex numbers
assigned as given zeros of that function that has to be determined. They form a special class of Weierstrass'
products\footnote{See (Remmert 1998, Chapter 4), (Lang 1974, Chapter 15) and (Lang 1999, Chapter XIII) for
technical details.}. Edmund Landau (see (Landau 1918)) reviewed Blaschke's work in 1918 and simplified the proof
by using a formula due to J.L.W. Jensen (see (Jensen 1898-99)). By means of the differentiation theorem of
products of holomorphic functions, in 1929 R. Ritt was able to give a factorization of an holomorphic function
at the origin, whose product is normally convergent into a disc about the origin (see (Remmert 1998, Chapter 1,
Section 2)). In the proceedings collected in (Mashreghi \& Fricain 2013), where remarkable applications of
Blaschke's products in pure and applied mathematics questions (amongst which one concerning approximation of
Riemann zeta function) are presented, we report what is said in the incipit of the Preface, according to which\\

\it<<Infinite Blaschke products were introduced by Blaschke in 1915. However, finite Blaschke products, as a
subclass of rational functions, has existed long before without being specifically addressed as finite Blaschke
products. In 1929, R. Nevanlinna introduced the class of bounded analytic functions with almost everywhere
unimodular boundary values. Then the term inner function was coined much later by A. Beurling in his seminal
study of the invariant subspaces of the shift operator. The first extensive study of the properties of inner
functions was made by W. Blaschke, W. Seidel and O. Frostman. The Riesz technique in extracting the zeros of a
function in a Hardy space is considered as the first step of the full canonical factorization of such elements.
The disposition of zeros of an inner function is intimately connected with the existence of radial limits of the
inner function and its derivatives. For almost a century, Blaschke products have been studied and exploited by
mathematicians. Their boundary behaviour, the asymptotic growth of various integral means of Blaschke products
and their derivatives, their applications in several branches of mathematics in particular as solutions to
extremal problems, their membership in different function spaces and their dynamics are examples from a long
list of active research domains in which they show their face>>.\\\\\rm Following (Borel 1900, Chapter I), the
major difficult in applying the Weierstrass theorem is the determination of the exponential factors $\exp G(x)$,
a hindrance that the next Hadamard work coped with success, whose pioneering work will turn out to be extremely
useful also in physics: amongst all the possible applications to which such a work has given rise, we here only
mention the use of entire function theory (following (Boas 1954)) made by Tullio Regge in achieving some notable
properties of the \it analytic $S$ matrix \rm of potential scattering theory (see (Regge 1958)), which are
closely connected with the distribution of the zeros of entire functions. In particular, Regge cleverly uses
infinite product expansions of entire functions, amongst which the Hadamard expansion, in finding analytic
properties of the analytic \it Jost functions \rm as particular asymptotic solutions to non-relativistic
Schr\"{o}dinger equation (\it $S$ waves\rm). Finally, following (Maz'ya \& Shaposhnikova 1998, Chapter 1,
Section 1.10), we also notice that remarkable applications of some results of entire function theory, amongst
which some results due to Hadamard, were also considered by Poincaré in his celebrated three volume work \it Les
Méthodes Nouvelles de la Mécanique Céleste \rm (see (Poincaré 1892-1899): to be precise, in (Poincaré 1892-1899,
Tome II, Chapter XVII, Section 187), the author considers some entire function factorization theorems in solving
certain linear differential equations also making reference to the well-known 1893 Hadamard results.

We wish to report some very interesting historical remarks made by Hermann Weyl in one of his last works, the
monograph on meromorphic functions wrote in 1943 and reprinted in 1965 with the collaboration of his son, F.
Joachim Weyl (see (Weyl \& Weyl 1965)). Weyl says that the main motif underlying the drawing up of this his
monograph was the work of Lars V. Ahlfors on meromorphic curves on complex plane, dating back to the late 1930s,
and that Weyl wanted to reformulate extending it to a general Riemann surface. In (Weyl \& Weyl 1965,
Introduction), Weyl states an analogical parallel, that is to say, that meromorphic functions stand for entire
functions as rational functions stand for polynomials, pointing out that degree is the most important
characteristic of a polynomial, hence considering the usual decomposition into linear factors of a complex
polynomial in dependence on its roots. A complex polynomial with $n$ roots, say $a_1,...,a_n$, counted with
their multiplicity, may be written in the
form$$f(z)=kz^h\prod_{i=1}^{n-h}\Big(1-\frac{z_i}{a_i}\Big)\leqno(\star_1)$$if $h(\leq n)$ out of the $n$ roots
are equal to zero, and $k$ is a non-zero constant. Then, Weyl considers the type of growth of a polynomial of
degree $n$, given by an inequality of the form $|f(z)|\leq C|z|^n$, or, more precisely, by an asymptotic
equation of the type $|f(z)|/|z|^n\rightarrow C$ as $|z|\rightarrow\infty$, where $C\neq 0$ is a constant. This
means that $f(z)$ takes on the value $\infty$ with multiplicity $n$ at $z=\infty$. Then, Weyl asks whether it is
possible to make statements about entire functions on the basis of what is known about polynomials. Weyl points
out that the perfect analogical extension is not possible simply because there exist entire functions which have
no zeros, like $e^z$ to mention the simplest one. Weyl hence goes on considering the problem of building up an
entire function knowing its zeros ordered according to their nondecreasing modulus which are in a finite number
in every finite region of complex plane, hence observing that this problem (named \it Weierstrass' problem\rm)
was first solved by Weierstrass in a paper of 1876 which is the starting point of many other investigations on
entire and meromorphic functions. Then, Weyl observes that the next problem of determining the growth of an
entire function through its canonical decomposition into primary factors according to Weierstrass, is not
solvable by the simple knowledge of such a decomposition because it is related an arbitrary but finite region of
complex plane. So, it was Poincaré, in 1883, to approach and solve, for the first time, such a problem in some
special cases connected with the convergence of certain series related to the zeros of the given entire
function. This last problem was then approached and solved, in more general cases, by E.N. Laguerre and E.
Borel, introducing the notions of \it genus \rm(or \it genre\rm) and \it order \rm of an entire function, hence
Hadamard, in 1893, gave a converse to Poincaré theorem for entire functions of finite genus, whose form was
later improved and extended by R. Nevanlinna. Just in regard to the novelties due to Hadamard, Weyl points out
that the driving force for these Hadamard's investigations was the wish to obtain sufficient information about
the zeros of the Riemann zeta function for establishing the asymptotic law for the distribution of prime
numbers. This law states that the number $\pi(n)$ of primes less than $n$ becomes infinite with
$n\rightarrow\infty$ exactly as strongly as $n\ln n$, that is to say$$\frac{\pi(n)\ln n}{n}\rightarrow
1\qquad\mbox{\rm as}\quad n\rightarrow\infty,\leqno(\star_2)$$Riemann having shown how this prime number problem
crucially depends on the zeros of his zeta function. In 1896, either Hadamard and de la Vallée-Poussin,
independently of each other, were able to draw the conclusion $(\star_2)$ from 1893 Hadamard's results
concerning entire functions. Afterwards, besides the problem to determine zeros of an entire function $f(z)$,
Weyl considers too the problem to determine the distribution of the points $z\in\mathbb{C}$ satisfying the
equation $f(z)=c$ for any preassigned complex value $c$, which Weyl calls \it $c$-places, \rm the former problem
being therefore the one determining the $0$-places of $f(z)$. Weyl quotes E. Picard results in this direction,
dating back 1880, hence the next results of G. Valiron, A. Wiman and Nevanlinna brothers of 1920s, until up L.
Ahlfors results of 1930s. Once meromorphic function theory was established, by various research papers from
1920s onward and mainly due, among others, to E. Borel, A. Bloch, P. Montel, R. Nevanlinna, F. Nevanlinna, H.
Cartan, T. Shimizu, O. Frostman, H. Weyl, J.F. Weyl, E. Ullrich, G. H\"{a}llstr\"{o}m and J. Dufresnoy, started
the so-called \it theory of meromorphic curves, \rm which originated by the rough idea to consider homogeneous
coordinates $x_0,x_1,...,x_k$ of a $k$-dimensional projective space, as meromorphic functions
$x_1/x_0,x_2/x_0,...,x_k/x_0$ depending on a certain complex parameter $z$ ranging over the whole complex plane
except $z=\infty$ (see (Weyl \& Weyl 1965, Chapter II, Section 2; Chapter III, Sections 2-4) and references
therein), so that a meromorphic function $f=x_1/x_2$ may be considered as a meromorphic curve in a
two-dimensional projective space.

\newpage\section*{2. Outlines of history of entire function theory}
\addcontentsline{toc}{section}{2. Outlines of history of entire function theory}

Following (Borel 1897), the Weierstrass' work on the decomposition of entire functions into primary factors, has
greatly contributed to the study of the distribution of zeros of the entire functions. The notion of \it genus
\rm of an entire function, introduced by Laguerre, will turn out to be of fundamental importance to this end, as
well as the analogous notion of \it order \rm of an entire function, which nevertheless will turn out to be much
more useful and precise than the former, above all thanks to the contributions of Poincaré, Hadamard and Picard
(see (Borel 1900)). Above all, Hadamard's work will provide new avenues to the theory of entire functions and
the related distribution laws of their zeros. Following (Bergweiler \& Eremenko 2006), the theory of entire
functions begins as a field of research in the works of Laguerre (see (Laguerre 1898-1905)), soon after the
Weierstrass product representation became available. Laguerre then introduced the first important classification
of entire functions, according to their genera. Following (Gil' 2010, Preface), one of the most important
problems in the theory of entire functions is the problem of the distribution of the zeros of entire functions.
Many other problems in fields close to the complex function theory, lead to this problem. The connection between
the growth of an entire function and the distribution of its zeros was investigated in the classical works of
Borel, Hadamard, Jensen, Lindel\"{o}f, Nevanlinna and others. On the other hand, following (Gonchar et al. 1997,
Part I, Chapter 1), the infinite product representation theory of entire functions marked the beginning of the
systematic study of their properties and structure, with the first works by Weierstrass and Hadamard. Following
(Marku\v{s}evi\v{c} 1966, Preface), entire functions are the simplest and most commonly encountered functions:
in high schools, we encounter entire functions (like polynomials, the exponential function, the sine and cosine,
and so forth), meromorphic functions, that is, the ratios of two entire functions (like the rationale functions,
the tangent and cotangent, and so on), and, finally, the inverse functions of entire and meromorphic functions
(like fractional powers, logarithms, the inverse trigonometric functions, etc.). Following (Levin 1980, Chapter
I), an entire function is a function of a complex variable holomorphic in the whole of the complex plane and
consequently represented by an everywhere convergent power series of the type $f(z)=\sum_{i=0}^{\infty}a_iz^i$,
these functions forming a natural generalization of the polynomials, and are therefore close to polynomials in
their properties. The theorem of Weierstrass on the expansion of entire functions into infinite products
provided the basic apparatus for the investigation of the properties of entire functions and it was the starting
point for their classification. This theorem plays a fundamental role in the theory of entire functions (see
(Saks \& Zygmund 1952, Chapter VII, Section 2)), being it, roughly, the analogue of the theorem on the
decomposition of polynomials into linear factors. Following (Tricomi 1968, Chapter IV, Section 8), this
Weierstrass theorem plays a central role in the whole of the theory of entire functions whose even most recent
developments are, more or less directly, reconnected to it. At approximately the same time as this celebrated
work of Weierstrass, Laguerre studied the connection between entire functions and polynomials, and introduced
the important concept of genus of an entire function. Since then, the theory of entire and meromorphic functions
underwent to a notable development, becoming one of the many wide chapters of complex analysis, assuming an
autonomous status. Amongst the many contributions to the theory, which will be briefly recalled below, the
classical investigations of Borel, Hadamard and Lindel\"{o}ff dealt with the connection between the growth of an
entire function and the distribution of its zeros. The rate of growth of a polynomial as the independent
variable goes to infinity is determined, of course, by its degree. Thus, the more roots a polynomial has, the
greater its growth is. This connection between the set of zeros of the function and its growth can be
generalized to arbitrary entire functions, the content of most of the classical theorems of the theory of entire
functions consisting just in establishing relations between the distribution of the roots of an entire function
and its asymptotic behavior as $z\rightarrow\infty$, to measure the growth of an entire function and the density
of its zeros, a special growth scale having been introduced. Following (Evgrafov 1961, Chapter II, Section 1),
the basic task of the theory of entire functions (at least, from the point of view of its applications to other
domains of analysis) is to establish connections between the different characterizing elements of an entire
function as, for example, between the coefficients, the behavior at infinity, and the zeros. It would hardly be
mistaken to say that the most important task of all is to establish such connections for entire functions that
are in some sense regular, that is, have regularly decreasing coefficients, or regularly distributed zeros, or a
simple integral representation, or else a simple functional equation. However, the study of entire functions
under such strong hypotheses is a very complicated task, and it is necessary to know those simpler and more
general laws that are less exact but which hold under weaker hypothesis. Amongst all the elements of an entire
function, it is customary to single out three as the most important ones: these are the Taylor coefficients, the
zeros of the function and its behavior at infinity. The simplest characteristics of these elements are the rate
of decrease of the coefficients, the number of zeros in the sphere $|z|<r$, and the rate of growth of the
logarithm of the maximum modulus of the function in the ball $|z|\leq r$. It is customary to compare the
logarithm of the maximum modulus of the function with certain very smooth functions, called \it orders of
growth. \rm In what follows, we shall try to treat these latter facts and notions from a deeper historical
viewpoint.

Following (Burkhardt et al. 1899-1927, Dritter Teil, erste H\"{a}lfte, C.4, Nr. 26-36; Zweiter Teil, B.1.III)
once again, the starting point of entire transcendental functions is just the 1876 Weierstrass paper (see
(Weierstrass 1876)) in which, from well-known special cases treated by Cauchy\footnote{See (Cauchy 1829, pp.
174-213), namely the chapter entitled \it Usage du calcul des résidus par l'évaluation ou la transformation des
produits composés d'un nombre fini ou infini de facteurs, \rm as well as (Cauchy 1827, pp. 277-297), in which a
method of decomposing a meromorphic function into simple fractions had already been given before
Mittag-Leffler's work - see also (Saks \& Zygmund 1952, Chapter VII, Section 4) and (Sansone 1972, Chapter IV,
Section 8).} and Gauss regarding $\Gamma$ function and trigonometric functions, an infinite product expansion of
non-constant entire rational and transcendental functions was given. Therefore, the factorization theorems of
entire functions have opened the way to a new chapter of complex function theory\footnote{Following (Della Sala
et al. 2006), the term \it complex analysis \rm is quite recent because it has been used, for the first time, in
the International Mathematical Union Congress held at Vancouver in 1974, where a section specifically devoted to
Complex Analysis was considered for exposing researches in the theory of holomorphic functions of one or more
variables.}, that regarding the entire functions. As we have seen above, the historical pathways of Riemann zeta
function theory and of entire function theory intertwined among them, for the first time, just with the
introduction of the Riemann $\xi$ function, and, thenceforth, there were other similar intersection points along
the history of mathematics and its applications that we wish to consider in what follows. Therefore, it is
needful to briefly outline the main historical points concerning entire function theory from Weierstrass onward.
Almost all treatises on entire function theory start with a first chapter devoted to Weierstrass' factorization
theorem: in this regard, for instance, the first monograph on the subject, that is to say (Borel 1900), just
begins with a first chapter recalling the main points concerning Weierstrass' work on factorization product of
entire functions, hence Borel goes on with a second chapter devoted to explain the Laguerre works upon what
previously made by Weierstrass, and in which, among other things, the fundamental notions of \it genus \rm and
\it order \rm of an entire function were introduced starting from the Weierstrass factorization theorem (see
also (Sansone 1972, Chapter V, Section 8)). With respect to these appreciated Laguerre works and on the wake of
those made, above all, by E. Cesaro, G. Vivanti, A. Bassi and D. Pizzarello\footnote{Domenico Pizzarello was
born in Scilla (Messina, IT) on August 3, 1873 from Gaetano and Teresa Bellantoni. He was graduated in
Mathematics from the University of Rome on November 12, 1899. Then, he was assistant at the Infinitesimal
Calculus chair of Professor Giulio Vivanti at the University of Messina. Afterwards, he taught in various
Italian high schools until 1924, when he was appointed head of the Francesco Maurolico classical high school at
Messina, where he passed away on July 23, 1943.} on those entire functions having arbitrary genus but devoid of
exponential factors (see (Vivanti 1928) for a most complete bibliographical account of the contributions of
these last authors), the third chapter of Borel's monograph deals with the fundamental 1883 Poincaré's work on
entire functions, until up the celebrated Hadamard work outlined in the next chapter IV, to end with the
Picard's contribution delineated in the final chapter V. As the author himself says, a natural continuation of
Borel's monograph is (Blumenthal 1910), where a central chapter, the fourth one with a final Note II, deals with
a general theory of canonical products as it turned out be until 1910s. Furthermore, O. Blumenthal himself
contributed to the theory of entire functions (see (Valiron 1949, Chapter II, Section 3)).

In what follows, we mainly refer to (Borel 1900), (Vivanti 1928), (Sansone 1972, Chapter V), (Levin 1980,
Chapter I) and references therein. Retaking into consideration the above mentioned Weierstrass' theorem,
Laguerre (see (Laguerre 1882a,b,c; 1883; 1884)), from 1882 onwards, published some short but remarkable papers
on certain concepts and properties of entire functions, amongst which the notion of genus. To be precise,
Laguerre first defines $j$ as the \it genus \rm of the Weierstrass' factors $E_j(z)$, letting
$\gamma(E_j(z))=j$, then he calls \it genus \rm (or \it rank\rm) of the entire function $f(z)$ as given by (8),
the number $p=\max\{\partial eg\ g(z),\sup\{\gamma(E_j(z)\}\}$, which may also be $\infty$ when
$\sup\{\gamma(E_j(z)\}=\infty$ or, otherwise, when $g(z)$ is a transcendental entire function (so $\partial eg\
g(z)=\infty$). The importance of the natural numbers $\partial eg\ g(z)$ and $\sup\{j;E_j(z)\}$ with respect to
the Weierstrass decomposition (8), had already been recognized by Weierstrass himself, but it was Laguerre the
first who understood that their maximum value has instead more importance and usefulness from a formal
viewpoint. Most of Laguerre's work was pursued on entire functions of genus zero and one as well as on the study
of the distribution of the zeros of an entire function and its derivatives, taking constantly into account the
comparison between polynomials and entire functions on the wake of what had already known about the
determination of the zeros of the former. Following (Gonchar et al. 1997, Part I, Chapter 1, Section 1), entire
functions are a direct generalization of polynomials but their asymptotic behavior has an incomparably greater
diversity. The most important parameter characterizing properties of a polynomial is its degree. A
transcendental entire function that can be expanded into an infinite power series can be viewed as a kind of
polynomial of infinite degree, and the fact that the degree is infinite brings no additional information to the
statement that an entire function is not a polynomial. That is why, to characterize the asymptotic behavior of
an entire function, one must use other quantities and new notions, like those of order, genus, the maximum
modulus $M_f(r)$, and so forth.

According to (Burkhardt et al. 1899-1927, Dritter Teil, erste H\"{a}lfte, C.4, Nr. 26-36), (Fou\"{e}t 1904-07,
Tome II, Chapter IX, Section II, Number 283), (Marden 1949; 1966) and (P\'{o}lya \& Szeg\H{o} 1998a, Part III;
1998b, Part V), just upon the possible analogical transfer of the known results about the theory of polynomials
(above all results on their zeros, like Rolle's and Descartes' theorems - see (Marden 1949; 1966)) towards
entire functions, the next work of Poincaré, Hadamard, Borel, as well as of E. Schou, E. Cesaro, E. Fabry, E.
Laguerre\footnote{See (Laguerre 1898-1905, Tome I, p. 168) in which he retakes a notable result achieved by
Hermite who, in turn, used previous methods found by G.A.A. Plana, A. Genocchi e F. Chiò on the zeros of
algebraic equations.}, G.A.A. Plana, F. Chiò, A. Genocchi, C. Runge, C. Hermite, E. Maillet, E. Jaggi, C.A.
Dell'Agnola, J. von Puzyna, M.L.M. de Sparre, C. Frenzel, M. Petrovitch, A. Winternitz and others, will be
oriented (see (Vivanti 1906)) since the early 1900s till to the 1920s with pioneering works of E. Lindwart, R.
Jentzsch, G. Grommer, N. Kritikos and, above all, G. P\'{o}lya. The first notable results in this direction were
obtained both by E. Picard in the late 1870s, who dealt with the values of an arbitrary entire function, and by
Poincaré in the early 1880s (see (Poincaré 1882; 1883) and (Sansone 1972, Chapter V, Section 14)), who
established some first notable relations between the modulus of an entire function, its genus and the variations
of the magnitude of its coefficients; Poincaré was too the first one to apply entire function theory methods to
differential equations. Following (Marku\v{s}evi\v{c} 1966, Preface), the so-called \it Picard's little theorem
\rm roughly asserts that the equation $f(z)=a$, where $f(z)$ is a transcendental entire function and $a$ is a
given complex number, has in general, an infinite set of roots. This theorem clearly can be regarded as the
analog, to the infinite degree, of the Gauss' fundamental theory of algebra, according to which the number of
roots of the equation $p(x)=a$, where $p(x)$ is a polynomial, is equal to the degree of the polynomial.
Following (Vivanti 1928, Part III, Section 184), the Poincaré theorems were underestimated up to the 1892-93
Hadamard work (that will be discussed later), notwithstanding their importance for having opened the way to the
study of the relations between the distribution of the zeros of an entire function and the sequence of its
coefficients. The relevance of the zeros of an entire function is simply due to the fact that this last is
determined by factorization theorems of the Weierstrass' type. With Poincaré, the notion of order of an entire
function is introduced as follows. First, Poincaré proved that, if $f(z)$ is an entire function of genus $p$ (as
defined above) and $\rho$ is a positive integer greater than $p$ such that $\sum_{n\in\mathbb{N}}r_n^{-\rho}$ is
convergent, then, for every positive number $\alpha$, there exists an integer $r_0(\alpha)>0$ such that, for
$|z|=r\geq r_0(\alpha)$, we have $|f(z)|<e^{\alpha r^{\alpha}}$ in $|z|=r$. Then, if $f(z)$ is an entire
function, to characterize the growth of an entire function, we introduce a not-decreasing function as follows:
let $M_f(r)=\max_{|z|=r}|f(z)|$ be the maximum value of $|f(z)|$ on the sphere having center into the origin and
radius $r$. $M_f(r)$ is a continuous not-decreasing monotonic function of $r$, tending to $+\infty$ as
$r\rightarrow\infty$. For a polynomial $f$ of degree $n$, the following asymptotic relation holds $\ln
M_f(r)\sim n\ln r$, so that $n=\lim_{r\rightarrow\infty}\ln M_f(r)/\ln r$, i.e., the degree of a polynomial is
closely related to the asymptotics of $M_f(r)$. The ratio $\ln M_f(r)/\ln r$ tends to $\infty$ for all entire
transcendental functions. That is why the growth of $\ln M_f(r)$ is characterized by comparing it, not with $\ln
r$, but with faster growing functions, the most fruitful comparison being that with power functions. Thus, in
order to estimate the growth of transcendental entire functions, one must choose comparison functions that grow
more rapidly than powers of $r$. If one chooses functions of the form $e^{r^{k}}\ k\in\mathbb{N}$, as comparison
functions, then an entire function $f(z)$ is said to be of \it finite order \rm if there exist $k\in\mathbb{N}$
and $r_0(k)\in\mathbb{R}^+$ such that the inequality $M_f(r)<e^{r^k}$ is valid for sufficiently large values of
$r>r_0(k)$, the greatest lower bound of such numbers $k$, say $\rho$, being said the \it order \rm of the entire
function $f(z)$; finally, further indices, introduced by E.L. Lindel\"{o}ff, H. Von Schaper, A. Pringsheim and
E. Borel in the early 1900s (with further contributions due to S. Minetti in 1927 - see (Vivanti 1906; 1928,
Section 203)), and often called \it Lindel\"{o}ff indices, \rm have been introduced to estimate the rapidity of
variation of the modulus of the zeros, of the coefficients and of the function $M_f(r)$ of a given entire
function $f(z)$ (see (Borel 1900, Chapter III), (Vivanti 1928, Part III, Section 176) and (Levin 1980, Chapter
I)).

With the pioneering works of Jacques Hadamard (see (Hadamard 1892; 1893)), deepening of the previous results, as
well as new research directions, were pursued. If Poincaré was the first to apply the early results of entire
function theory to the study of differential equations, so Hadamard was the first to explicitly consider
applications of the theory of entire functions to the number theory, just working upon what previously made by
Riemann on the same subject. Following (Borel 1900, Chapter III) and (Maz'ya \& Shaposhnikova 1998, Chapter 1,
Section 1.10; Chapter 9, Section 9.2), the 1893 work of Hadamard roughly consisted in finding relations between
the behavior of the coefficients and the distribution of the zeros of an entire function as well as in providing
more explicit formulas of the Weierstrass type for functions growing slower than $\exp(|z|^{\lambda})$, so
becoming easier to prove the absence of exponential factors in the case of the Riemann $\xi$ function. Following
(Maz'ya \& Shaposhnikova 1998, Chapter 9, Section 9.2), the 1893 Hadamard memoir is divided into three parts.
The first one, after having improved some previous results achieved by Picard (and mentioned above), is mostly
devoted to the relationships between the rate of growth of $M_f(r)$ and the decreasing law of the coefficients
$c_n$ of the Taylor expansion of the given entire function $f(z)$. At the beginning, Hadamard found a majorant
for $M_f(r)$ described in terms of the sequence of the coefficients $c_n$, noting, for example, that if
$|c_n|<(n!)^{-1/\alpha},\ \alpha>0$, then $M_f(r)<e^{Hr^{\alpha}}$ for some constant $H$. Then, he considered
the inverse problem, already approached by Poincaré, to find the law of decreasing of the coefficients departing
from the law of growth of the function, extending Poincaré method in order to include functions satisfying
$M_f(r)<e^{V(r)}$, where $V(r)$ is an arbitrary positive increasing unbounded function. Then, as the central
goal of the paper in the aim of the author, Hadamard deals with an improvement of the Picard theorem, but it
will be Borel, in 1896, to give a general prove of it, valid for every entire function. Following (Gonchar et
al. 1997, Chapter 5, Section 1), 1879 Picard famous theorem is concerned with the problem of the distribution of
the values of entire functions and it may be considered as one of the starting points of the theory of the
distribution of the values of meromorphic functions which then began to develop only in the 1920s with the
pioneering works of R. Nevanlinna, albeit its very early starting point was the following formula
$$\log\frac{r^n|f(0)|}{|z_1...z_n|}=\frac{1}{2\pi}\int_0^{2\pi}\log|f(re^{i\theta})|d\theta,$$due either to J.L.W.
Jensen (see (Jensen 1898-99)) and J. Petersen (see (Petersen 1899)) but already known to Hadamard since the
early 1890s (see\footnote{With this historical remark, we might answer to a query expressed in (Davenport 1980,
Chapter 11, p. 77, footnote $^1$) about the use of Jensen's formula in proving Hadamard factorization theorem,
where textually the author says that \it <<strangely enough, Jensen's formula was not discovered until after the
work of Hadamard>>. \rm Also H.M. Edwards, in (Edwards 1974, Chapter 2, Section 2.1, footnote$^1$), about the
Hadamard proof of 1893 memoir, affirms that \it <<A major simplification is the use of Jensen's theorem, which
was not known at the time Hadamard was writing>>. \rm Nevertheless, there are historical proves which state the
contrary, amongst which a witness by a pupil of Hadamard, Szolem Mandelbrojt (1899-1983), who, in (Mandelbrojt
1967, p. 33), states that Hadamard was already in possession of Jensen's formula before Jensen himself, but did
not not publish it, since he could not find for it any important application (see also (Narkiewicz 2000, Chapter
5, Section 5.1, Number 1)). The first part of the Volume 13, Issue 1, of the year 1967 of the review \it
L'Enseignement Mathématique, \rm was devoted to main aspects of Hadamard mathematical work, with contributions
of P. Lévy, S. Mandelbrojt, B. Malgrange and P. Malliavin.} (Maz'ya \& Shaposhnikova 1998, Chapter 9, Section
9.2)), where $z_1,z_2,...,z_n,...$ are the zeros of $f(z)$, $|z_1|\leq |z_2|\leq...$ and $|z_n|\leq r\leq
|z_{n+1}|$. This formula was called \it Poisson-Jensen formula \rm by R. Nevanlinna around 1920s who, later,
will give an extended version of it, today known as \it Jensen-Nevanlinna formula \rm (see (Zhang 1993, Chapter
I)).

Following (Maz'ya \& Shaposhnikova 1998, Chapter 9, Section 9.2), in the second part of the 1893 memoir, as we
have already said above, Hadamard considers a question converse to the one treated by Poincaré, that is to say,
what information on the distribution of zeros of an entire function can be derived from the law of decreasing of
its coefficients? In particular, he shows that the genus of the entire function is equal to the integer part
$[\lambda]$ of $\lambda$ provided by $|c_n|(n!)^{-1/(\lambda+1)}\underset{n\rightarrow\infty}{\longrightarrow}0$
with $\lambda$, in general, not integer\footnote{Poincaré proved that, if $f(z)$ is an entire function of genus
$p$ such that $f(z)=\sum_{n\in\mathbb{N}_0}c_nz^n$, then
$(n!)^{1/(1+p)}c_n\underset{n\rightarrow\infty}{\longrightarrow}0$ (see (Sansone 1972, Chapter V, Section 9)).
Further studies on entire functions of non-integral order were also attained by L. Leau in 1906.}. From this
statement, one concludes that a function $f(z)$ has genus zero if $M_f(r)<e^{Hr^{\alpha}}$ holds with
$\alpha<1$. Hadamard's theorem, nevertheless, is less precise for the case when $\lambda$ is integer, because,
in this case, the function may have genus either $\lambda$ or $\lambda+1$. Hadamard's result was improved by
Borel in 1897 (see also (Borel 1900)), who used two important characteristic parameters of an entire function,
namely the order $\rho$ (that, d'après Borel, he called \it apparent order\rm) and the exponent of convergence
of zeros $p$ (said to be the \it real order\rm, in his terminology borrowed by Von Schaper). The order is the
upper lower bound $\rho$ of the numbers $\alpha$ such that $M_f(r)<e^{Hr^{\alpha}}$, its explicit expression
being given by
$$\rho=\limsup_{r\rightarrow\infty}\frac{\ln\ln M_f(r)}{\ln r}$$which might therefore be taken as the definition of the
order of the function $f$; the quantity instead $\lambda_{\rho}\doteq\liminf_{r\rightarrow\infty}\ln\ln
M_f(r)/\ln r$ is said to be the \it lower order \rm of $f$. For a polynomial we have $\rho=0$, while for the
transcendental functions $\exp z,\sin z,\exp(\exp z)$ the order is respectively $1,1$ and $\infty$. If we have
$\rho<\infty$, then the quantity $\sigma\doteq\limsup_{r\rightarrow\infty}r^{-\rho}\ln M_f(r)$ is called the \it
type value \rm of the entire function $f$. The \it exponent of convergence \rm of the zeros, say $p$, is defined
as the upper lower bound of those $\lambda>0$ for which the series $\sum_n|z_n|^{-(\lambda+1)}$ converges. One
can also check that the exponent of convergence is also provided by $\mu=\limsup(\ln n/\ln|z_n|)$. Hadamard
proved that $\rho\geq p$, often said to be the \it first Hadamard theorem \rm (see (Sansone 1972, Chapter V,
Section 3)). If an entire function has only a finite number of zeros, then we say that it has exponent of
convergence zero. Thus, while the order characterizes the maximal possible growth of the function, the exponent
of convergence $p$ is an indicator of the density of the distribution of the zeros of $f(z)$. Therefore, the
Hadamard's refinement of Weierstrass formula (11) by using the notion of order, states that, if $f$ is an entire
function of finite order $\rho$, then the entire function $g(z)$ in (11) is a polynomial of degree not higher
than $[\rho]$. As we have been said above, Borel obtained a kind of converse to this result by showing how the
order can be found from the factorization formula, his theorem stating that, if $\mu<\infty$ and $g(z)$ is the
polynomial appearing in (11), then $f(z)$ is an entire function of order $p=\max\{\mu,q\}$. Finally, the lst
third part his memoir, Hadamard applied his results on the genus of an entire function, achieved in the first
part, to the celebrated Riemann zeta function. To be precise, Riemann reduced the study of the zeta function to
that of the even entire function $\xi$ defined by $\xi(z)=z(z-1)\Gamma(z/2)\zeta(z)/2\pi^{z/2}$, and writing
$\xi$ as the series $\xi(z)=b_0+b_2z^2+b_4z^4+...$, Hadamard proved the inequality
$|b_m|<(m!)^{-1/2-\varepsilon},\varepsilon>0$, thus verifying, for $1/\alpha=1/2+\varepsilon$, the following
estimate $|c_n|<(n!)^{-1/\alpha},\alpha>0$ deduced in the first part of his memoir and briefly mentioned above,
so that it follows that the genus of $\xi$, as a function of $z^2$, is equal to zero and that
$$\xi(z)=\xi(0)\prod_{k=1}^{\infty}\Big(1-\frac{z^2}{\alpha_k^2}\Big)$$where the $\alpha_k$ are the zeros of
$\xi$, this last property, as is well-known, having already been provided by Riemann, in his celebrated 1859
paper, but without a rigorous proof.

Following (Gonchar et al. 1997, Part I, Chapter 1, Section 1), the classical Weierstrass theorem is well-known
on the representation of an entire function with a given set of zeros in the form of an infinite product of
Weierstrass primary factors. In the works of Borel and Hadamard on entire function of finite order, the
Weierstrass theorem was significantly improved, showing that the genus of the primary factors could be one and
the same, in the representation of an entire function only a finite number of parameters being not defined by
the set of zeros. As early as the turn of the 20th century, the theory of factorization of entire functions was
regarded as fully completed, albeit in a series of works started in 1945, M. Dzhrbashyan and his school
constructed a new factorization theory, as well as H. Behnke and K. Stein extended, in 1948, factorization
theorem to arbitrary non-compact Riemann surfaces (see (Remmert 1998, Chapter 4, Section 2)). The remarkable
work of Behnke and Stein (see (Behnke \& Stein 1948)) revaluated the role of the so-called \it Runge sets \rm in
the theory of non-compact Riemann surfaces, demonstrating a Runge type theorem. Following (Maurin 1997, Part V,
Chapters 3 and 6), Carl Runge (1856-1927) gave fundamental contributions, between 1885 and 1889, to the theory
of complex functions, proving a basic result, in which he introduced particular sets later called \it Runge's
sets, \rm regarding the approximation of holomorphic functions by a sequence of polynomials, almost in the same
years in which Weierstrass gave his as much notable theorem on the approximation of a function on interval by
polynomials. From Runge outcomes, hence also from Behnke-Stein ones, it follows much of the representation
theorems for meromorphic functions due to Weierstrass and Mittag-Leffler. To point out the central result of the
1893 Hadamard paper, we recall, following (Levin 1980, Chapter I), the Weierstrass theorem, namely that every
entire function $f(z)$ may be represented in the form
$$f(z)=z^me^{g(z)}\prod_{n=1}^{\omega}G\Big(\frac{z}{a_n};p_n\Big)\qquad (\omega\leq\infty)$$where $g(z)$ is an
entire function, $a_n$ are the non-zero roots of $f(z)$, $m$ is the order of the zero of $f(z)$ at the origin,
and $G(u;p)=(1-u)\exp(u+u^2/2+...+u^p/p)$ is the generic \it primary factor. \rm The sequence of numbers $p_n$
is not uniquely determined and, therefore, the function $g(z)$ is not uniquely determined either. After Laguerre
work, the representation of the function $f(z)$ is considerably simpler if the numbers $a_n$ satisfy the
following supplementary condition, that is, the series $\sum_{n\in\mathbb{N}}|a_n|^{-(\lambda+1)}$ converges for
some positive $\lambda$. In this case, let $p$ denote the smallest integer $\lambda>0$ for which the series
$\sum_{n\in\mathbb{N}}|a_n|^{-(\lambda+1)}$ converges. Thus, also the infinite product
$\prod_{n\in\mathbb{N}}G(z/a_n;p)$ converges uniformly: it is called a \it canonical product, \rm and the number
$p$ is called, following B.J. Levin, the \it genus \rm of the canonical product, or else, following Borel, the
exponent of convergence of the zeros $a_n$. If $g(z)$ is a polynomial, $f(z)$ is said to be an entire function
of \it finite genus. \rm If $q$ is the degree of the polynomial $g(z)$, the largest of the numbers $p$ and $q$
is called the \it genus \rm of $f(z)$. If $g(z)$ is not a polynomial or if the series
$\sum_{n\in\mathbb{N}}|a_n|^{-(\lambda+1)}$ diverges for all the values of $\lambda>0$, then the genus is said
to be \it infinite. \rm The representation of an entire function as an infinite product makes it possible to
establish a very important dependence between the growth of the function and the density of distribution of its
zeros. As a measure of the density of the sequence of the points $a_n$, having no finite limit point, we
introduce, d'après Borel (see (Borel 1900, Chapter II)), the convergent exponent of the sequence
$a_1,a_2,...,a_n,...$, with $a_n\neq 0$ definitively and $\lim_{n\rightarrow\infty}a_n=\infty$, which is defined
by the greatest lower bound of the numbers $\lambda>0$ for which the series
$\sum_{n\in\mathbb{N}}(1/|a_n|^{\lambda+1})$ converges. Clearly, the more rapidly the sequence of numbers
$|a_n|$ increases, the smaller will be the convergent exponent, which may be also zero. A more precise
description of the density of the sequence $\{a_n\}_{n\in\mathbb{N}}$, than the convergence exponent is given by
the growth of the function $n(r)$, said to be \it zero-counting function, \rm equal to the number of points of
the sequence in the circle $|z|<r$, so that by the \it order \rm of this monotone function we mean the number
$\rho_1=\limsup_{r\rightarrow\infty}(\ln n(r)/\ln r)$, and by the \it upper density \rm of the sequence
$\{a_n\}$, we mean the number $\Delta=\limsup_{r\rightarrow\infty}(n(r)/r^{\rho_1})$; if the limit exists, then
$\Delta$ is simply called the \it density \rm of the sequence $\{a_n\}$. Classical results on the connection
between the growth of an entire function and the distribution of its zeros mainly describe the connection
between $\ln M_f(r)$ and the zero-counting function $n(r)$. If $f$ is a polynomial, then
$\lim_{r\rightarrow\infty}n(r)=n$ if and only if $\ln M_f(r)\sim n\ln r$, whereas no simple connection exists
between the asymptotic behavior of $\ln M_f(r)$ and $n(r)$ for entire transcendental functions. It is possible
to prove that the convergent exponent of the sequence $\{a_n\}$, with $\lim_{n\rightarrow\infty}|a_n|=\infty$,
is equal to the order of the corresponding function $n(r)$. Borel moreover proved that the order $\rho$ of the
canonical product $\Pi(z)=\prod_{n\in\mathbb{N}}G(z/a_n;p)$, does not exceed the convergence exponent $\rho_1$
of the sequence $\{a_n\}$, even better $p=\rho_1$ (\it Borel theorem\rm; see also (Sansone 1972, Chapter V,
Section 6)). Hadamard's factorization theorem is a refinement concerning the representation of entire functions
of finite order, and is one of the classical theorems of the theory of entire functions. This theorem states
that an entire function $f(z)$ of finite order $\rho$ and genus $p$, can be represented in the form
$$f(z)=z^me^{P(z)}\prod_{n=1}^{\omega}G\Big(\frac{z}{a_n};p\Big)\qquad(\omega\leq\infty),$$where $a_n$ are the
non-zero roots of $f(z)$, $p\leq\rho$, $P(z)$ is a polynomial whose degree $q$ does not exceed $[\rho]$, and $m$
is the multiplicity of the zero at the origin. This theorem, hence, states that the genus of an entire function
does not exceed its order. Sometimes, the factor $e^{P(z)}$ is also called \it external exponential factor \rm
(see (Vivanti 1928) and (Sansone 1972, Chapter V, Section 5)). Following (Maz'ya \& Shaposhnikova 1998, Chapter
9, Section 9.2), Borel obtained as well a sort of converse to this result by showing how the order can be found
from the factorization formula, stating as follows: if $p<\infty$ and $P(z)$ is a polynomial of degree $q$, then
$f(z)$ is a function of order $\rho=\max\{p,q\}=\max\{\rho_1,q\}$ (via Borel theorem). Finally, we recall that
in this 1893 Hadamard memoir, further estimates for the minimum of the modulus of an entire function were also
established (forming the so-called \it second Hadamard theorem\rm), upon which, then, Borel (see (Borel 1900)),
P. Boutroux, E. Maillet, A. Kraft, B. Lindgren, G. Faber, A. Denjoy, F. Schottky, E. Lindel\"{o}f, J.L.W.
Jensen, J.E. Littlewood, G. Hardy, W. Gross, R. Mattson, G. Rémoundos, O. Blumenthal, R. Mattson, E. Landau, C.
Carathéodory, A. Wahlund, G. P\'{o}lya, A. Wiman, P. Fatou, P. Montel, T. Carleman, L. Bieberbach, F. Iversen,
E. Phragmèn, A. Pringsheim, E.F. Collingwood, R.C. Young, J. Sire, G. Julia, A. Hurwitz, G. Valiron and others
will work on, at first providing further improvements to the estimates both for the minimum and the maximum of
the modulus of an entire function and its derivatives (see (Burkhardt et al. 1899-1927, Dritter Teil, erste
H\"{a}lfte, C.4, Nr. 26-36) and (Sansone 1972, Chapter V, Sections 4, 13 and 16)), till to carry out a complete,
rich and autonomous chapter of complex analysis. Later studies on entire functions having integral order were
also accomplished, in the early 1900s, above all by A. Pringsheim as well as by E. Lindel\"{o}f and E. Phragmèn
who defined what is known as \it Phragmèn-Lindel\"{o}f indicator \rm of an entire function which will be the
basic characteristic of growth of an entire function of finite order (see (Ostrovski\v{i} \& Sodin 1998, Section
3)). Anyway, the description of the state-of-the-art of the theory of entire functions until 1940s, may be found
above all in the treatise (Valiron 1949), as well as in the last editions of the well-known treatise (Whittaker
\& Watson 1927). Furthermore, it is also useful to look at the notes by G. Valiron, to the 1921 second edition
of Borel's treatise on entire functions (i.e., (Borel 1900)), that is to say (Borel 1921), where, at the
beginning of the Note IV, Valiron says that\\

\it<<La théorie des fonctions entières a fait l'object d'un tr\^{e}s grand nomhre de travaux depuis la
publication des Mémoires fondamentaux de J. Hadamard et E. Borel. Plus de cent cinquante Mémoires ou Notes ont
été publiés entre 1900, dale de la première édition des Le\c{c}ons sur les fonctions entières, et 1920; beaucoup
de ces travaux ont leur origine dans les suggestions de E. Borel. On peut répartir ccs recherches en quatre
groupes: 1$^o$. Étude de la relation entre la croissance du module maximum et la croissance de la suite des
coefficients de la fonction et démonstrations élémentaires du théorème de Picard; 2$^o$. Études directes de la
relation entre la suite des zéros et la croissance du module maximum; 3$^o$. Recherches sur les fonctions
inverses et généralisations du théorème de Picard; 4$^o$. Recherches de nature algébrique et étude des fonctions
d'ordre fini considérées comme fonctions limites d'une suite de polynomes. Il eût été difficile de donner dans
quelques pages un aperçu des travaux particuliers de chaque auteur, certaines questions ayanl été traitées
simultanément ou d'une fa\c{c}on indipendante par plusieurs mathématiciens \rm [...]\it>>.\\\\\rm Afterwards,
Valiron briefly exposes the main results achieved by those mathematicians whose names have been just recalled
above, a more detailed treatment being given in his treatise (Valiron 1949) which covers the European area until
up mid-1900s. After such a period, a great impulse to the theory of entire functions was given by Russian school
which grew up around Boris Yakovlevich Levin (1906-1993) whose scientific and human biography may be found in
the preface to (Levin 1996). Herein, we give a very brief flashing out on the research work on entire function
theory achieved by Russian school, referring to (Ostrovskii \& Sodin 1998; 2003) for a deeper knowledge. The
fundamental problem in the theory of entire functions is the problem of the connection between the growth of an
entire function and the distribution of its zeros, a basic characteristic of growth of an entire function of
finite order being the so-called \it Phragmén-Lindel\"{o}f indicator \rm defined by
\begin{equation}h(\varphi,f)=\limsup_{r\rightarrow\infty}r^{-\rho(r)}\ln|f(re^{i\varphi})|,\qquad\varphi\in[0,2\pi].\end{equation}
The systematic study of the connection of the indicator with the distribution of zeros, started in the 1930s
with the Russian school leaded by Levin and Mark G. Kre\v{i}n. Following (Levin 1980, Chapter VIII), the
representation of an entire function by a power series shows the simple fact that any entire function is the
limit of a sequence of polynomials which converges uniformly in every bounded domain. If we impose on the
polynomials which are approaching uniformly the given entire function the additional requirement that their
zeros belong to a certain set, then the limit functions will form a special class, depending on the set. The
first notable results in this direction were due to Laguerre (see (Laguerre 1898-1905, Tome I, pp. 161-366)),
who gave a complete characterization of the entire functions that can be uniformly approximated by polynomials,
distinguishing two chief cases: the first one (I) in which the zeros of these polynomials are all positive, and
the second one (II) in which these zeros are all real. In this latter case, a proof of his theorem was later
given by G. P\'{o}lya (see (P\'{o}lya 1913)), while a more complete investigation of the convergence of
sequences of such polynomials was carried out by E. Lindwart and P\'{o}lya (see (Lindwart \& P\'{o}lya 1914)),
showing, in particular, that in the two above just mentioned cases I and II (as well as in more general cases),
the uniform convergence of a sequence of polynomials, in some disk $|z|<R$, implies its uniform convergence on
any bounded subset of the complex plane. Now, the main results achieved in the theory of representation of an
entire function by a power series, namely that any entire function is the limit of a sequence of polynomials
which converges uniformly in every bounded domain, in turn refer to the theory of approximation of entire
functions by polynomials whose zeros lie in a given region, say $G$, of the open or closed upper complex
half-plane. Besides important results achieved by E. Routh and A. Hurwitz in the 1890s, the basic algebraic fact
in this domain is a theorem stated by C. Hermite in 1856 (see (Hermite 1856a,b)) and C. Biehler (see (Biehler
1879)), and nowadays known as \it Hermite-Biehler theorem, \rm which provides a necessary and sufficient
condition for a polynomial of the type $\omega(z)=P(z)+iQ(z)$, with $P$ and $Q$ real polynomials, not have any
root in the closed lower half-plane $\Im z\leq 0$, imposing conditions just on $P$ and $Q$. In 20th century, the
Russian school achieved further deep results along this direction and, in carrying over the Hermite-Biehler
criterion to arbitrary entire functions, an essential role is played by particular classes of entire functions,
introduced by M.G. Kre\v{i}n in a 1938 work devoted to the extension of some previous Hurwitz criteria for zeros
of entire functions (see (Ostrovski\v{i} 1994)), and said to be \it Hermite-Biehler classes (HB \rm and \it
$\overline{HB}$ classes). \rm An entire function $\omega(z)$ is said to be a function of class $HB$
[respectively $\overline{{HB}}$] if it has no roots in the closed [open] lower half-plane $\Im z\leq 0$, and if
$|\omega(z)/\bar{\omega}(z)|<1$ [$|\omega(z)/\bar{\omega}(z)|\leq 1$] for\footnote{Here we understand by
$\bar{\omega}(z)$ the entire function obtained from $\omega(z)$ by replacing all the coefficients in its Taylor
series by their conjugates.} $\Im z>0$. On the basis of results achieved by M.G. Kre\v{i}n, N.N. Me\v{i}man,
Ju.I. Ne\v{i}mark, N.J. Akhiezer, L.S. Pontrjagin, B.Ja. Levin, N.G. \v{C}ebotarev and others, around 1940s and
1950s, simple criteria for an entire function to belong to the class $HB$, as well as representation theorems
for elements of this class of entire functions using special infinite products, were provided (see (Levin 1980,
Chapter VII)). A polynomial which has no zeros in open lower half-plane will be called an \it H-polynomial. \rm
Then, the so-called \it Laguerre-P\'{o}lya class (LP class) \rm is given by a particular class of entire
functions obtained as limit of a sequence of $H$-polynomials uniformly converging in an angular
$\delta$-neighborhood of the origin, hence in an arbitrary bounded domain (see (Levin 1980, Chapter VIII))
through a criterion called \it Laguerre-P\'{o}lya theorem \rm due to previous outcomes obtained by Laguerre in
the late 1890s. The classical Laguerre-P\'{o}lya theorem asserts that an entire function $f$ belongs to this
class if and only if
\begin{equation}f(z)=e^{-\gamma z^2+\beta z+\alpha}z^m\prod_n\Big(1-\frac{z}{z_n}\Big)e^{\frac{z}{z_n}}\end{equation}
where all $z_n$, $\alpha$ and $\beta$ are real, $\gamma\leq 0$, $m\in\mathbb{N}_0$ and
$\sum_n|z_n|^{-2}<\infty$. Following (Bergweiler et al., 2002), in passing we recall that Laguerre-P\'{o}lya
class $LP$ coincides with the closure of the set of all real polynomials with only real zeros, with respect to
uniform convergence on compact subsets of the plane. This is just what was originally proved by Laguerre in
(Laguerre 1882c) for the case of polynomials with positive zeros and by P\'{o}lya in (P\'{o}lya 1913) in the
general case. It follows that $LP$ class is closed under differentiation, so that all derivatives of a function
$f\in LP$ have only real zeros. P\'{o}lya, in (P\'{o}lya 1913), also asked whether the converse is true, that is
to say, if all derivatives of a real entire function $f$ have only real zeros then $f\in LP$. This conjecture
was later proved by S. Hellerstein and J. Williamson in 1977 (see (Bergweiler et al., 2002) and references
therein). Other notable results for entire functions belonging to $LP$ class were achieved by E. Malo in the
late 1890s and by G. P\'{o}lya, J. Eger\'{a}dry, E. Lindwart, A.I. Marku\v{s}evi\v{c}, I. Schur, J.L.V. Jensen,
O. Szàsz, J. Korevaar, M. Fekete, E. Meissner, E. B\'{a}lint, D.R. Curtiss, J. Grommer, M. Fujiwara, E. Frank,
S. Benjaminowitsch, K.T. Vahlen, A.J. Kempner, I. Schoenberg, S. Takahashi, N. Obrechkoff and others, between
the 1910s and the 1950s (see (Levin 1980, Chapter VIII) and (Marden 1949)). For other interesting historical
aspects of entire function theory see also (Korevaar 2013) and references therein.

Finally, a notable work on after the mid-1900s entire function theory developments has surely been the one
achieved by Louis de Branges since 1950s with his theory of Hilbert spaces of entire functions, culminated in
the treatise (de Branges 1968). In the intention of the author expressed in the Preface to the latter, anyone
approaches Hilbert spaces of entire functions for the first time will see the theory as an application of the
classical theory of entire functions. The main tools are drawn from classical analysis, and these are the
Phragmén-Lindel\"{o}f principle, the Poisson representation of positive harmonic functions, the factorization
theorem for functions of P\'{o}lya class, Nevanlinna's theory of functions of bounded type, and the
Titchmarsh-Valiron theorem relating growth and zeros of entire functions of exponential type. The origins of
Hilbert spaces of entire functions are found in a theorem of Paley and Wiener that characterizes finite Fourier
transforms as entire functions of exponential type which are square integrable on the real axis. This result has
a striking consequence which is meaningful without any knowledge of Fourier analysis. The identity
$$\int_{-\infty}^{+\infty}|F(t)|^2dt=\frac{\pi}{a}\sum_{-\infty}^{+\infty}|F(n\pi/a)|^2$$
which holds for any entire function $F(z)$ of exponential type at most a which is square integrable on the real
line. The formula is ordinarily derived from a Fourier series expansion of the Fourier transform of $F(z)$. In
the fall of 1958, de Branges discovered an essentially different proof which requires nothing more than a
knowledge of Cauchy's formula and basic properties of orthogonal sets. The identity is a special case of a
general formula which relates mean squares of entire functions on the whole real axis to mean squares on a
sequence of real points. Certain Hilbert spaces, whose elements are entire functions, enter into the proof of
the general identity. Since such an identity has its origins in Fourier analysis, de Branges conjectured that a
generalization of Fourier analysis was associated with these spaces, spending the years 1958-1961 to verify this
conjecture. The outlines of this de Branges theory are best seen by using the invariant subspace concept. The
theory of invariant subspaces sprung out of some early studies of the end of 19th Century on the zeros of
polynomials and their generalization by C. Hermite and T.J. Stieltjes, just after the Riemann conjecture (see
(de Branges 1968; 1986)). The next axiomatization of integration just due to Stieltjes in the last years of 19th
century, greatly contributed to settling up these studies, above all thanks to the work of Hilbert. A
fundamental problem is to determine the invariant subspaces of any bounded linear transformation in Hilbert
space and to write the transformation as an integral in terms of invariant subspaces: this is one of the main
problems of spectral analysis. A similar problem can be stated for an unbounded or partially defined
transformation once the invariant subspace concept is clarified. To this purpose, it may help to say that there
exist invariant subspaces appropriate for a certain kind of transformation, the theory of Hilbert spaces of
entire functions being the best behaved of all invariant subspace theories. Moreover, nontrivial invariant
subspaces always exist for nontrivial transformations; invariant subspaces are totally ordered by inclusion. The
transformation admits an integral representation in terms of its invariant subspaces, this representation being
stated as a generalization of the Paley-Wiener theorem and of the Fourier transformation. Hilbert spaces of
entire functions also have other applications, an obvious area being the approximation by polynomials of entire
functions of exponential type. On the other hand, it was just through such problems that de Branges discovered
such spaces. Although it is easy to construct entire functions with given zeros, it is quite difficult to
estimate the functions so obtained. To this end, de Branges used the extreme point method to construct
nontrivial entire functions whose zeros lie in a given set and whose reciprocals admit absolutely convergent
partial fraction decompositions. A classical problem is to estimate an entire function of exponential type in
the complex plane from estimates on a given sequence of points, so de Branges constructed Hilbert spaces of
entire functions of exponential type with norm determined by what happens on a given sequence of real points.

\newpage\section*{3. On some applications of the theory of entire functions. Applications to the case of Lee-Yang theorem}
\addcontentsline{toc}{section}{3. On some applications of the theory of entire functions. Applications to the
case of Lee-Yang theorem}

\bf 3.1. On the applications of entire function theory to Riemann zeta function: the works of J. Hadamard, H.
Von Mangoldt, E. Landau, G. P\'{o}lya, and others. \rm Following (Valiron 1949, Chapter I), the early origin of
the general theory of entire functions, that is to say of functions which are regular throughout the finite
portion of the plane of the complex variable, is to be found in the work of Weierstrass. He shown that the
fundamental theorem concerning the factorization of a polynomial can be extended to cover the case of such
functions, and that in the neighborhood of an isolated essential singularity the value of a uniform function is
indeterminate. These two theorems have been the starting point of all subsequent research. Weierstrass himself
did not complete his second theorem, this having been done in 1879 by Picard who proved that in the neighborhood
of an isolated essential singularity a uniform function actually assumes every value with only one possible
exception. Much important work, the earliest of which was due to Borel, has been done in connection with
Picard's theorem; and the consequent introduction of new methods has resulted in much light being thrown on
obscure points in the theory of analytic functions. The notion of the genus of a Weierstrassian product was
introduced and its importance first recognized by Laguerre, but it was not until after the work of Poincaré and
Hadamard had been done that any substantial advance was made in this direction. Here also Borel has enriched the
theory with new ideas, and his work has done much to reveal the relationship between the two points of view and
profoundly influenced the trend of subsequent research. The theory of entire functions, or more generally of the
functions having an isolated singularity at infinity, may be developed in two directions. On the one hand, we
may seek to deduce from facts about the zeros information concerning the formal factorization of an entire
function; on the other hand, regarding the problem from the point of view of the theorems of Weierstrass and
Picard, we may endeavor to acquire a deeper insight into the nature of the function by investigating the
properties of the roots of an equation of the type $f(z)-a=0$, where $f(z)$ is an entire function. The study of
the zeros of these functions thus serves a double purpose, since it contributes to advance the theory along both
these avenues. The first step consists in giving all the theorems due to Hadamard and Borel concerning the
formal factorization of an entire function, and then proceed towards a direct investigation of the moduli of its
zeros by the methods provided by Borel, the resulting outcomes bringing out very clearly the close relationships
existing between these two points of view. Along this treatment, the Jensen work plays a fundamental role. Apart
Weierstrass' work, the Hadamard one on factorization of entire functions started from previous work of Poincaré
but was inspired by Riemann 1859 paper, to be precise by problematic raised by Riemann $\xi$ function and its
properties. The next work of Borel, then, based on Hadamard one. In this section, nevertheless, we outline only
the main contributions respectively owned to Hadamard, Edmund Landau and George P\'{o}lya, the only ones
who worked on that meeting land between the theory of Riemann zeta function and the theory of entire functions.\\\\
$\bullet$ \it The contribution of J. Hadamard. \rm In section 5.1 of the previous chapter, we have briefly
outlined the main content of the celebrated 1893 Hadamard memoir, where in the last and third part he deals with
Riemann $\xi$ function. Now, in this section, we wish to start with an historical deepening of this memoir, to
carry on then with other remarkable works centered on the applications of entire function theory to Riemann zeta
function issues, amongst which those achieved by H. Von Mangoldt, E. Landau, G. P\`{o}lya, and others. Following
(Narkiewicz 2000, Chapter 5, Section 5.1, Number 1), the last twenty years of the 19th century seen a rapid
progress in the theory of complex functions, summed up in the monumental works of \'{E}mile Picard and Camille
Jordan. The development of the theory of entire functions, started with pioneering 1876 Weierstrass work and
rounded up by Hadamard in 1893, revived the interest in Riemann's memoir and forced attempts to use these new
developments to solve questions left open by Riemann. This led to the first proofs of the \it Prime Number
Theorem \rm (PNT), early conjectured in the late 1700s by Gauss and Legendre independently of each other, in the
form (d'après E. Landau) $\theta(x)=\sum_{p\leq x}\ln x=(1+o(1))x$ obtained independently by Hadamard and de la
Vallée-Poussin in 1896. They both started with establishing the non-vanishing of $\zeta(s)$ on the line $\Re
s=1$ but obtained this result in completely different ways, but with equivalent results (as pointed out by Von
Schaper in his PhD dissertation of 1898). Also the deduction of the Prime Number Theorem from that result is
differently done by them, even if, according to (Montgomery \& Vaughan 2006, Chapter 6, Section 6.3) and
(Bateman \& Diamond 1996), the methods of Hadamard\footnote{Furthermore, in (Ayoub 1963, Chapter II, Notes to
Chapter II), the author says that it is worthwhile noting that Hadamard based his proof on that given by E.
Cahen in (Cahen ), where it is assumed the truth of the Riemann hypothesis, ascribing the ideas of his proof to
G.H. Halphen.} and de la Vallée-Poussin depended on the analytic continuation of $\zeta(s)$, on bounds for the
size of $\zeta(s)$ in the complex plane, and on Hadamard theory of entire functions. Anyway, also (Ayoub 1963,
Chapter II, Section 6) claims that the original 1896 proof by de la Vallée-Poussin made use of the product
formula provided by Hadamard work of 1893, in deducing an expression for $\zeta'(s)/\zeta(s)$. Likewise, in
(Chen 2003, Chapter 6, Section 6.1), the author comments that Hadamard product representation played an
important role in the first proof of prime number theorem. Hadamard). Finally, as also pointed out in (It\^{o}
1993, Article 429, Section B), almost all the outcomes delineated above concerning entire functions, originated
in the study of the zeros of the Riemann zeta function and constitute the beginning of the theory of entire
functions. Therefore, due to its fundamental importance, we shall return back again in discussing upon this 1893
Hadamard work.

The members of the evaluation's commission of the annual \it grand prix des sciences mathématiques \rm raffled
by the French Academy of Sciences, namely Jordan, Poincaré, Hermite, Darboux and Picard, decided to award the
celebrated 1893 Hadamard paper for having put attention to certain apparently minor questions treated by Riemann
in his famous paper on number theory, from which arose new and unexpected results of entire function theory, as
already said in the previous sections. In their report (see (Jordan et al. 1892)), published at the pages
1120-1122 of the Tome CXV, Number 25 of the \it Comptes Rendus de l'Academie des Sciences de Paris \rm in the
year 1892, the relevance of the new complex function $\zeta(s)$ for studying number theory issues, introduced by
Riemann in his 1859 celebrated seminal paper, have been pointed out together some its chief properties, just by
Riemann himself but without providing any rigorous proof of them. In 1885, George H. Halphen (1844-1889) (see
(Narkiewicz 2000, Chapter 4, Section 4.3) and references therein), referring to the latter unsolved Riemann
questions, wrote that\\

\it <<Avant qu'on sache établir le théorème de Riemann (et il est vraisemblable que Riemann ne l'a pas su
faire), il faudra de nouveaux progrès sur une notion encore bien nouvelle, le genre des transcendantes
entières>>.\\\\\rm Thus Hadamard, within the framework of the new entire function theory and in agreement with
the above Halphen's consideration\footnote{Although Hadamard never quoted Halphen in his 1893 paper.}, proved
one of these, determining the genus of the auxiliary $\xi(s)$ function which is as an entire function of the
variable $s^2$ having genus zero but, at the same time, from an apparently minor issue (drawn from number
theory), opening the way to new and fruitful directions in entire function theory (see also (Jordan et al. 1892,
p. 1122)). Therefore, as it has been many times said above, the theory of entire function has plainly played -
and still plays - a crucial and deep role in Riemann's theory of prime numbers whose unique 1859 number theory
paper has been therefore one of the chief input for the development, on the one hand, of the number theory as
well as, on the other hand, of the entire function theory itself with the next study of the Riemann $\xi$
function and its infinite product factorization, opened by Hadamard work. Hence, we go on with a more
particularized historical analysis of the 1893 celebrated Hadamard paper starting from the 1859 Riemann original
memoir. Before all, we briefly recall the main points along which the Hadamard memoir lays down, referring to
the previous chapter for more information. To be precise, Hadamard starts with the consideration that the
decomposition of an entire function $f(x)$ into primary factors, achieved via Weierstrass' method as follows
\begin{equation}f(x)=e^{g(x)}\prod_{j=1}^{\infty}\Big(1-\frac{x}{\xi_j}\Big)e^{Q_j(x)},\end{equation}leads to the
notion of genus of an entire function, taking into consideration a result as early as achieved by Poincaré in
1883 (and already mentioned in the previous chapter), namely that, given an entire function of genus $p$, then
the coefficient of $x^m$, say $c_m$, multiplied by $\sqrt[p+1]{m!}$, tends to zero as $m\rightarrow\infty$, as
well as outcomes achieved by Picard, Hadamard expresses the intention to complete this Poincaré result, trying
to find the general possible relations between the properties of an entire function and the laws of decreasing
of its coefficients. In particular, as we have already been said in the previous chapter, Hadamard proves that,
if $c_m$ is lower than $({m!})^{-\frac{1}{p}}$, then it has, in general, a genus lower than $p$. Hence, in the
first and second parts of the memoir, Hadamard proceeds finding relations between the decreasing law of the
coefficients of the Taylor expansion of the given entire function $f(x)$ and its order of magnitude for high
values of the variable $x$. Then, in the third and last part of his memoir, Hadamard applies what has been
proved in the previous parts, to Riemann $\xi$ function. Precisely, Hadamard reconsiders the 1859 Riemann paper
in which he first introduces the function $\zeta(s)$ to study properties of number theory, from which he then
obtained a particular entire function $\xi(s)$ defined by
\begin{equation}\xi(x)=\frac{1}{2}-\big(x^2+\frac{1}{4}\big)\int_1^{\infty}\Psi(t)t^{-\frac{3}{4}}
\cos\big(\frac{x}{2}\ln t\big)dt\end{equation}with $\Psi(t)=\sum_{n=1}^{\infty}e^{-n^2\pi t}$. Therefore, next
Riemann's analysis lies on the main fact that such a function $\xi$, considered as a function of $x^2$, he says
to have genus zero, but without providing right prove of this statement. The proof will be correctly given by
Hadamard in the final part of his memoir through which a new and fruitful road to treat entire functions,
through infinite product expansion, was opened.

From a historical viewpoint, the first extended treatises on history of number theory appeared in the early
1900s, with the two-volume treatise of Edmund Landau (1877-1933) (see (Landau 1909)) and the three-volume
treatise of Leonard Eugene Dickson (1874-1954) (see (Dickson 1919-23)). In the preface to the first volume of
his treatise, Landau highlights the great impulse given to the analytic number theory with the first rigorous
outcomes achieved by Hadamard since the late 1880s on the basis of what exposed by Riemann in his 1859
celebrated paper, emphasizing the importance of the use of entire function method in number theory. For
instance, in (Landau 1909, Band I, Erstes Kapitel, § 5.III-IV; Zweites Kapitel, § 8; F\"{u}nfzehntes Kapitel),
infinite product expansions à la Weierstrass are extensively used to factorize Riemann $\xi$ function, until
Hadamard work. But, the truly first notable extended report (as called by Landau who quotes it in the preface to
the first volume of his treatise) on number theory was the memoir of Gabriele Torelli (1849-1931), the first
complete survey on number theory that was drawn up with a deep and wide historical perspective not owned by the
next treatise on the subject. Also G.H. Hardy and E.M. Wright, in their monograph (Hardy \& Wright 1960, Notes
on Chapter XXII), state that \it <<There is also an elaborate account of the early history of the theory in
Torelli, Sulla totalità dei numeri primi, Atti della R. Acad. di Napoli, (2) 11 (1902) pp. 1-222>>, \rm even if
then little attention is paid to Riemann's paper (with which analytic number theory officially was born (see
(Weil 1975)), to which a few sections of chapter XVII are devoted. Therefore, herein we will recall the
attention on this work, focusing on those arguments which are of our historical interest. Following (Marcolongo
1931) and (Cipolla 1932), Torelli\footnote{Not to be confused with Ruggero Torelli (1884-1915), his son, who
gave remarkable contributions to complex algebraic geometry amongst which, for example, the celebrated \it
Torelli theorem \rm on projective algebraic curves (see (Maurin 1997)).} started his academic career at the
University of Palermo in 1891, as a teacher of Infinitesimal Calculus and Algebraic Analysis. Afterwards, he
moved to the University of Naples in 1907, as a successor of Ernesto Cesàro, until up his retirement in 1924. He
chiefly made notable researches in algebra and infinitesimal calculus, upon which he wrote valid treatises. But,
the most notable achievement of Torelli was memoir, entitled \it Sulla totalità dei numeri primi fino ad un
limite assegnato, \rm which is a monograph on the subject drawn up by the author when he was professor at the
University of Palermo, for a competition called by the \it Reale Accademia delle Scienze fisiche e matematiche
di Napoli. \rm To be precise, as himself recall in the second cover of this monograph (with a dedication to
Francesco Brioschi), such a work was drawn up to ask to the following question:\\

\it<<Esporre, discutere, e coordinare, in forma possibilmente compendiosa, tutte le ricerche concernenti la
determinazione della totalità dei numeri primi, apportando qualche notevole contributo alla conoscenza delle
leggi secondo le quali questi numeri si distribuiscono tra i numeri interi>>.\\\\\rm[\it <<Explain, discuss and
coordinate, in a compendious manner, the state of the art of all the researches concerning the determination of
the totality of prime numbers, also personally concurring with some notable contribution to the knowledge of the
distribution laws of prime numbers>>.\rm]\\\\ \rm Until the publication of Edmund Landau treatise, Torelli
report was the only monograph available at that time which covered the subject from Legendre work onwards. This
work, for his novelty and importance, won the competition announced by \it Reale Accademia delle Scienze fisiche
e matematiche di Napoli \rm, hence it was published in the related Academy Acts (see (Torelli 1901)). As we have
already said above, this monograph puts much attention and care to the related historical aspects, so that it is
a valuable historiographical source for the subject. For our ends, we are interested in the chapter VIII, IX and
X, where the works of Riemann and Hadamard are treated with carefulness.

Since Euclid times, one of the central problems of mathematics was to determine the totality of prime numbers
less than a given assigned limit, say $x$. In approaching this problem, three main methods were available: a
first one consisting in the effective explicit enumeration of such numbers, a second one which tries to
determine this totality from the knowledge of a part of prime numbers, and a third one consisting in building up
a function of $x$, say\footnote{It will be later called \it prime number counting function, \rm and seems to
have been introduced by Tchebycheff (see (Fine \& Rosenberger 2007, Chapter 4, Section 4.3)). To be precise,
$\pi(x)=\sum_{p\leq x}1$ is the counting function for the set of primes not exceeding $x$, while
$\theta(x)=\sum_{p\leq x}\ln x$ (see (Nathanson 2000, Chapter 8, Section 8.1)) is a Tchebycheff function.}
$\theta(x)$, without explicitly knowing prime numbers, but whose values provide an estimate of the totality of
such numbers less than $x$. To the first method, which had an empirical nature, little by little was supplanted
by the other two methods, which were more analytical in their nature. Around the early 1800s, the method for
determining the number of the primes through the third method based on $\theta(x)$ function, was believed the
most important one even if it presented great formal difficulties of treatment. After the remarkable profuse
efforts spent by Fermat, Euler, Legendre, Gauss, Von Mangoldt and Lejeune-Dirichlet, it was P.L.
Tchebycheff\footnote{Besides, under advice of the mathematician Giuseppe Battagliani (1826-1894), in 1891 an
Italian translation of an important monograph of Tchebycheff, with Italian title \it Teoria delle congruenze \rm
(see (Tchebycheff 1895)), was undertaken by Iginia Massarini, the first Italian woman to be graduated in
mathematics from the University of Naples in 1887. Such a monograph, many times is quoted by Torelli in his
memoir.} the first one to provide a powerful formal method for the determination of the function $\theta(x)$,
even if he gave only asymptotic expressions which were unable to be used for finite values, also at approximate
level. The function $\theta(x)$ plays a very fundamental role in solving the problem of determining the
distribution laws of prime numbers, so that an explicit albeit approximate expression of it, was expected. This
problem, however, constituted a very difficult task because such a function was extremely irregular, with an
infinite number of discontinuities, and with the typical characteristic that, such a formula for $\theta(x)$,
couldn't explicitly provide those points in which the prime numbers were placed. It was Riemann, in his famous
1859 paper, to give, through Cauchy's complex analysis techniques, a formula thanks to which it was possible, in
turn, to deduce a first approximate expression for the $\theta$ function, valid for finite values of the
variable. In any case, leaving out the formal details, Riemann was induced to introduce a complex function, that
is to say the $\zeta$ function, to treat these number theory issues, bringing back the main points of the
question to the zeros of this function. To be precise, for the determination of the non-trivial zeros of this
function, Riemann considered another function obtained by $\zeta$ through the functional equation to which it
satisfies, the so-called Riemann $\xi$ function, namely
\begin{equation}\xi(t)=\frac{s(s-1)}{2}\pi^{-\frac{s}{2}}\Gamma\Big(\frac{s}{2}\Big)\zeta(s)\end{equation}which
is an even entire function of $t$ if one puts $s=1/2+it$. About such a function, Riemann enunciates the
following three propositions\begin{enumerate}\item the number of zeros of $\xi(t)$, whose real part is comprised
into $[0,T]$ with $T\in\mathbb{R}^{+}$, is approximately given by $(T/2\pi)\log(T/2\pi)-(T/2\pi)$,\item all the
zeros $\alpha_i$ of $\xi(t)$ are real,\item we have a decomposition of the type
$$\xi(t)=\xi(0)\prod_{\nu=1}^{\infty}\Big(1-\frac{t^2}{\alpha_i^2}\Big),$$\end{enumerate}but without giving a
correct proof of them. The points 1. and 3. will be proved later by Von Mangoldt and Hadamard, while the point
2. is the celebrated \it Riemann hypothesis \rm which still resist to every attempt of prove or disprove. But,
notwithstanding that, Riemann, assuming as true such three propositions, goes on in finding an approximate
formula for the prime number counting function $\theta(x)$. For the proof of many other results of number theory
discussed in his memoir, many times Riemann makes reference to such a $\xi$ function and its properties, but
without giving any detailed and corrected development of their prove, so constituting a truly seminal paper upon
which a whole generation of future mathematicians will work on, amongst whom is Hadamard.

Soon after the appearance of Riemann memoir, Angelo Genocchi (see (Genocchi 1860)) published a paper in which
contributed to clarify some obscure points of 1859 Riemann paper as well as observed some mistakes in Riemann
memoir about $\xi$ function, giving a detailed development of its expression as an entire function as deduced
from a (Jacobi) theta function\footnote{That is to say, $\vartheta(u)=\sum_{n\in\mathbb{Z}}e^{-\pi n^2u}$, while
$\omega(u)=\sum_{n\in\mathbb{N}}e^{-\pi n^2u}$. These two function were introduced by Tchebychev in the 1850s
(see (Goldstein 1973)).} transformation of the $\zeta$ function, hence pointing out some remarks concerning its
infinite product factorization and related properties, referring to the well-known Briot and Bouquet treatise
(see (Briot \& Bouquet 1859)) as regards the infinite product factorization of the $\xi$ function from the
knowledge of the sequence of its zeros $\alpha_i$. Nevertheless, this Genocchi's remark is not enough to give a
rigorous and complete proof of the proposition 3. of above, thing that will be accomplished later by Hadamard in
1893. The words of Torelli (see (Torelli 1901, Chapter IX, Section 74)), in this regard, are very meaningful.
Indeed, he says that Riemann assumed to be valid only statements 1. and 3. of above, while the statements 2. was
considered to be uninteresting to the ends that Riemann wished to pursue. Hadamard was the first one to cope the
very difficult task to solve the Riemann statements 1. and 3., achieving this with success in 1892 with a memoir
presented to the Academy of Sciences of Paris. Indeed, he was able to brilliantly prove statement 3., from which
he derived also of statement 1. as a corollary stated in the last part of the next paper \it \'{E}tude sur les
propriétés des fonctions entières et en particulier d'une fonctions considérée par Riemann, \rm which was
published in 1893. As the title itself shows, the Hadamard work is a very new chapter of complex analysis in
which are treated general properties of entire functions, so opening new directions in the theory of entire
functions. In 1898, Hans Von Schaper, in his dissertation entitled \it \"{U}ber die Theorie der Hadamardschen
Funktionen und ihre Anwendung auf das Problem der Primzahlen, \rm under the supervision of Hilbert, as well as
Borel, in his important work \it Le\c{c}ons sur les fonctions entières \rm (see (Borel 1900)), reconsider this
notable Hadamard paper as a starting point for a further deepening of the theory of entire functions of finite
order, in particular simplifying the original proof of the above Riemann statement 3., as given by Hadamard in
his 1893 paper. And in his report, Torelli considers Von Schaper proof of this statement, adopting the following
terminology. If $a_1,a_2,...,a_{\nu},...$ are non-zero complex numbers arranged according to a non-decreasing
modulus sequence tending to $\infty$, such that there the integer number $k+1$ is the lowest one such that
$\sum_{\nu=1}^{\infty}(1/|a_{\nu}|^{k+1})$ converges, then we may consider the following absolutely and
uniformly convergent infinite product
\begin{equation}G(z)=\prod_{\nu=1}^{\infty}\Big(1-\frac{z}{a_{\nu}}\Big)
e^{\sum_{j=1}^{k}\frac{1}{j!}\frac{z^j}{a^j_{\nu}}}.\end{equation} Every entire function $F(z)$ having zeros
$a_1,a_2,...,a_n,...$, must have the form $e^{H(z)}G(z)$, where $G(z)$ is given by (17) and $H(z)$ is also an
entire function which may be a polynomial as well. When $H(z)$ is a polynomial of degree $q$, then the integer
number $p=\max\{q,k\}<\infty$ is called, by Borel, the \it genus \rm of the entire function $F(z)$, while Von
Schaper speaks of \it height \rm of $F(z)$; the entire functions of finite order are called \it Hadamard's
functions \rm by Von Schaper. If it is not possible to reduce $H(z)$ to a polynomial, or if the sequence of
zeros $a_{\nu}$ of the given entire function $F(z)$ is such that the above integer $k+1$ does not exist, then we
will say that $F(z)$ has \it infinite genus. \rm

We are interested in entire functions of finite order, of which other two parameters have to be defined as
follows. The upper lower bound of the integer numbers\footnote{Which will be denoted by the same letter.} $k$
(or $\lambda$) such that, for any arbitrarily fixed $\varepsilon\in\mathbb{R}^+$,
$\sum_{\nu=1}^{\infty}(1/|a_{\nu}|^{k+\varepsilon})$ converges and
$\sum_{\nu=1}^{\infty}(1/|a_{\nu}|^{k-\varepsilon})$ diverges, is said to be (after Von Schaper) the \it
exponent of convergence \rm of the sequence of the zeros $a_{\nu}$, or (after Borel) the \it real order \rm of
the function $F(z)$. The upper lower bound of the integer numbers\footnote{Which will be denoted by the same
letter.} $\rho$ such that, for any arbitrarily fixed $\varepsilon\in\mathbb{R}^+$, we have
$|F(z)|<\exp(|z|^{\rho+\varepsilon})$ from a certain value of $|z|$ onwards, and
$|F(z)|<\exp(|z|^{\rho-\varepsilon})$ into an infinite number of points $z$ arbitrarily far\footnote{Taking into
account the modern notions of lower and upper limit of a function, these last notions are nothing but those
exposed in section 5.1.}, is called (after Borel) the \it apparent order \rm of the function $F(z)$, while Von
Schaper says that $F(z)$ is of the \it exponential type $\exp(|z|^{\rho})$. \rm Afterwards, Von Schaper and
Borel proved a series of notable properties concerning the possible relationships between the above defined four
parameters $p,q,k,\rho$, which are summarizable as follows. We have the following properties\begin{itemize}\item
$p\leq\rho$;\item If $\rho$ is not an integer number, then $k=\rho$ and $p=[\rho]$ (= integer part of $\rho$);
\item If $\rho$ is an integer number, the the genus $p$ is equal to $\rho$ or to $\rho-1$. We have $p=\rho-1$ is
and only if $q\leq\rho-1$ and $\sum_{\nu=1}^{\infty}(1/|a_{\nu}|^{k})<\infty$; \item
$\rho=\max\{k,q\}$.\end{itemize}From all that, Torelli retraces the original Hadamard treatment of Riemann $\xi$
function taking into account the just quoted above results achieved by Von Schaper and Borel, reaching to prove
that the following even entire function in the variable $t$
\begin{equation}\xi(t)=\frac{s(s-1)}{2}\pi^{-\frac{s}{2}}\Gamma\Big(\frac{s}{2}\Big)\zeta(s)\end{equation}
is an entire function that, with respect to $t^2$, has an apparent order which cannot exceed $1/2$, so that it
follows that it has genus zero with respect to the variable $t^2$, if one puts $s=1/2+it$. Therefore, the
infinite product expansion of the above Riemann statement 3., is now proved. Furthermore, the absence of
exponential factors in this expansion into primary factors of the function $\xi(t)$, implies the existence of,
at least, one root for $\xi(t)=0$, while the expansion into an infinite series of increasing powers of $t^2$,
implies too that such roots are in an infinite number. We have already said that Hadamard himself, in his
celebrated 1893 memoir, also proved the above Riemann statement 1., even if further improvements were achieved
later by H. Von Mangoldt (see (Von Mangoldt 1896)), J. Franel (see (Franel 1896)) and Borel (see (Borel 1897)).
In conclusion, the pioneering Hadamard work contained in his 1893 memoir, has finally proved two out of the
three above Riemann statements, namely the 1. and 3., to which other authors have later further contributed with
notable improvements and extensions.

Following (Torelli 1909, Chapter VIII, Sections 71 and 72; Chapter IX, Section 77), as regards, instead, Riemann
statement 2., that is to say, what will be later known as the \it Riemann hypothesis, \rm first attempts to
approach the solution of the equation $\xi(t)=0$, were made by T.J. Stieltjes (see (Stieltjes 1885)), J.P. Gram
(see (Gram 1895)), F. Mertens (see (Mertens 1897) and J.L.W. Jensen (see (Jensen 1898-99)). We are interested in
the Jensen's work for its historical role played in the development of entire function theory, of which we have
already said something about this in section 5.1. where it has been pointed out what fundamental role played
this Jensen's work in the early developments of the theory of value distribution of entire and meromorphic
functions as opened by R. Nevanlinna work. To be precise, in this Jensen's work of 1898, the author considers a
meromorphic function, say $f(z)$, defined into a region of complex plane, say $D$, containing the zero and where
such a function is neither zero nor infinite. Let $a_1,...,a_n$ be the zeros and $b_1,...,b_m$ be the poles of
the function $f(z)$, counted with their respective multiplicity and supposed to be all included into a circle,
say $\mathcal{C}_r$, given by $|z|\leq r$ centered in $0$ and with radius $r$ such that $\mathcal{C}_r\subseteq
D$. Then, Jensen easily proves the following formula
\begin{equation}\frac{1}{2\pi}\int_0^{2\pi}\ln|f(re^{i\theta})|d\theta=\ln|f(0)|+
\ln\frac{r^{n-m}|b_1\cdot...\cdot b_m|}{|a_1\cdot...\cdot a_n|}.\end{equation}Now, Jensen argues that, if $f(z)$
is an entire function, then $r$ may be chosen arbitrarily large in such a manner $\mathcal{C}_r$ does not
contain any zero, so that the second term in the right hand side of (19) reduces only to the first, constat
term. In doing so, we have thus a simple criterion for deciding on the absence or not of zeros within a given
circle of complex plane. Once Jensen stated that, he finishes the paper announcing to have proved, through his
previous researches on Dirichlet's series, that the function $\xi(t)$ does not have any zero within an arbitrary
circle centered into the finite imaginary axis and comprehending the zero, which implies that $\xi(t)=0$ has
only real zeros, as Riemann conjectured. On the other hand, as we have already pointed out in the previous
section 5.1., it is just from this Jensen formula that started the theory of value distribution of entire and
meromorphic functions which was built by the pioneering work of Rolf Nevanlinna of 1920s. Often, in many
treatise on entire and meromorphic functions, Jensen formula is the first key element from which to begin.
Indeed, following (Zhang 1993, Chapter I), the theory of entire and meromorphic functions starts with Nevanlinna
theory which, in turn, is based either on a particular transformation of the formula (19), called \it
Poisson-Jensen formula \rm by Nevanlinna\footnote{Because either the original Jensen formula (19) and another
formula of potential theory (\it Poisson formula\rm) due to D.S. Poisson, are special cases of this formula due
to Nevanlinna.}, and on the previous works made by Poincaré, Hadamard and Borel on entire functions. Therefore,
another central starting point of the theory of entire and meromorphic functions, as the one just examined above
and due to Jensen, relies on the prickly problematic raised by Riemann $\xi$ function. Lastly, the appreciated
Torelli's monograph comes out with the following textual words\\

\it <<Come conclusione di questo capitolo e dell'intero lavoro, si può senza alcun dubio affermare che la
memoria di Riemann, insieme alle esplicazioni e i complimenti arrecati da Hadamard, Von Mangoldt, e de la
Vallèe-Poussin, resta tuttora come il faro, che guidar possa nella scoperta di quanto ancora v'è di ignoto nella
Teoria dei Numeri primi>>.\\\\\rm[\it<< As a conclusion of this chapter as well as of the whole work, surely we
may state that the Riemann memoir, together all the explications and complements due to Hadamard, Von Mangoldt,
and de la Vallèe-Poussin, still remains as that lighthouse which can drive towards the discovery of what yet is
unknown in the theory of prime numbers>>\rm].\\\\Following (Ingham 1964, Introduction, 6.), as we have already
said above Riemann enunciated a number of important theorems concerning the zeta function - i.e., the above
Riemann statements 1., 2., and 3. - together with a remarkable relatioship connecting the prime number counting
function with its zeros, but he gave in most cases only insufficient indications of proofs. These problems
raised by Riemann memoir inspired, in due course, the fundamental researches of Hadamard in the theory of entire
functions, the results of which at last removed some of the obstacles which for more than thirty years had
barred the way to rigorous proofs of Riemann statements. The proofs sketched by Riemann were completed (in
essentials), in part by Hadamard himself in 1893, and in part by Von Mangoldt in 1894. Following (Ingham 1964,
Introduction, 6.), as we have already said above Riemann enunciated a number of important theorems concerning
the zeta function - i.e., the above Riemann statements 1., 2., and 3. - together with a remarkable relatioship
connecting the prime number counting function with its zeros, but he gave in most cases only insufficient
indications of proofs. These problems raised by Riemann memoir inspired, in due course, the fundamental
researches of Hadamard in the theory of entire functions, the results of which at last removed some of the
obstacles which for more than thirty years had barred the way to rigorous proofs of Riemann statements. The
proofs sketched by Riemann were, in essentials, completed in part by Hadamard himself in 1893, and in part by
Von Mangoldt in 1894. These discoveries due to Hadamard prepared the way for a rapid advance in the theory of
the distribution of primes. The so-called \it prime number theorem, \rm according to which $\theta(x)\sim x/\ln
x$, was first proved in 1896 by Hadamard himself and by de la Vallée-Poussin, independently and almost
simultaneously, the proof of the former having used the results achieved in his previous 1893 memoir. Out of the
two proofs, Hadamard one is the simpler, but de la Vallée-Poussin, in another paper published in 1899 (see (de
la Vallée-Poussin 1899-1900)), studied in great detail the question of closeness of approximation as well as
gave further improvements to prime number theorem. Finally, either de la Vallée-Poussin work (see (de la
Vallée-Poussin 1896)) and Von Mangoldt one (see (Von Mangoldt 1896) and (Von Mangoldt 1905)), used the results
on entire function factorization achieved by Hadamard in his 1893 memoir (see (Maz'ya \& Shaposhnikova 1998,
Chapter 10, Section 10.1)). Following (Kudryavtseva 2005, Section 7), Riemann's paper is written in an extremely
terse and difficult style, with huge intuitive leaps and many proofs omitted. This led to (in retrospect quite
unfair) criticism by E. Landau and G.H. Hardy in the early 1900s, who commented that Riemann had only made
conjectures and had proved almost nothing. The situation was greatly clarified in 1932 when C.L. Siegel (see
(Siegel 1932)) published his paper, representing about two years of scholarly work studying Riemann's left over
mathematical notes at the University of G\"{o}ttingen, the so-called \it Riemann's Nachlass. \rm From this
study, it became clear that Riemann had done an immense amount of work related to his 1859 memoir and that never
appeared in it. One conclusion is that many formulas that lacked sufficient proof in 1859 paper were in fact
proved in these notes. A second conclusion is that the notes contained further discoveries of Riemann that were
never even written up in the original memoir. One such is what is now called the \it Riemann-Siegel formula, \rm
which Riemann had written down and that Siegel (with great difficulty) was able to prove (see (Karatsuba 1994)
and (Edwards 1974)). This formula, in essentials, arises from an Hankel integral type expression for $\xi(s)$,
and gives a refined method to calculate $\xi(1/2+it)$, in comparison to previous ones. In any way, after
Hadamard work on Riemann $\xi$ function, only a few authors have put considerable attention to it, amongst whom
are E. Landau and G. P\'{o}lya. Notwithstanding that, in the following sections we wish briefly to retrace the
historical path which gather those main works on Riemann $\xi$ function which will lead to a particular
unexpected result of mathematical physics, to be precise belonging to statistical mechanics which, in turn, has
opened a new possible line of approach Riemann hypothesis.

In conclusion, we wish to textually report what retrospectively Hadamard himself says about his previous work on
entire function, following (Hadamard 1901, Chapter I), that is to say\\

\it <<Les formules démontrées dans ma thèse relativement aux singularités polaires\footnote{To be precise, we
report an excerpt of the last part of the Hadamard discussion about his 1892 thesis on Taylor development of
functions (see (Hadamard 1892)), from which he starts to discuss his next work on entire functions. He states
that \it <<A partir de la publication de ma Thèse, l'attention des géomètres s'est portée sur ce sujet. Grâce
aux travaux de MM. Borel, Fabry, Leau, Lindel\"{o}f, et à la découverte de M. Mittag-Leffler, cette théorie, qui
n'existait pas en 1892, forme aujourd'hui un chapitre assez important de la Théorie des fonctions, celui de tous
(avec la théorie des Fonctions entières dont il va \^{e}tre question plus loin) qui, dans ces dernières années,
a acquis à la Science le plus grand nombre de résultats. Une très grande partie de ceux-ci ont d'ailleurs été
obtenus par le développement des méthodes m\^{e}mes que j'avais indiquées. Je n'ai pas perdu de vue, dans la
suite, cette catégorie de questions, et, en 1897, j'ai démontré, également par la considération d'une intégrale
définie, un théorème qui fait connaître les singularités possibles de la série $\sum a_ib_ix^i$ quand on connaît
celles de la série $\sum a_ix^i$ et de la série $\sum b_ix^i$. Celte proposition dérive évidemment du m\^{e}me
principe que le théorème mentionné en dernier lieu; comme lui, elle offre cet avantage de s'appliquer à toute
l'étendue du plan. Aussi, ce travail a-t-il attiré l'attention des géomètres sur le principe en question et
provoqué une nouvelle série de recherches ayant pour objet d'en obtenir de nouvelles applications>>.} ont trouvé
une application immédiate dans un Mémoire auquel l'Académie a décerné, en 1892, le grand prix des Sciences
mathématiques. La question posée par l'Académie, et qui portait sur une fonction employée par Riemann, soulevait
un problème général: celui du genre des fonctions entières. Ou sait que la notion de genre est liée au théorème
de Weierstrass d'après lequel toute fonction entière $F(x)$ peut être mise sous forme d'un produit de facteurs
(facteurs primaires)$$F(x)=e^{G(x)}\prod_n\Big[\Big(1-\frac{x}{a_n}\Big)e^{P_n\big(\frac{x}{a_n}\big)}\Big]$$
(où $G(x)$ est une nouvelle fonction entière et les $P(x)$ des polynômes): décomposition analogue à celle d'un
polynôme en ses facteurs linéaires.

Si l'on peut s'arranger les polynômes $P(x)$ soient de degré $E$ au plus, la fonction $G(x)$ se réduisant
elle-m\^{e}me à un polynôme de degré égal ou inférieur à $E$, la fonction $F(x)$ est dite (Il est sous-entendu
que $E$ doit \^{e}tre le plus petit entier satisfaisant aux conditions indiquées) de genre $E$. Il est
nécessaire pour cela (mais non suffisant) que les racines $a_1, a_2, ..., a_n$ de l'équation $F(x)=0$ ne soient
pas trop rapprochées les unes des autres: la série $\sum(1/|a_n|)^{E+1}$ doit \^{e}tre convergente. M. Poincaré
a donné, en 1883 (Bull. de la Soc. math. de France), une condition nécessaire pour qu'une fonction $F(x)$ soit
de genre $E$; cette condition est que les coefficients du développement de $F(x)$ décroissent au moins aussi
vite que les valeurs successives de $1/(m!)^{1/(E+1)}$. Cette condition nécessaire était-elle la condition
nécessaire et suffisante pour que la fonction fût au plus de genre $E$? Etant donnée la manière compliquée dont
les racines d'une équation dépendent de ses coefficients, il semblait hautement improbable que la réponse fût
aussi simple, ni surtout qu'elle fût aisée à obtenir. C'était elle qui avait manqué à Halphen pour continuer les
recherches qu'il avait commencées en 1883, sur les travaux de Riemann. La Commission (M. Picard, rapporteur)
chargée déjuger le concours de 1892 rappelait, dans son Rapport, l'exemple d'Halphen et faisait observer combien
il semblait peu vraisemblable au premier abord que l'on pût donner une réciproque au théorème de M. Poincaré. De
son côté, ce dernier, dans le Mémoire précédemment cité, après s'\^{e}tre posé une question étroitement liée à
la précédente, celle de savoir si la dérivée d'une fonction de genre $E$, ou la somme de deux fonctions de genre
$E$, est également du m\^{e}me genre, ajoutait: «Ces théorèmes, en admettant qu'ils soient vrais, seraient très
difficiles à démontrer».

Le problème qui consiste à déterminer le genre d'une fonction entière donnée par son développement en série de
puissances se rattache d'une manière évidente aux recherches dont j'ai parlé jusqu'ici, puisque celles-ci ont
pour objet général l'étude d'une série de Maclaurin donnée a priori. J'ai pu effectuer celle détermination on
toulc rigueur dans le Mémoire soumis au jugement de l'Académie. Désormais, la théorie des fonctions entières
est, au point de vue des zéros, toute parallèle à celle des polynômes. Le genre (ou, plus généralement, l'ordre
décroissance) joue le rôle du degré, la distribution des zéros de la fonction étant en général réglée par ce
genre comme le nombre des zéros d'un polynome par son degré. Dans un article ultérieur \rm [see (Hadamard
1896c)], \it j'ai précisé et simplifié la loi qui donne la croissance du module de la fonction lorsqu'on donne
la suite des coefficients et qui joue un rôle important dans ces recherches. \rm [...] \it Quant aux questions
posées par M. Poincaré et relatives à la conservation du genre dans la dérivation ou dans les combinaisons
linéaires, elles ne sont pas, il est vrai, résolues d'une façon tout à fait complète par les théorèmes dont je
viens de parler, et ne sauraient, d'ailleurs, l'\^{e}tre par des méthodes de cette nature. Mais on peut dire
qu'elles sont résolues en pratique. D'une part, en effet, les cas qui échappent aux méthodes précédentes sont
tout exceptionnels, d'autre part, l'hésitation ne peut jamais \^{e}tre que d'une unité sur le genre
cherché.\\\\\rm Again, Hadamard goes on, recalling what follows\\

\it<<Du théorème relatif au rayon de convergence d'une série entière découle celte conséquence: la condition
nécessaire et suffisante pour qu'une série de Maclaurin représente une fonction entière est que la racine
m$^{ieme}$ du coefficient de $x^m$ tende vers 0. Les propriétés les plus importantes de la fonction entière sont
liées à la plus ou moins grande rapidité avec laquelle a lieu cette décroissance des coefficients. L'étude de
ces propriétés consiste tout d'abord dans l'établissement de relations entre cette loi de décroissance et les
deux éléments suivants : 1$^o$ L'ordre de grandeur du module maximum de la fonction pour les grandes valeurs du
module de la variable; 2$^o$ La distribution des zéros et la valeur du genre, laquelle est étroitement liée à
cette distribution. Une partie de ces relations avait été établie dans le Mémoire cité de M. Poincaré: une
limite supérieure des coefficients successifs avait pu \^{e}tre trouvée, connaissant l'une ou l'autre des deux
lois qui viennent d'\^{e}tre énumérées. Mais on n'avait pas pu, depuis ce moment, obtenir les réciproques,
c'est-à-dire déduire d'une limite supérieure supposée connue pour chaque coefficient les conséquences qui en
découlent, d'une part quant à l'ordre de grandeur de la fonction ellem\^{e}me, d'autre part quant à la
distribution de ses zéros. C'est à l'établissement de ces conséquences qu'est principalement consacré le Mémoire
couronné par l'Académie en 1892 et publié en 1893 au Journal de Mathématiques. J'ai ensuite précisé les
premières dans la Note \rm [see (Hadamard 1896c)] \it insérée au Bulletin de la Société Mathématique de France
et dont j'ai également parlé dans l'Introduction. Quant aux zéros, les résultats contenus dans ma Thèse
fournissaient aisément à leur égard cette conclusion simple: La loi de croissance des racines de la fonction
entière $\sum a_mx^m$ est au moins aussi rapide que celle des quantités $1/\sqrt[m]{|a_m|}$. Pour étudier le
facteur exponentiel, de nouvelles déductions ont, au contraire, été nécessaires. Ces déductions m'ont, en
particulier, permis de démontrer, avec une extr\^{e}me simplicité, le théorème de M. Picard sur les fonctions
entières, pour toutes les fonctions de genre fini. La démonstration ainsi donnée s'étend d'elle-m\^{e}me,
moyennant une restriction analogue, au théorème plus général du m\^{e}me auteur sur le point essentiel, ainsi
que je l'ai montré depuis \rm [see (Hadamard 1896b)]. \it On sait que mon Mémoire de 1893 a été le point de
départ des si importants travaux de M. Borel, consacrés à la démonstration du premier théorème de M. Picard sans
restriction, et aussi de ceux de MM. Schou et Jensen. Outre les applications à la fonction $\zeta(s)$ et aux
fonctions analogues, dont il me reste à parler, la proposition fondamentale de ce Mémoire a été utilisée par M.
Poincaré dans une question relative aux déterminants infinis qui s'introduisent en Astronomie (Les méthodes
nouvelles de la Mécanique céleste, t. II)>>.\\\\\rm Whereupon, in a brief but complete manner, Hadamard
discusses his work on Riemann $\zeta$ function, stating first that\\

\it<<Le dernier anneau de la chaîne de déductions commencée dans ma Thèse et continuée dans mon Mémoire couronné
aboutit à l'éclaircissement des propriétés les plus importantes de la fonction $\zeta(s)$ de Riemann. Par la
considération de cette fonction, Riemann détermine la loi asymptotique de fréquence des nombres premiers. Mais
son raisonnement suppose : 1$^o$ que la fonction $\zeta(s)$ a des zéros eu nombre infini; 2$^o$ que les modules
successifs de ces zéros croissent à peu près comme $n\ln n$; 3$^o$ que, dans l'expression de la fonction
auxiliaire $\xi(t)$ en facteurs primaires, aucun facteur exponentiel ne s'introduit. Ces propositions étant
restées sans démonstration, les résultats de Riemann restaient complètement hypothétiques, et il n'en pouvait
\^{e}tre recherché d'autres dans cette voie. De fait, aucun essai n'avait été tenté dans cet ordre d'idées
depuis le Mémoire de Riemann, à l'exception : 1$^o$ de la Note précédemment citée d'Halphen, qui était, en
somme, un projet de recherches pour le cas où les postulats de Riemann seraient établis; 2$^o$ d'une Note de
Stieltjes, où ce géomètre annonçait une démonstration de la réalité des racines de $\xi(t)$, démonstration qui
n'a jamais été produite depuis. Or les propositions dont j'ai rappelé tout à l'heure l'énoncé ne sont qu'une
application évidente des théorèmes généraux contenus dans mon Mémoire.

Une fois ces propositions établies, la théorie analytique des nombres premiers put, après un arr\^{e}t de trente
ans, prendre un nouvel essor; elle n'a cessé, depuis ce moment, de faire de rapides progrès. C'est ainsi que la
connaissance du genre de $\zeta(s)$ a permis, tout d'abord, à M. Von Mangoldt d'établir en toute rigueur le
résultat final du Mémoire de Riemann. Auparavant, M. Cahen avait fait un premier pas vers la solution du
problème posé par Halphen; mais il n'avait pu arriver complètement au but: il fallait, en effet, pour achever de
construire d'une façon inattaquable le raisonnement d'Halphen, prouver encore que la fonction $\zeta$, n'avait
pas de zéro sur la droite $R(s)=1$. J'ai pu vaincre cette dernière difficulté en 1896, pendant que M. de la
Vallée-Poussin parvenait de son côté au m\^{e}me résultat. La démonstration que j'ai donnée est d'ailleurs de
beaucoup la plus rapide et M. de la Vallée-Poussin l'a adoptée dans ses publications ultérieures. Elle n'utilise
que les propriétés les plus simples de $\zeta(s)$. En m\^{e}me temps, j'étendais le raisonnement aux séries de
Dirichlet et, par conséquent, déterminais la loi de distribution des nombres premiers clans une progression
arithmétique quelconque, puis je montrais que ce raisonnement s'appliquait de lui-m\^{e}me aux formes
quadratiques à déterminant négatif. Les m\^{e}mes théorèmes généraux sur les fonctions entières ont permis,
depuis, à M. de la Vallée-Poussin d'achever ce cycle de démonstrations en traitant le cas des formes a $b^2-ac$
positif>>.\rm\\\\Then, Hadamard recalls furthermore that\\

\it<<La détermination du genre de la fonction $\zeta(s)$ - et c'était d'ailleurs l'objet môme de la question
posée par l'Académie - était nécessaire pour l'éclaircissement des points principaux du Mémoire principal de
Riemann ''Sur le nombre des nombres premiers inférieurs à une grandeur donnée''. Cette détermination, qui avait
été jusque-là cherchée en vain, s'eirectue sans aucune difficulté à l'aide des principes précédemment établis
sur les fonctions entières. Aussi M. Von Mangoldt put-il peu après établir avec une entière rigueur les
résultats énoncés par Riemann. Un seul point restait à élucider: la question de savoir si, conformément à une
assertion énuse, en passant, par ce grand géomètre, les racines imaginaires de l'équation $\zeta(s)=0$ sont
toutes de la forme $1/2+it$, $t$ étant réel. Cette question n'a pas encore reçu de réponse décisive (le Mémoire
dans lequel M. Jensen annonce qu'il donnera ce résultat n'ayant pas encore paru); mais j'ai pu en 1896 \rm [see
(Hadamard 1896a) and references therein] \it établir que la partie réelle des racines dont il s'agit, laquelle
n'est évidemment pas supérieure à l'unité, ne peut non plus, pour aucune d'elles, \^{e}tre égale à 1. Or ce
résultat suffit pour établir les principales lois asymptotiques de la théorie des nombres premiers, de m\^{e}me
que le résultat complet de Riemann conduirait à montrer (voir un Mémoire récent de M. Helge von Koch) que ces
lois sont vraies à une erreur près, laquelle n'est pas seulement d'ordre inférieur à celui de la quantité
considérée $x$, mais est tout au plus comparable à $\sqrt{x}$.

De plus le mode de démonstration que j'emploie n'utilise (pie les propriétés les plus simples de la fonction
$\zeta(s)$. Il en résulte que ce mode de démonstration s'étend sans grande difficulté aux séries analogues qui
ont été ulilisées dans la théorie des nombres. J'ai fait voir en particulier, dans le même travail, qu'il
s'applique aux séries qui servent à étudier la distribution des nombres premiers représentables soit par une
forme linéaire (séries de Dirichlet\footnote{\it Pour démontrer le théorème relatif à la distribution des
nombres premiers dans une progression arithmétique, j'ai utilisé, en la complétant sur un point, la proposition
de M. Lipschitz qui établit, pour les séries de Dirichlet, une relation fonctionnelle analogue à celle de
Riemann-Schl\"{o}milch. J'ai été conduit, depuis le travail \rm [see (Hadamard 1897)], \it à simplifier la
démonstration de ce théorème.}), soit par une forme quadratique définie. M. de la Vallée-Poussin parvenait en
m\^{e}me temps au m\^{e}me résultat, mais par une voie moins rapide. Depuis, ce savant (tout en simplifiant son
Analyse par l'emploi du mode de raisonnement que j'avais indiqué) a étendu ses recherches au cas des formes
quadratiques indéfinies et aussi à celui où l'on donne à la fois une forme linéaire et une forme quadratique; de
sorte que les m\^{e}mes principes relatifs aux fonctions entières servent de base à la solution générale de
toutes les questions qui s'étaient posées relativement à la distribution des nombres premiers. Ce ne sont
d'ailleurs pas les seules questions de Théorie des nombres pour la solution desquelles les théorèmes qui
viennent d'\^{e}tre rappelés se soient montrés d'une importance essentielle. Je me contente de signaler, à cet
égard, les Mémoires récents de MM. von Mangoldt, Landau, etc>>.\\\\\rm In any event, Hadamard work until early
1900s was always influenced by 1859 Riemann's memoir: indeed, as remembers (Mandelbrojt \& Schwartz 1965), for instance\\

\it<<Hadamard's theorem on composition of singularities was proved in 1898. When stated without much rigour, it
reads as follows. $\sum a_nb_nz^n$ has no other singularities than those which can be expressed as products of
the form $\alpha\beta$, where $\alpha$ is a singularity of $\sum a_nz^n$ and $\beta$ a singularity of $\sum
b_nz^n$. The theorem is proved by the use of Parseval's integral, which Hadamard adapted to Dirichlet series (in
works of 1898 and 1928), not for the research of the singularities of the composite series, but for the study of
interesting relationships between the values of Riemann's $\zeta$ function at different points, or between
different types of $\zeta$ functions>>.\\\\\rm Then, Mandelbrojt and Schwartz recall further that\\

\it<<The year 1892 is one of the richest in the history of Function Theory, since then not only did Hadamard's
thesis appear, but also his famous work on entire functions, which enabled him, a few years later (in 1896), to
solve one of the oldest and most important problems in the Theory of Numbers. The general results obtained,
establishing a relationship between the rate of decrease of the moduli of the coefficients of an entire function
and its genus (the converse to Poincaré's theorem), applied to the entire function $\xi(z)$, related to
$\zeta(s)$, shows that its genus, considered as a function of $z^2$, is (as stated, but not proved correctly, by
Riemann) zero. This relationship (for general entire functions) between the moduli of the zeros of an entire
function and the rate of decrease of its coefficients is obtained by using the results of the Thèse \rm [see
(Hadamard 1892)], \it and concerning the determinants $D_{n,m}$ of a suitable meromorphic function (the
reciprocal of the considered entire function). This paper on entire functions was written for the Grand Prix de
l'Académie des Sciences in 1892 for studying the function $\pi(x)$. As a matter of fact, the mathematical world
in Paris was sure that Stieltjes would get the prize, since Stieltjes thought that he had proved the famous
''Riemannische Vermutung'', and it is interesting, I believe, to quote a sentence from Hadamard's extremely
famous paper of 1896 with the suggestive title, ''Sur la distribution des zéros de la fonction $\zeta(s)$ et ses
conséquences arithmétiques''. Hadamard writes: ''Stieltjes avait démontré, conformément aux prévisions de
Riemann, que ces zéros sont tous de la forme $1/2+it$ (le nombre $t$ étant réel), mais sa démonstration n'a
jamais été publiée, et il n'a m\^{e}me pas été établi que la fonction $\zeta$ n'ait pas de zéros sur la droite
$R(s)=1$. C'est cette dernière conclusion que je me propose de démontrer.

The ''modesty'' and the grandeur, of the purpose: to prove that $\zeta(s)\neq 0$ for $\sigma=l$ $(s=\sigma+it)$,
after the assertion that Stieltjes had ''proved'' the Riemannische Vermutung, are remarkably moving. The more so
that, due to this proof, Hadamard could prove, in the same paper of 1896, the most important proposition on the
distribution of primes: $\pi(x)$ being the number of primes smaller than $x$ $(x>0)$, $\pi(x)\sim x/\ln x$
$(x\rightarrow\infty)$. The event had certainly a great historical bearing. The assumption was made, at the
beginning of the last century, by Legendre (in the form $\pi(x)=x/(\ln x-A(x))$, with $A(x)$ tending to a finite
limit). Tchebycheff had shown that $0.92129\leq\pi(x)\ln x/x\leq 1.10555$ ..., but did not prove that the
expression tends to a limit, and there was no hope that his method could yield any such proof. Many
mathematicians, Sylvester among them, were able, in using the same methods as Tchebycheff, to improve these
inequalities. But there was nothing fundamentally new in these improvements. Let us quote Sylvester (in 1881) on
this matter (quotation given by Landau). ''But to pronounce with certainty upon the existence of such
possibility ($\lim\pi(x)\ln x/x= 1$) we should probably have to wait until someone is born into the world as far
surpassing Tchebycheff in insight and penetration as Tchebycheff proved himself superior in these qualities to
the ordinary run of mankind''. And, as Landau says, when Sylvester wrote these words Hadamard was already born.
It should be pointed out that independently, and at the same time, de La Vallée-Poussin also proved the
non-vanishing of $\zeta$ on $\sigma=1$ and, thus, the prime-number theorem; however, Hadamard's proof is much
simpler. Hadamard's study of the behavior of the set of zeros of $\zeta(s)$ is based on his result quoted above
(proved in his paper of 1892, written for the Grand Prix), on the genus of $\xi(z)$. It seems to me of
importance to insist upon the ''chain of events'' in Hadamard's discoveries: relationship between the position
of the poles of a meromorphic function and the coefficients of its Taylor series; this result yields later the
genus of an entire function by the rate of decrease of its Taylor coefficients; and from there, four years
later, the important properties of $\zeta(s)$, and finally, as a consequence, the prime-number theorem. Clearly,
one of the most beautiful theories on analytic continuation, so important by itself, and so rich by its own
consequences, seems to have been directed in Hadamard's mind, consciously or not, towards one aim: the
prime-number theorem. He proved also the analogous theorems on the distribution of primes belonging to a given
arithmetical progression, since by his methods he was able to study Dirichlet series which, with respect to
these primes, play the same role as the $\zeta$ function plays with respect to all primes>>.\\\\\rm We have
above reported only two of the numerous witnesses about the important Hadamard work, namely (Mandelbrojt \&
Schwartz 1965) and (Mandelbrojt 1967), just those two strangely enough not quoted in (Maz'ya \& Shaposhnikova
1998) (to which we however refer for the bibliographical richness and completeness on the subject), however the
most complete and updated human and scientific biography devoted to the ''living legend'' in mathematics as
Hardy defined Hadamard when introduced him to the London Mathematical Society in 1944 (see (Kahane 1991)).
Finally, in (Ayoub 1963, Chapter II, Section 4), the author says that, relatively to (18) (see also next (54)),
such a formula was not proved rigorously until about 1892 when Hadamard constructed his general theory of entire
functions, and that Landau showed that it is possible to avoid this theory. Moreover, in (Meier \& Steuding
2009), the authors, in relation to this Hadamard work of 1893, simply note that his general theory for zeros of
entire functions forms now an important part of complex analysis, while (Rademacher 1973, Chapter 7, Section 56)
points out that the existence of infinitely many non-trivial zeros of $\zeta(s)$ is usually proved through the
application of Hadamard's theory of entire functions to the function $s(s-1)\Phi(s)$ where
$\Phi(s)=\pi^{-s/2}\Gamma(s/2)\zeta(s)$, whereas usual arguments connect it with the existence of infinitely
many primes. Of course Hadamard's theory yields much more than merely the existence of infinitely many
non-trivial zeros. Riemann conjectured, and Von Mangoldt proved in 1905 (see later), that the number $N(T)$ of
zeros $\rho=\beta+i\gamma$ with $0<\gamma\leq T$ is$$N(T)=\frac{T}{2\pi}\Big(\ln\frac{T}{2\pi}-1\Big)+O(\ln
T),$$and Hardy showed first in 1914 (see (Hardy 1914)) that infinitely many of these zeros lie on the middle
line $\sigma=1/2$ of the critical strip, while a better estimate in this direction was later provided by A.
Selberg's theorem according to which the number $N(T)$ of zeros on $\sigma=1/2$ satisfies $N_0(T)>AT\ln T$ for a
certain positive $A$. Selberg's result states too that among all non-trivial zeros, those on the middle-line
have a positive density.\\\\ $\bullet$ \it The contributions of H. Von Mangoldt, E. Landau, and others. \rm
After the pioneering 1893 work of Hadamard on Riemann $\xi$ function, and its next fruitful application to prove
prime number theorem by Hadamard himself, de la Vallée-Poussin and Von Mangoldt in the late 1890s, a few works
were carried out on Riemann $\xi$ function, except some researches which date back to early 1900s. Amongst the
latter are the works of Edmund Landau (1877-1938), namely (Landau 1909a) and (Landau 1927). These works were
accomplished by Landau after the publication of his famous treatise on the theory of prime numbers, namely
(Landau 1909b), where, therefore, these are not quoted but, nevertheless, a deep and complete treatment of
product formulas for $\xi(s)$ function is given (see (Landau 1909b, Band I, §§ 67-81)), mainly reporting what
was known until then, that is to say, Hadamard work, as well as its applications to number theory issues by
Hadamard himself, de la Vallée-Poussin and Von Mangoldt. In particular, Landau showed that, to estimate
$\zeta'(s)/\zeta(s)$, it was enough to study behavior of $\zeta(s)$ for $\Re s>0$ considering, as done by
Riemann, the integral$$\frac{1}{2\pi i}\int_{a-i\infty}^{a+i\infty}\ln\zeta(s)\frac{x^s}{s}ds,$$where $a>1$ and
$x$ is not an integer, avoiding entire function theory (see (Ayoub 1963, Chapter II, Sections 4 and 6)), applied
to $\xi$ function, according to 1893 Hadamard route. In any case, almost all the subsequent works on Riemann
$\xi$ function were turned towards applications to number theory, above all to estimate the number of zeros of
Riemann $\zeta$ function within a given finite region of complex plane. Following (Burkhardt et al. 1899-1927,
II.C.8, pp. 759-779), such estimates were achieved considering the following product formula
\begin{equation}(s-1)\zeta(s)=\frac{1}{2}e^{bs}\frac{1}{\Gamma\Big(\frac{s}{2}+1\Big)}\prod_{\rho}\Big(1-\frac{s}{\rho}\Big)
e^{\frac{s}{\rho}}\end{equation}from which it is possible to deduce the following one
\begin{equation}\frac{\zeta'(s)}{\zeta(s)}=b-\frac{1}{s-1}-\frac{1}{2}\frac{\Gamma\Big(\displaystyle 1+\frac{s}{2}\Big)}
{\Gamma\Big(\displaystyle 1+\frac{s}{2}\Big)}+\sum_{\rho}\Big(\frac{1}{\rho}+\frac{1}{s-\rho}\Big)\end{equation}
which will play a fundamental role in the subsequent analytic number theory researches. As an ansatz already
given by Riemann in his 1859 memoir and correctly proved, for the first time, by Hadamard, in 1893, through
(20), it was possible to estimate the number of zeros of $\zeta(s)$ as follows (see (Landau 1909a, Section 2))
\begin{equation}N(T)=\frac{1}{2\pi}T\ln T-\frac{1+\ln 2\pi}{2\pi}T+O(\ln T)\end{equation}where $N(T)$ is the
number of zeros $s+it$ of $\zeta(s)$ with $0<\sigma<1$ and $0<t\leq T$. Subsequently, Von Mangoldt (see (Von
Mangoldt 1905)) improved this estimate through properties of either the $\zeta$ functional equation and the
Gamma function, proving that\begin{equation}N(T)=\frac{1}{2\pi}T\ln T-\frac{1+\ln 2\pi}{2\pi}T+\frac{1}{2\pi
i}\int^{2+iT}_{-1+iT}\frac{\zeta'(s)}{\zeta(s)}ds+O(1),\end{equation}an estimate which will be further improved
later by E. Landau in the years 1908-09 (see (Landau 1908; 1909a)) as well as by C. Hermite and J. Stieltjes in
the early 1905 (see (Landau 1909a) and references therein). In any event, either Von Mangoldt and Landau have
based their works on the Hadamard one upon entire functions, whereas the next works of R. Backlund (see
(Backlund 1914; 1918)) gave too further improvements of this estimate without appealing to Hadamard work but
simply on the basis of approximation properties of $\zeta(s)$. Further contributions to this subject, were also
given later by J.P. Gram, H. Cramer, H. Bohr, J.L.W. Jensen, J.E. Littlewood, F. Nevanlinna and R. Nevanlinna
(see (Borel 1921, Chapter IV), (Nevanlinna \& Nevanlinna 1924) and references therein). Von Mangoldt, in the
long-paper (Von Mangoldt 1896), considered Hadamard's work of 1893 for estimating the number of zeros of zeta
function into a given finite region of the complex plane, starting from the previous estimate already provided
by Hadamard himself, which will be extended and improved by Von Mangoldt, then deepening and extending the
various number theory issues considered by Riemann in his 1859 original memoir, together to what will be
achieved in the next papers (Von Mangoldt 1898; 1905) in which the author takes into account the well-known
Hadamard's and de la Vallée-Poussin's works of 1896 even to number theory issues related to estimates of the
number of zeros of Riemann $\zeta$ function via $\xi$ function. Afterwards, Edmund Landau began to consider the
previous work of Hadamard, de la Vallée-Poussin and Von Mangoldt on analytic number theory, drawing up a first
long-paper in 16 sections (see (Landau 1908)) in which are gathered a great number of methods and applications
of the entire function theory to number-theoretic questions and where, in particular, a detailed treatment of
the Riemann $\xi$ function is achieved, taking into account the related properties of entire functions - as, for
example, exposed in the 1906 German translation of the G. Vivanti treatise (see (Vivanti 1901)), which is quoted
in (Landau 1908, p. 199, footnote $^{52}$) - even in view of their applications to prime number distribution
theory questions, but with a considerable attention to Hadamard's papers of 1893 and 1896. This 1908 Landau
paper may be considered as a first little monograph on analytic number theory applied to the distribution of
prime numbers, which will be shortly afterwards followed by the more consistent treatise (Landau 1909b); in such
a paper, Landau, amongst other things, improves and rectifies previous Von Mangoldt's formula to estimate the
number of Riemann zeta function zeros through the ratio $\zeta'/\zeta$, as well as provides a proof of the
so-called \it explicit formula \rm for the difference $\pi(x)-$li$(x)$ between the prime-counting function
$\pi(x)$ and the function li$\displaystyle(x)=\int_0^x\frac{dt}{\ln t}$. In the next Landau's paper of 1909 (see
(Landau 1909a)), the author starts with a consideration of a product expansion of the type (22) till to reach
expression (23), on the basis of what already made in his previous paper of 1908. Landau begins with the
consideration of some estimates achieved by Hadamard and Von Mangoldt, extending this to the case of a general
Dirichlet's series. In the subsequent paper of 1927 (see (Landau 1927)), Landau starts with the consideration of
a work due to the Indian mathematician K. Ananda-Rau (1893-1966), namely (Ananda-Rau 1924), saying that he was
the first to consider the following type of Hadamard's infinite product expansion of Riemann $\xi$ function
\begin{equation}(s-1)\zeta(s)=e^{H(s)}\prod_{\rho}\Big(1-\frac{s}{\rho}\Big)e^{\frac{s}{\rho}}\end{equation}where
$H(s)$ is an arbitrary linear function, then applying this case study to Riemann's $\xi$ function and for
further estimations of its zeros. The paper (Aranda-Rau 1924) mainly is centered around the getting of the
equation (22) by means of Jensen's formula and some previous results achieved by Landau himself (to which
Ananda-Rau refers mentioning (Landau 1909b)), rather that Hadamard's theory of entire functions (see (Narkiewicz
2000, Chapter 5, Section 5.1, Number 2)), with formal procedure which will be further improved by Landau himself
in (Landau 1927). Finally, following (Titchmarsh 1986) and references therein, further works on Riemann $\xi$
function were pursued by G.H. Hardy, E.K. Haviland, N. Koshlyakov, O. Miyatake, S. Ramanujan, A. Wintner (see
(Wintner 1935; 1936; 1947)) and N. Levinson, most of them even referring to the previous pioneering 1893 work of
J. Hadamard on entire functions.\\\\ $\bullet$ \it The contribution of G. P\'{o}lya. \rm After the contributions
by Von Mangoldt and Landau to Riemann $\xi$ function as sketchily delineated above, for our historical ends we
should further deepen the next contribution to Riemann $\xi$ function due to George P\'{o}lya (1887-1985) in
1926, with the paper (P\'{o}lya 1926), and that will be the truly joint point between the entire function theory
and the so-called Lee-Yang theorems. This paper, entitled \it Bemerkung \"{u}ber die Integraldarstellung der
Riemannschen $\xi$-Funktion, \rm has been considered a minor contribution of P\'{o}lya to the Riemann zeta
function theory but, as we will see later, it instead has contributed to open a new avenue in mathematical
physics with very interesting methods and outcomes which, successively, will turn out to be useful also for some
attempts to solve Riemann conjecture itself. Also following (Hejhal 1990, Introduction), P\'{o}lya's paper
starts with the consideration of a Fourier integral representation of $\xi(1/2+it)$ to develop a very
tantalizing result in the general direction of the Riemann hypothesis. To be precise, P\'{o}lya starts with the
following transformation of the Riemann $\xi$ function (see also (Titchmarsh 1986, Chapter X, Section 10.1))
\begin{equation}\xi(iz)=\frac{1}{2}\Big(z^2-\frac{1}{4}\Big)\pi^{-\frac{z}{2}-\frac{1}{4}}\Gamma\Big(\frac{z}{2}+
\frac{1}{4}\Big)\zeta\Big(\frac{1}{2}+z\Big),\end{equation}hence he considers the following integral
transformation\begin{equation}\xi(z)=2\int_0^{\infty}\Phi(u)\cos zudu\end{equation}where
\begin{equation}\Phi(u)=2\pi e^{\frac{5u}{2}}\sum_{n=1}^{\infty}\big(2\pi e^{2u}n^2-3\big)n^2e^{-n^2\pi
e^{2u}}\end{equation}with\begin{equation}\Phi(u)\sim 4\pi^2e^{\frac{9u}{2}-\pi e^{2u}}\qquad\mbox{\rm as}\quad
u\rightarrow+\infty,\end{equation}so that, due to the parity condition of this last asymptotic approximation, we
have as well
\begin{equation}\Phi(u)\sim 4\pi^2\Big(e^{\frac{9u}{2}}+e^{-\frac{9u}{2}}\Big)e^{-\pi(e^{2u}+e^{-2u})}\qquad\mbox{\rm as}\quad
u\rightarrow\pm\infty.\end{equation}Afterwards, P\'{o}lya deals with an approximation formula for $\xi$ obtained
from (29) considering a finite sum of $N$ terms rather than an infinite series, and with most of exponentials
replaced by hyperbolic cosines (see (Haglund 2009, Section 1) and (Balazard 2010)). Through this ansatz,
P\'{o}lya proves that the resulting integral (28) has only real zeros when $N=1$ (this result will
asymptotically extended to an arbitrary $N$ finite by D.A. Hejhal in (Hejhal 1990)). P\'{o}lya also showed that
if we replace $\Phi(u)$ by any function which is not an even function of $u$, then the resulting integral (28)
has only finitely many real zeros. In regard to the so-called \it Riemann Vermutung \rm (i.e., the Riemann
hypothesis), taking into account the asymptotic conditions (30) and (31), P\'{o}lya, on the basis of a personal
discussion with E. Landau had in 1913, argues on the possible existence or not of real zeros of such an
approximation of the $\xi$ function and, to this end, he considers (28) with a finite approximation for $\Phi$
given by $N=1$, under the asymptotic conditions (29) and (30), so obtaining
\begin{equation}\xi^*(z)=8\pi^2\int_0^{\infty}\Big(e^{\frac{9u}{2}}+e^{-\frac{9u}{2}}\Big)e^{-\pi(e^{2u}+e^{-2u})}
\cos zudu,\end{equation}whence questions inherent real zeros of $\xi(z)$ reduce to questions inherent real zeros
of $\xi^*$. Then, P\'{o}lya argues that asymptotically we have $\xi(z)\sim\xi^*(z)$ within an infinite angular
sector comprehending the real axis, with vertex in 0 and symmetrically opening with respect to the real axis.
Under this hypothesis, if $N(r)$ [respectively $N^*(r)]$ is the number of zeros of $\xi$ [respectively of
$\xi^*$] within this angular sector, the we have $N(r)\sim N^*(r)$ with $N(r)-N^*(r)=O(\ln r)$. P\'{o}lya,
therefore, reduces the study of real zeros of $\xi(z)$ to the study of real zeros of $\xi^*(z)$ (or to the study
of imaginary zeros of $\xi^*(iz)$), at least for the case $N=1$. To be precise, he considers $\xi^*$ and, to
analyze this function, he introduces an auxiliary entire function, namely the following one
\begin{equation}\mathfrak{G}(z)=\mathfrak{G}(z;a)\doteq\int_{-\infty}^{+\infty}e^{-a(e^u+e^{-u})+zu}du,\end{equation}where
$a$ is an arbitrary parameter even having positive values. Since we have
\begin{equation}\xi^*(z)=2\pi^2\Big\{\mathfrak{G}\Big(\frac{iz}{2}-\frac{9}{4};\pi\Big)+\mathfrak{G}
\Big(\frac{iz}{2}+\frac{9}{4};\pi\Big)\Big\},\end{equation}it is evident that the study of the $\xi^*$ function
may be reduced to the study of the auxiliary function $\mathfrak{G}$, to be precise, the above question about
the imaginary zeros of $\xi^*$ function is reduced to the study of real zeros of the function
$\mathfrak{G}(iz)$. Hence, P\'{o}lya keeps on with a detailed analysis of the various formal properties and
possible functional representations of this auxiliary function $\mathfrak{G}$ in view of their applications to
Riemann $\xi^*$ function and its zeros, starting to consider the case in which
\begin{equation}\xi^*(z)=2\int_0^{\infty}\Phi^*(u)\cos zudu\end{equation}and
\begin{equation}\Phi^*(u)=\sum_{n=1}^Ne^{-2\pi n^2\cosh(2u)}[8\pi^2n^4\cosh(9u/2)-12\pi
n^2\cosh(5u/2)]\end{equation}as an approximation to $\Phi(u)$ for $N$ finite. In particular, the case $N=1$ is
quite interesting, and P\'{o}lya was able to prove that all zeros of $\xi^*$ function are real just in this
case. P\'{o}lya also succeeded to find a Riemann-like estimate of the number of zeros of $\mathfrak{G}$, in the
form\begin{equation}\frac{y}{\pi}\ln\frac{y}{a}-\frac{y}{\pi}+O(1).\end{equation}Afterwards, P\'{o}lya
introduces and proves two general lemmas which are need just for proving the main aim of the paper, that is to
say, to evaluate the nature of the zeros of an approximation of the Riemann $\xi$ function: the first one
concerns general analytic function theory, while the second one regards instead entire functions. Due to the
importance of the latter, we stress what P\'{o}lya says in this regard, and, let us say immediately, he attains
a point in which necessarily intervenes Hadamard's factorization theorem of entire functions applied to certain
representations of $\mathfrak{G}$, just to find its zeros. These P\'{o}lya lemmata will be just those formal
outcomes which will lead to the prove of the theorems of Lee and Yang. To be precise, his first lemma,
called \it Hilfssatz I, \rm states that\\

\it<<Let $F(u)$ be an analytic function which has real values for each real value of $u$, and furthermore let
$$\lim_{u\rightarrow\infty}u^2F^{(n)}(u)=0\leqno(a)$$ for $n=0,1,2,...$. Then, when $F(u)$ is not an even function, then we
have$$G(x)=\int_0^{\infty}F(u)\cos xudu\leqno(b)$$for values of $x$ great enough>>.\\\\\rm Hence, P\'{o}lya
considers the case in which the function $F(u)$ of Hilfssatz I is of the type (31), so that the corresponding
function $G(x)$ of $(b)$ is an entire function to which is now applicable the well-known Hadamard factorization
theorem of entire functions to find its zeros. To this end, extending to entire functions a previous result
achieved by C. Hermite and C. Biehler for polynomials\footnote{Following (P\'{o}lya \& Szeg\H{o} 1998a, Part
III, Exercise 25), this result, due to Hermite and Biehler, is as follows. We assume that all the zeros of the
polynomial $P(z) = a_0z^n+a_1z^{n-1}+...+a_{n-1}z+a_n$ are in the upper half-plane $\Im z>0$. Let
$a_{\nu}=\alpha_{\nu}+i\beta_{\nu}$, with $\alpha_{\nu},\beta_{\nu}$ real, $\nu=0,1,2, ...,n$, and $U(z) =
\alpha_0z^n +\alpha_1z^{n-1}+\alpha_{\nu-1}z+\alpha_0$, $V(z) =\beta_0z^n
+\beta_1z^{n-1}+\beta_{\nu-1}z+\beta_0$. Then the polynomials $U(z)$ and $V(z)$ have only real zeros.} (and
already quoted in the previous chapter about $HB$ class - see (Biehler 1879) and (Hermite 1856a,b; 1873)),
P\'{o}lya considers a second lemma, called \it Hilfssatz II, \rm which states that\\

\it<<Let $A$ be a positive constant and let $G$ be an entire function of order 0 or 1, which has real values for
real values of $z$, has, at least, on real zero, and has real zeros only. Then the function $G(z-ia)+G(z+ia)$
has real zeros only>>.\\\\\rm Due to the importance of this P\'{o}lya Hilfssatz II, herein we report the brief
and elegant proof given by P\'{o}lya, following (Cardon 2002, Section 1). Applying Hadamard's factorization
theorem to the entire function $G(z)$, we have
$$G(z)=cz^qe^{\alpha z}\prod_n\Big(1-\frac{z}{\alpha_n}\Big)e^{\frac{z}{\alpha_n}}$$where
$c,\alpha_1,\alpha_2,...$ are real constant, $\alpha_n\neq 0$ for each $n$, $q\in\mathbb{N}_0$ and
$\sum_n|\alpha_n|^2$ is convergent. When $z=x+iy$ is a zero of the function $G(z-ia)+G(z+ia)$, then we have
$|G(z-ia)|=|G(z+ia)|$, so that, by means of the above Hadamard factorization
$$1=\Big|\frac{G(z-ia)}{G(z+ia)}\Big|^2=\Big(\frac{x^2+(y-a)^2}{x^2+(y+a)^2}\Big)^q\prod_n\frac{(x-\alpha_n)^2+(y-a)^2}
{(x-\alpha_n)^2+(y+a)^2}.\leqno(\circ)$$Now, if it were $y>0$, then the right hand side of the last expression
would be lesser than 1, whereas if it were $y<0$, then the right hand side of the last expression would be
greater than 1, and both of these cases are impossible, so $y=0$ whence $G(z+ia)+G(z-ia)$ has only real zeros.
Accordingly, it follows that $\mathfrak{G}(iz/2)$ has not real zeros when we apply Hilfssatz II to the function
$G(z)=\mathfrak{G}(iz/2;\pi)$. From all this, related considerations follow for $\xi^*(z)$. In conclusion, we
may say that Hadamard factorization theorem has played a crucial role in proving Hilfssatz II which, in turn,
has helped in achieving the main aim of this P\'{o}lya's paper. Furthermore, a good part of the next literature
on the argument, like that related to the work achieved by D.A. Cardon and co-workers (see, for example (Adams
\& Cardon 2007)), makes wide and frequent use of factorization theorems of the type Weierstrass-Hadamard. Later
on, P\'{o}lya improved this result for finite values of $N>1$ in (P\'{o}lya 1927a), this line of research
results having been vastly generalized later by D.A. Hejal in (Hejal 1990), whilst Hilfssatz II was generalized
by D.A. Cardon in (Cardon 2002). Following (Newman 1976), the problem of determining whether a Fourier transform
has only real zeros arises in two rather disparate areas of mathematics: number theory and mathematical physics.
In number theory, the problem is intimately associated with the Riemann hypothesis (see (Titchmarsh 1980,
Chapter X)), while in mathematical physics it is closely connected with Lee-Yang theorems of statistical
mechanics and quantum field theory. Finally, this 1926 work of P\'{o}lya was then commented by Mark Kac
(1914-1984), when it was inserted into the \it Collected Papers \rm of P\'{o}lya, of which we shall talk about
in the next section.\\\\\bf 3.2. On the history of the theorems of T.L. Lee and C.N. Yang. \rm In regard to the
just above mentioned P\'{o}lya's paper, Kac says that, although this beautiful paper takes one to within an
hair's breadth of Riemann's hypothesis, it didn't seem to have inspired much further work, and references to it,
in the subsequent mathematical literature, were rather poor. Nevertheless, Kac says that, because of this, it
may be of interest to refer that P\'{o}lya's paper did play a small, but perhaps not wholly negligible, part in
the development of an interesting and important chapter in Statistical Mechanics, as we will see later. Instead,
according to us, P\'{o}lya's paper played a notable, and not simply a small, role (though quite implicit) not
only as regard the pioneering work of T.D. Lee and C.N. Yang in statistical mechanics of the early 1950s, but
also for some next developments of Riemann zeta function theory, as witnessed by the latest researches on the
subject as, for instance, those made by D.A. Cardon and co-workers (see (Adams \& Cardon 2007) and references
therein), in which Polya work is put into an interesting and fruitful relationship with Lee-Yang theorems in
view of its applications just to Riemann zeta function. However, to begin in delineating the history of the
Lee-Yang theorem, we first report the related comment and witness due to Yang himself and drawn from (Yang 2005,
pp. 14-16)\\

\it<<In the fall of 1951, T.D. Lee came to the Institute for Advanced Study, and we resumed our collaboration.
The first problem we tackled was the susceptibility of the two-dimensional Ising model. As stated in a previous
Commentary, the Onsager-Kaufman method yielded information about all eigenvectors of the transfer matrix. I had
used some of that information to evaluate the magnetization, and I thought we might be able to use more of that
information to evaluate the susceptibility by a second-order perturbation method, one order beyond that used to
obtain equation (14) of the previous paper \rm The Spontaneous Magnetization of a Two-Dimensional Ising Model,
The Physical Review, 85, 808 (1952). \it This led to a formula that was, so to speak, an order of magnitude more
difficult to evaluate than the magnetization. After a few weeks of labor we gave up and turned our attention to
the lattice gas, then to J. Mayer's theory of gas-liquid transitions, and finally to the unit-circle theorem.
These considerations led to papers (Yang \& Lee 1952) and (Lee \& Yang 1952). The idea of the lattice gas was
more or less in the minds of many authors (see reference 2 of (Lee \& Yang 1952)). We firmed it up and
elaborated on it because with the result of (Yang \& Lee 1952), we were able to construct the exact two-phase
region of a simple two-dimensional lattice gas. (We were especially pleased by the ''law of constant diameter'',
which resembled the experimental ''law of rectilinear diameter''). The two-phase region consists of flat
portions of the $P-V$ diagram, bounded by the liquid and gas phases. We were thus led very naturally to the
question of why Mayer's theory of condensation gave isotherms that stayed flat into the liquid phase, instead of
becoming curves in the liquid phase. Besides, Mayer theory of condensation was a milestone in equilibrium
statistical mechanics, for it broke away from the mean field type of approach to phase transitions. It caused
quite a stir at the Van der Waals Centenary Congress on November 26, 1937. Mayer's theory led to a number of
papers by Mayer himself, by B. Kahn and G.E. Uhlenbeck, and by others in succeeding years. In the early 1940s I
had attended a series of lectures by J.S. Wang in Kunming on these developments and had been very much
interested in the subject ever since. Using the lattice gas model, for which we had a lot of exact information,
Lee and I examined Mayer's theory as applied to this case. This led to a study of the limiting process in the
evaluation of the grand partition function for infinite volume. Paper (Yang \& Lee 1952) resulted from this
study. It clarified the limiting process and made transparent the relationship between the various portions of
an isotherm and the limiting process. In late 1952, after (Yang \& Lee 1952) had appeared in print, Einstein
sent Bruria Kaufman, who was then his assistant, to ask me to see him. I went with her to his office, and he
expressed great interest in the paper. That was not surprising, since thermodynamics and statistical mechanics
were among his main interests. Unfortunately I did not get very much out of that conversation, the most
extensive one I had with Einstein, since I had difficulty understanding him. He spoke very softly, and I found
it difficult to concentrate on his words, being quite overwhelmed by the nearness of a great physicist whom I
had admired for so long. Back in the fall of 1951, Lee and I, in familiarizing ourselves with lattice gases,
computed the partition function for several small lattices with 2, 3, 4, 5, etc. lattice points. We discovered
to our amazement that the roots of the partition functions, which are polynomials in the fugacity, are all on
the unit circle for attractive interactions. We were fascinated by this phenomenon and soon conjectured that it
was a general theorem for a lattice of any size with attractive interactions. The theorem, later called the
circle theorem, became the main element that was exploited in (Lee \& Yang 1952) to discuss the thermodynamics
of a lattice gas. Our attempt at proving the conjecture was a struggle, which I described in a letter to M. Kac,
dated September 30, 1969, when he was writing for the Collected Papers of George P\'{o}lya. I quote now from
that letter \rm

\begin{quotation}When Lee arrived at Princeton, in the fall 1951, I was just recovering from my computation of the
magnetization of the Ising model. I realized that the Ising model is equivalent to the concept of a lattice gas.
So, we worked on that and finally produced our paper (Yang \& Lee 1952). In the process of doing that, we
discovered, by working on a number of examples, the conjectured unit circle theorem. We then formulated a
physicist's ''proof'' based on no double roots when the strength of the couplings were varied. Very soon we
recognized this was incorrect; and for, I would guess, at least six weeks we were frustrated in trying to prove
the conjecture. I remember our checking into Hardy's book \it Inequalities, \rm our talking to Von Neumann and
Selberg. We were, of course, in constant contact with you all along (and I remember with pleasure your later
help in showing us Wintner's work, which we acknowledged in our paper). Sometime in early December, I believe,
you showed us the proof of the special case when all the couplings are there and are of equal strength, the case
that you are now writing about in connection with P\'{o}lya's collected works. The proof was fine, but we were
still stuck on the general problem. Then one evening around December 20, working at home, I suddenly recognized
that by making $z_l,z_2,...$ independent variables and studying their motions relative to the unit circle one
could, through an induction procedure, bring to bear a reasoning similar to the one used in your argument and
produce the complete proof. Once this idea was there, it took only a few minutes to tighten up all the details
of the argument. The next morning I drove Lee to pick up some Christmas trees, and I told him the proof in the
car. Later on, we went to the Institute; and I remember telling you about the proof at a blackboard. I remember
these quite distinctly because I'm quite proud of both the conjecture and the proof. I t is not such a great
contribution, but I fondly consider it a minor gem.\end{quotation}\it The unit circle theorem was later
generalized and extended to very interesting additional types of interactions\footnote{The theory of phase
transitions and its rigorous results, was then improved, generalized, enlarged and extended in many respects,
through the works of D. Ruelle (see, for instance, (Ruelle 1969; 1994; 2000; 2010)), B.M. McCoy, T.T. Wu, T.
Asano, M. Suzuki, M.E. Fischer, C.M. Newman, J.L. Lebowitz, R.B, Griffiths, E.R. Speer, B. Simon (see (Simon
1974) and references therein), E.H. Lieb, O.J. Heilmann, A.D. Sokal, D.G. Wagner, R.L. Dobrushin, G. Gallavotti,
S. Miracle-Sole, D.W. Robinson, J. Fr\"{o}hlich, P-F. Rodriguez (see (Frohlich \& Rodriguez 2012)), J. Borcea,
P. Br\"{a}ndén (see (Borcea \& Br\"{a}ndén 2008; 2009a,b), (Br\"{a}ndén 2011) and references therein), M.
Biskup, C. Borgs, J.T. Chayes, R. Kotecky, L.J. Kleinwaks, L.K. Runnels, J.B. Hubbard, A. Hinkkanen, C. Gruber,
A. Hintermann, D. Merlini, and others. See (Georgii 2011, Bibliographical Notes) as well as (Gruber et al.
1977), (Baracca 1980, Appendice), (Lebowitz et al. 2012) and references therein. Of the interesting work of
these authors, we shall deal with another, next paper.}. With the unit circle theorem, it appeared to Lee and me
in early 1952 that we could somehow figure out or guess at the root-distribution function $g(\theta)$ on the
unit circle (see (Lee \& Yang 1952, Section V)) for the two-dimensional Ising model. We thought that, with the
exact expressions for the free energy and the magnetization already known, we had powerful handles on the
structure of $g(\theta)$. Unfortunately these handles were not powerful enough, and the exact form of
$g(\theta)$ remains unknown today (the exact form of $g(\theta)$ is of course transformable into the exact
partition function of the Ising model in a magnetic field). But our efforts in this direction did lead to two
useful results. In listening to a seminar around the end of February, 1952, I learned about the new, ingenious
combinatorial method of M. Kac and J. Ward for solving the Ising problem without a magnetic field. It occurred
to me during the seminar that, by a slight modification of the Kac-Ward method, one could find the partition
function for the king model with an imaginary magnetic field $H=i\pi/2$. This requires the evaluation of an
$8\times 8$ matrix, which Lee and I carried out in the next couple of days, arriving at equation (48) of (Lee \&
Yang 1952) for the free energy with $H=i\pi/2$. Comparing this expression with the known L. Onsager result for
the same quantity for the case $H=0$, Lee and I observed that they are very similar except for some sign changes
and related alterations. Thus it seemed that the change $H=0+H=i\pi/2$ is altogether minor. We therefore tried
similar minor changes on the magnetization for $H=0$ and tested the results by checking whether they were in
agreement with the first few terms of a series expansion of the magnetization for $H=i\pi/2$. This was a very
good method, and we soon arrived at equation (49) of (Lee \& Yang 1952), which we knew was correct, but did not
succeed in proving. It was finally proved by B.M. McCoy and T.T. Wu\footnote{In (McCoy \& Wu 1967a), using
higher mathematical techniques of complex analysis, like Wiener-Hopf method, Szeg\H{o}'s theorems for $N\times
N$ Toeplitz determinants applied to determine magnetization $M(i\pi/2)$ as $N\rightarrow\infty$ (for $\beta=1$),
etc. See also (McCoy \& Wu 1966; 1967b), (Cheng \& Wu 1967) and (McCoy \& Wu 1973).} in 1967>>.\\\\\rm Following
(Huang 1987, Chapter 9), after pioneering work of Lee and Yang, phase transitions are manifested in experiments
by the occurrence of singularities in thermodynamic functions, such as the pressure in a liquid-gas system, or
the magnetization in a ferromagnetic system, with $N$ particles. Huang asks: How is it possible that such
singularities arise from the partition function, which seems to be an analytic function of its arguments? Huang
says that the answer lies in the fact that a macroscopic system is close to the idealized thermodynamic limit -
i.e., the limit of infinite volume with particle density held fixed. As we approach this limit, the partition
function can develop singularities, simply because the limit function of a sequence of analytic functions need
not be analytic. Yang and Lee just proposed a definite framework for the occurrence of singularities in the
thermodynamic limit. Due to its formal character, it belongs to a chapter of statistical physics sometimes known
as ''rigorous statistical mechanics'' (see the 1969 Ruelle's monograph). Following (Ma 1985, Chapter 9), if the
number of particles $N$ of a given thermodynamical system is finite, then the calculation of the various
thermodynamic potentials does not pose any problem. Although $N$ is not infinite, it is nevertheless a very
large number like $10^{23}$ (Avogadro's number), hence the problem of the $N\rightarrow\infty$ limit becomes a
very important problem for the application of the basic assumption in thermodynamics, i.e. the so-called problem
of the \it macroscopic limit. \rm The rigorous mathematical analysis of this limit is a branch of statistical
mechanics. The pioneering work in this topic is just the Yang-Lee theorem of 1950s, which was originally
proposed for phase transitions. Following them, many have applied rigorous mathematical analyzes to describe
phase transitions, because the model problems of phase transitions are not easily solvable and less than
rigorous analysis is not reliable. However the application of the Yang-Lee theorem is quite universal. Following
(McCoy \& Wu 1973), the analyticity properties of the grand canonical partition function for Ising models
correspond to qualitative features that appear in the thermodynamic limit which are not possible in a system
with a finite number of particles. These analytic properties are intimately related to the physical notion of
phase transition. The major reason for studying the two-dimensional Ising model (as, for example, masterfully
exposed in (McCoy \& Wu 1973)) is to attempt to make this connection more precise. Following (Domb \& Green
1972, Chapter 2, II. and IV.), a mathematically ''sharp'' phase transition can only occur in the thermodynamical
limit. It is also true in general that, only in the thermodynamic limit, the different statistical ensembles
(i.e., microcanonical, canonical, and grand canonical) yield equivalent thermodynamic functions. Hence this
limit permits a mathematically precise discussion of the question of phase transitions. The problem of proving
the existence of a thermodynamic limit for the thermodynamic properties of a system of interacting particles
seems first to have been discussed by L. Van Hove in 1949 in the case of a continuum classical gas with hard
cores in the canonical ensemble, although the proof is incomplete due to an error in the appendix of the paper.
Later, Yang and Lee, in 1952, considered the same system in the grand ensemble and L. Witten, in 1954, extended
their proof with a relaxation of the condition of hard cores. Then, D. Ruelle, in 1963, proved the existence of
limits in both the canonical and grand canonical case under a ''strong-tempering'' condition on the potential,
and extended the results to quantum gases. Hence, R.L. Dobrushin and M.E. Fisher, in 1964, showed how Ruelle
arguments could be extended to a more general class of potentials, and Fisher considered in some detail the
possible class of domains tending to infinity for which a limit exists. The thermodynamic limit for lattice
systems was discussed by R.B. Griffiths in 1964 for both classical and quantum systems. Additional results have
been obtained  in 1967 by G. Gallavotti and S. Miracle-Sole for classical lattice systems and by D.W. Robinson
for quantum lattice systems (see (Domb \& Green 1972, Chapter 2) for detailed bibliographical information).  Lee
and Yang opened the way to the rigorous theory of phase transitions, then included into the wider chapter of
algebraic methods of statistical mechanics (see (Lavis \& Bell 1999, Volume 2, Chapter 4)); moreover, for most
recent formal developments of advanced statistical mechanics see also (McCoy 2010).

From a direct analysis of the original papers (also following the modern treatment given by (Huang 1987, Chapter
9)), it turns out that already in (Yang \& Lee 1952, Section III) the authors make use of simple polynomial
factorizations of the Weierstrassian type. Indeed, in (Yang \& Lee 1952, Section II), it is considered a system
of $N$ particles filling a region of finite volume $V$, undergoing a two-body interaction by means of a
potential of the type $U$, whose partition function, defined on the grand canonical ensemble in the complex
variable $y=A\exp(\mu/kT)$ (fugacity), is given by
\begin{equation}\mathcal{Q}_y=\sum_{N=0}^{M}\frac{Q_N}{N!}y^N\end{equation}where
$$Q_N=\int...\int_V\exp(-\frac{U}{kT})d\tau_1...d\tau_N$$
where $U=\Sigma_{ij}u(r_{ij})=\sum_{ij}u(|r_i-r_j|)$ is the sum of the various interaction potentials between
the $i$-th and $j$-th particles, which undergo particular restrictions, and $M=M(V)$ is the maximum number of
particles which can be crammed into the finite volume $V$. Then Yang and Lee consider the following limits for
infinite volume
\begin{equation}\frac{p}{kT}=\lim_{V\rightarrow\infty}\frac{1}{V}\ln\mathcal{Q}_y,\qquad\rho=\lim_{V\rightarrow\infty}
\frac{\partial}{\partial\ln y}\frac{1}{V}\ln\mathcal{Q}_y\end{equation}reaching two main theorems thanks to
which it is possible to study these limits for potentials of the type\footnote{It is said to be an \it hard-core
potential. \rm Such a potential is related to an impenetrable sphere of radius $a$ surrounded by an attractive
potential with action radius $r_0$ and maximal deep $\epsilon$. The occurrence of such an impenetrable potential
implies that, for each fixed value of the total volume $V$, only a finite number of particles may be considered,
and if $N_{max}(V)=M$ is the maximum of such a number, then we have that, when $N>M$, at least two particles are
in touch, so the potential $U$ is infinite, and $\mathcal{Q}_y=0$. Therefore, $\mathcal{Q}_y$ is a polynomial of
degree $M$ (see (41)). Nevertheless, the thermodynamics of physical systems is ruled by the logarithm of the
partition function, so that its zeroes broke analyticity of thermodynamic functions, so giving rise to
singularities which, on its turn, are related to the occurrence of phase transitions.}
$$U(r)=\left\{\begin{array}{ll}
\infty & \mbox{\rm for}\ \ r<a, \\
-\infty<U(r)<-\epsilon & \mbox{\rm for}\ \ a<r<r_0,\\
0 & \mbox{\rm for}\ \ r>r_0,
\end{array}\right.$$
which is a reasonable approximation of a potential of the Lennard-Jones type. In such a case, $Q_N$ converges
and $\mathcal{Q}_y$ is a polynomial in $y$ whose degree depends on $V$ and whose coefficients are analytic
functions of $\beta=1/kT$, defined to be positive for real values of $\beta$. Accordingly, the zeros of
$\mathcal{Q}_y$, in the complex plane $y$, are in a finite number and lie out of the positive real axis. Only in
the thermodynamic limit $V\rightarrow\infty$ (or infinite volume limit), the zeros of $\mathcal{Q}_y$ are
infinite and may approach positive real axis\footnote{In this case, we feel allowable to refer to many theorems
on the distribution of the zeros of entire functions like, for example, expounded in (Levin 1980, Chapters VII,
VIII) and mainly regarding $LP$ and $HB$ classes of entire functions, some of which just provide necessary
and/or sufficient conditions for zeros of certain entire functions approach positive real axis.}, with the
appearance of singularities in the thermodynamic potentials. To be precise, in both limits $N\rightarrow\infty$
and $V\rightarrow\infty$ in such a way that the specific volume $v=V/N$ is bounded, some of the zeros of
$\mathcal{Q}_y$ may approach the positive real axis, so giving rise to possible phase transitions. From these
considerations applied to particular physical systems (amongst which ferromagnetic spin systems), Lee and Yang
have worked out a phase transition model whose one of the main characteristics is having pointed out the close
relationships between the existence of phase transitions and general properties of the related potential on the
one hand, and between the thermodynamic limit and the occurrence of singularities of thermodynamic potentials on
the other hand. These latter relationships are, on its turn, related to the occurrence of zeros of the partition
function. In particular, for Ising models of ferromagnetic spin systems, the distribution of the zeros of the
partition function takes a well-determined geometrical shape by means of the deduction of certain general
theorems proved by Lee and Yang in their two seminal papers of 1952. To be precise, we are interested in two of
these theorems, namely the so-called Theorem 1, according to which, for all positive real values of $y$, the
first limit approaches, as $V\rightarrow\infty$, a limit which is independent of the shape of $V$, this limit
being moreover a continuous, monotonically increasing function of $y$, and the so-called Theorem 2, which states
that, if in the complex $y$ plane a region $R$ containing a segment of the positive real axis is always free of
roots, then in this region as $V\rightarrow\infty$, all the quantities
\begin{equation}\frac{1}{V}\ln\mathcal{Q}_y,\quad\frac{\partial}{\partial\ln y}\frac{1}{V}\ln\mathcal{Q}_y,\quad
\Big(\frac{\partial}{\partial\ln y}\Big)^2\frac{1}{V}\ln\mathcal{Q}_y,\quad...,\end{equation}approach limits
which are analytic with respect to $y$. To study the limit of $\displaystyle\frac{\partial}{\partial\ln
y}\frac{1}{V}\ln\mathcal{Q}_y$ we notice that $\mathcal{Q}_y$ is a polynomial in $y$ of finite degree $M$. This
is a direct consequence of the assumed impenetrable core of the atoms (roughly formalized by an hard-core
potential $U$ as done above). It is therefore possible to factorize $\mathcal{Q}_y$ and
write\begin{equation}\mathcal{Q}_y=\prod_{i=1}^M\Big(1-\frac{y}{y_i}\Big)\end{equation} where $y_1,...,y_M$ are
the roots of the algebraic equation $\mathcal{Q}_y(y)=0$. Evidently none of these roots can be real and
positive, since all the coefficients in the polynomial $\mathcal{Q}_y$ are positive. Following (Yang \& Lee
1952, Section IV), by Theorem 2 it follows that, as $V$ increases, these roots move about in the complex $y$
plane and their number $M$ increases (essentially) linearly with $V$. Their distribution in the limit
$V\rightarrow\infty$ gives the complete analytic behavior of the thermodynamic functions in the $y$ plane. On
the other hand, the problem of phase transition is intrinsically related to the form of the regions $R$
described in Theorem 2, and Lee and Yang discuss two main cases related to the geometry of this region $R$, and
the related roots of $\mathcal{Q}_y(y)=0$, around real $y$ axis, reaching the conclusion that phase transitions
of the system occur only at the points on the positive real $y$ axis onto which the roots of
$\mathcal{Q}_y(y)=0$ close in as $V\rightarrow\infty$ (which entails $M\rightarrow\infty$ in (41)). For other
values of the fugacity $y$, a single phase system is obtained. The study of the equations of state and phase
transitions can thus be reduced to the investigation of the distribution of roots of the grand partition
function. In many cases, as will be seen in (Lee \& Yang 1952), such distributions will turn out to have some
surprisingly simple regularities. The above mentioned theorems 1 and 2, will be proved respectively in Appendix
I and II of (Yang \& Lee 1952), considering arbitrary circles lying inside $R$. On the other hand, since the
degree $M$ of the polynomial $\mathcal{Q}_y$ is function of $V$, so $M\rightarrow\infty$ as
$V\rightarrow\infty$, following (Ruelle 1969, Chapter 3, Sections 3.2, 3.4; Exercise 3.E), under the hypothesis
that Hamiltonian operator of the physical system be bounded below (\it stability condition\rm), we have that the
partition function in the grand canonical ensemble\footnote{Following (Huang 1987, Chapter 7, Section 7.3), such
a function is often simply called the \it grand partition function.} is given by the following entire function
of the fugacity (or activity) $z$ (corresponding to $y$ of Lee and Yang notation) considered as a complex
variable (in the notations of Ruelle)
$$\Xi(\Lambda,z,\beta)=1+\sum_{n=1}^{\infty}\frac{z^n}{n!}\int...\int_{\Lambda^n}dx_1\ ...\ dx_n\exp[-\beta
U(x_1,...,x_n)]=$$ $$=\sum_{n=0}^{\infty}z^nQ(\Lambda,n,z),\qquad\mbox{\rm with}\quad Q(\Lambda,0,z)=1$$which is
of order at most 1, and of order 0 in the case of superstable potentials, that is, potentials $U$ satisfying,
into a cube $\Lambda$ of volume $V$, the condition $U(x_1,...,x_n)\geq n(-B+nC/V)$ for certain constants
$B,C>0$, in such last case Hadamard's factorization yielding$$\Xi(z)=\prod_i\Big(1-\frac{z}{z_i}\Big),$$where
$z_i$ are the zeros of $\Xi$, and reducing to a polynomial when $U$ is an hard-core potential (Lee-Yang case).
Nevertheless, the construction of the partition function of a given physical system is one of the tricky task of
statistical mechanics.

Thus\footnote{We here follow, almost verbatim, (Lee \& Yang 1952).}, in (Yang \& Lee 1952), the authors have
seen that the problem of a statistical theory of phase transitions and equations of state is closely connected
with the distribution of roots of the grand partition function. There, it was shown that the distribution of
roots determines completely the equation of state, and in particular its behavior near the positive real axis
prescribes the properties of the system in relation to phase transitions. It was also shown there that the
equation of state of the condensed phases as well as the gas phase can be correctly obtained from a knowledge of
the distribution of roots. While this general and abstract theory clarifies the problems underlying the
statistical theory of phase transitions and condensed phases, it is natural to ask whether it also provides us
with a means of obtaining practical approximation methods for calculating properties pertaining to phase
transitions and condensed phases. The problem is clearly that of seeking for the properties of the distribution
of roots of the grand partition function. At a first sight, this appears to be a formidable problem, as the
roots are in general complex and would naturally be expected to spread themselves for an infinite sample in the
entire complex plane, or at least regions of the complex plane, and make it very difficult to calculate their
distribution. We were quite surprised, therefore, to find that for a large class of problems of practical
interest, the roots behave remarkably well in that they distribute themselves not all over the complex plane,
but only on a fixed circle. This fact will be stated as a theorem in Section IV of (Lee \& Yang 1952) and proved
in the Appendix, while implications of the theorem are discussed in Section V. Lee and Yang return to the
general problem of the condensation of gases, and shall in the following apply the results of the previous paper
(Yang \& Lee 1952) to the problem of a lattice gas. There was then no loss of generality in confining their
attention to a lattice gas, as a real continuum gas can be considered as the limit of a lattice gas as the
lattice constant becomes infinitesimally small. The equivalence proved in Section II of (Lee \& Yang 1952)
states that the problem of a lattice gas is identical with that of an Ising model in a magnetic field, and that
the grand partition function in the former problem is proportional to the partition function in the latter
problem. It is convenient to introduce in the Ising model problem the variable $z=\exp(-2H/kT)$ which is
proportional to the fugacity $y$ of the lattice gas $y=\sigma z$, where $\sigma$ is a constant. In terms of $z$
the partition function $\exp(-NF/kT)$ of the Ising lattice is equal to $\exp(NH/kT)$ times a polynomial
$\mathcal{P}$ in $z$ of degree $N$, that is, $\exp(-NF/kT)=\mathcal{P}\exp(NH/kT)$ where
$\mathcal{P}=\sum_{n=0}^NP_nz^n$. The coefficients $P_n$, are the contribution to the partition function of the
Ising lattice in zero external field from configurations with the number of $\downarrow$ spin down equal to $n$.
It should be noticed that $P_n=P_{n'}$ if $n+n'=N$, with each $P_n$ real and positive. Furthermore, the roots of
the polynomial $\mathcal{P}$ are never on the positive real $z$ axis, and are in general complex. The results of
(Yang \& Lee 1952), show that if at a given temperature as $N$ approaches infinity, the roots of the polynomial
$\mathcal{P}$ do not close in onto the positive real axis in the complex $z$ plane, the free energy $F$ is an
analytic function of the positive real variable $z$. Physically this means that the Ising model has a smooth
isotherm in the $I$-$H$ diagram (where $I$ is the intensity of magnetization and $H$ is the magnetic field) and
that the corresponding lattice gas undergoes no phase transition at the given temperature. If, on the other
hand, the roots of the polynomial $\mathcal{P}$ do close in onto the positive real $z$ axis at the points $z=
t_l, t_2, ...$, each of these points would correspond to a discontinuity of the isotherm in the $I$-$H$ diagram
of the Ising lattice and to a phase transition of the lattice gas. To study the problem of phase transitions of
a lattice gas as well as of an Ising model related to ferromagnetic spin systems, one therefore needs only to
study the distribution in the complex $z$ plane of the roots of the polynomial $\mathcal{P}$. The surprising
thing is that under quite general conditions, this distribution shows a remarkably simple regularity, which may
be stated in the form a theorem, say Theorem 3 (see (Lee \& Yang 1952, Section IV)), stating that, if the
interaction $u$ between two gas atoms is such that $u=+\infty$ if the two atoms occupy the same lattice and
$u\leq 0$ otherwise, then all the roots of the polynomial $\mathcal{P}$ lie on the unit circle in the complex
$z$-plane. This theorem will be proved in Appendix II of (Lee \& Yang 1952). Thus, for the interaction of the
theorem 3, the roots of $\mathcal{P}$ lie on the unit circle, so its distribution as $N\rightarrow\infty$ may be
described as a function $g(\theta)$ so that $Ng(\theta)d\theta$ is the number of roots with $z$ between
$e^{i\theta}$ and $e^{i(\theta+d\theta)}$, with $g(\theta)=g(-\theta)$ and
$\displaystyle\int_0^{\pi}g(\theta)d\theta=1/2$. The average density of a finite lattice gas is easily seen to
be $\sum_k z/(z-\exp(i\theta_k))$ where $z=\exp(i\theta_k)$ are the zeros of the grand partition function. The
results of (Yang \& Lee 1952) show that this average density converges to an analytic function in $z$ both
inside and outside of the unit circle as the size of the lattice approaches infinity. It seems intuitively clear
from this that the distribution of these roots should also approach a limiting distribution on the unit circle
for an infinite lattice, this being indeed the case whose a rigorous mathematical proof exists in the literature
(see\footnote{This work mainly deals with mathematical properties of asymptotic distributions $\sigma$ of the
values of certain fast periodic and quasi-periodic functions (above all following many E. Helly and H. Bohr
works on this subject), by means of certain Cauchy transforms (see also (M\"{u}ller-Hartmann 1977, Section
II.B)). Yang and Lee pointed out what fundamental role has played the paper of Aurel Wintner of 1934, suggested
them again by M. Kac, in proving that roots of grand partition function are distributed along a unit circle.}
(Wintner 1934)), the authors acknowledging Professor Kac for have shown them the proof. After having considered
a certain number of specific physical cases, the authors finish stating that the previous results have direct
bearing on the distribution function $g(\theta)$ of the zeros of the partition function on the unit circle,
showing too that the motion of the roots deploys toward the right along the unit circle as the temperature
decreases. They also say that, since the relation between the distribution of roots of a polynomial and its
coefficients is mathematically a very complicated problem, it is therefore very surprising that the distribution
should exhibit such simple regularities as proved in Theorem 3 which applies under very general conditions, so
being tempted to generalize such outcomes. One cannot escape the feeling that there is a very simple basis
underlying the theorem, with much wider application, which still has to be discovered. Finally, the authors
express their gratitude to Professor M. Kac for many stimulating and very pleasant discussions from which they
learned much in mathematics. The paper (Lee \& Yang 1952) finishes with the Appendix II in which Theorem 3 is
proved in a detailed manner. Usually, all the theorems contained in (Yang \& Lee 1952) and (Lee \& Yang 1952),
are sometimes called \it Lee-Yang theorems \rm (or \it Yang-Lee theorems\rm), while some other times, Theorem 3
is the one to which is usually referred to the single expression \it Lee-Yang theorem, \rm when it is declined
in the singular. Often, the latter is also referred to as the \it Lee-Yang circle theorem \rm (or \it Lee-Yang
unit circle theorem\rm). To summarize, these theorems therefore imply that the zeros of a finite physical system
cannot lie in the positive real axis, with consequent absence of phase transitions. But a quite different
situation arises when we deal with infinite systems, in such a case being possible that the zeros of the
partition function may approach real axis and, in the thermodynamic limit, produce that catastrophic situation
given by a phase transition. In the special case of a two-dimensional lattice Ising ferromagnetic system, Lee
and Yang proved that such zeros laid all into a unitary circle of the complex plane of fugacity $y$, so that
real axis is cut in $y=1$ by this circle, corresponding therefore to a phase transition with zero magnetic
field, while all the other positive real values of $y$ are points of analyticity for the thermodynamical
functions. This pioneering idea of Lee and Yang, albeit related to a particular case, will be the subject-matter
of further researches meant to generalize or extend it.

But, as a further historical deepening of this case, we report the witness of M. Kac inserted into the comment
to 1926 P\'{o}lya paper included into the the second volume of the \it Collected Papers \rm of P\'{o}lya (see
(P\'{o}lya 1926)). To be precise, Kac says that, in the fall of 1951 and in the spring of 1952, Yang and Lee
were developing their theory of phase transitions which has since become justly celebrated. To illustrate the
theory, they introduced the concept of a ''lattice gas'' and they were led to a remarkable conjecture which (not
quite in its most general form) can be stated as follows. Let
\begin{equation}G_N(z)=\sum_{\mu_k}\exp\big(\sum_{k,l=1}^NJ_{kl}\mu_k\mu_l\big)\exp\big(iz\sum_{k=1}^N\mu_k\big)\end{equation}
where $J_{kl}\geq 0$ and the summation is over all $2^N$ sequences $(\mu_1,...,\mu_N)$ with each $\mu_k$
assuming only values $\pm 1$. Then, $G_n(z)$ has only real roots (\it Lee-Yang theorem\rm). Textually, Kac tells
that, when he first heard of this conjecture, he considered the simplest case $J_{k,l}=\nu/2$ for all $k,l$, and
somehow Hilfssatz II of P\'{o}lya's paper came into his mind. Then, Kac shows how, by a slight modification of
Polya proof, one can prove the Lee-Yang theorem in the above special case. First of all, for all $N$, $G_N(z)$
is an entire function of order 1 which assumes real values for real $z$. Note furthermore that
$$\Big(\frac{\nu}{2}\Big)\big(\sum_{k=1}^{N+1}\mu_k\big)^2+iz\sum_{k=1}^{N+1}\mu_k=$$ $$=\Big(\frac{\nu}{2}\Big)
\big(\sum_{k=1}^N\mu_k\big)^2+(\nu\mu_{N+1}+iz)\sum_{k=1}^N\mu_k+iz\mu_{N+1}+\Big(\frac{\nu}{2}\Big)$$and
therefore
\begin{equation}e^{-\nu/2}G_{N+1}(z)=e^{iz}G_N(z-i\nu)+e^{-iz}G_N(z+i\nu).\end{equation}If $z$ is a root of
$G_{N+1}$, we have\begin{equation}|e^{2iz}|^2=\Big|\frac{G_N(z+i\nu)}{G_N(z-i\nu)}\Big|^2,\end{equation}and if
we assume that $G_N$ has only real roots, say $\alpha_1,\alpha_2,...$, then, by Hadamard factorization theorem,
we have\begin{equation}G_{N}(z)=ce^{\alpha z}\prod_{n=1}^{\infty}(1-z/\alpha_n)e^{z/\alpha_n}\end{equation}where
$c$ and $\alpha$ (as well as $\alpha_1,\alpha_2,...$) are real. Equation (44) now becomes (upon setting
$z=x+iy$)
\begin{equation}e^{-4y}=\prod_{n=1}^{\infty}\frac{(\alpha_n-x)^2+(y+\nu)^2}{(\alpha_n-x)^2+(y-\nu)^2}.\end{equation}
Since $\nu>0$, each term of the product (and hence the product itself) is greater than 1 if $y>0$ and less than
1 if $y<0$. On the other hand, $\exp(-4y)$ is less than 1 for $y>0$ and greater than 1 for $y<0$. Thus (46) can
hold only if $y=0$, i.e., all roots of $G_{N+1}$ are also real. Since for $N=2$ a direct check shows that all
roots of $G_2$ are real, the theorem for all $N$ follows by induction. Then, Kac refers that he showed this
proof for the special case to Yang and Lee. A couple of weeks later, they produced their proof of the general
theorem (in (Lee \& Yang 1952, Appendix II)). Moreover, Kac also remembers Professor Yang telling him at the
time that Hilfssatz II of P\'{o}lya, in the form discussed above, was one essential ingredient in their proof,
as also recalled above. In any way, one immediately realizes that the key tool of the above Kac's argument, is
just Hadamard factorization theorem.

Therefore, P\'{o}lya works (see (P\'{o}lya 1926a,b)) have opened new fruitful avenues in pure and applied
mathematics. Indeed, according to (Dimitrov 2013), we consider the following question: suppose that $K$ is a
positive kernel which decays sufficiently fast at $\pm\infty$, supposing it belongs in the Schwartz class, and
its Fourier transform $\displaystyle\mathcal{F}(z;K)\doteq\int_{-\infty}^{+\infty}e^{-izt}K(t)dt$ is an entire
function. More generally, we consider positive Borel measures $d\mu$ with the property that
$\displaystyle\mathcal{F}_{\mu}(z)\doteq\int_{-\infty}^{+\infty}e^{-izt}d\mu(t)$ defines an entire function. The
problem to characterize the measures $\mu$  for which $\mathcal{F}_{\mu}$ has only real zeros has been of
interest both in mathematics, because of the Riemann hypothesis, and in physics, because of the validity of the
so-called general Lee-Yang theorem for such measures. It seems that P\'{o}lya was the first to formulate the
problem explicitly in his works (P\'{o}lya 1926a,b), beginning, mutatis mutandis, with the following issue: What
properties of the function $K(u)$ are sufficient to secure that the integral $\displaystyle
2\int_{-\infty}^{+\infty}K(u)\cos zudu=\mathcal{F}(z)$ has only real zeros? The origin of this rather artificial
question is the well-known hypothesis concerning the Riemann zeta function, as the author himself recognizes in
(P\'{o}lya 1926a). If we put $K(u)=\Phi(u)$ as given by (29), then $\mathcal{F}(z)$ is nothing but that the
Riemann $\xi$ function. Since $\Phi(u)$ is an even kernel which decreases extremely fast, the above definition
of P\'{o}lya for $\mathcal{F}$, in the case when $K$ is even, is exactly the one for the Fourier transform. The
Riemann's hypothesis, as formulated by P\'{o}lya himself, states that the zeros of $\xi$ are all real. The
efforts to approach Riemann hypothesis via $\xi$ defined as a Fourier transform, have failed despite of the
remarkable efforts due to P\'{o}lya, N.G. de Bruijn (see (de Bruijn 1950)) and many other mathematicians for two
chief reasons. The first one is that the above question of P\'{o}lya still remains open, whilst the second one
is that sufficient conditions for the kernels have turned out to be extremely difficult to be verified for
$\Phi$ or simply do not hold for it. Finally, the notable work of C.M. Newman (see (Newman 1974) as well as (Kim
1999), (Ki \& Kim 2003), (Ki et al. 2009), (Korevaar 2013) and references therein for a modern sight of the
question and related arguments) based on an extension and generalization of the above pioneering work of T.D.
Lee and C.N. Yang, has proved the latter to be equivalent to the above P\'{o}lya's question. Moreover, following
(Korevaar 2013), de Bruijn and J. Korevaar were both inspired by work of P\'{o}lya on the zeros of entire
functions. de Bruijn was fascinated by P\'{o}lya's results of 1926 on the zeros of functions given by
trigonometric integrals, while Korevaar was attracted by other P\'{o}lya's papers on the approximation of entire
functions by polynomials whose zeros satisfy certain conditions. All these articles by P\'{o}lya have been
reproduced with commentary in the second volume of his \it Collected papers. \rm Moreover, de Bruijn and
Korevaar both published extensions of P\'{o}lya's work in Duke Mathematical Journal, referring to (Korevaar
2013) for a deeper historical analysis of all that and for other notable aspects of the history of entire
function theory. Likewise, some works of D.A. Cardon and collaborators (see (Cardon \& Nielsen 2003), (Cardon
2002; 2005), (Adams \& Cardon 2007) and references therein) have fruitfully combined and fitted very well
together, on the basis of certain extensions, formal comparative analogies and possible generalizations, the
1952 Lee-Yang formal approach to phase transitions with the original 1926 P\'{o}lya approach to Riemann $\xi$
function revisited from the modern setting given by entire function theory as exposed in (Levin 1980), to be
precise, reformulating the P\'{o}lya results within either the Hermite-Biehler and Laguerre-P\'{o}lya classes of
entire functions with related distributions of zeros also using some tools drawn from stochastic and
probabilistic analysis. Finally, very interesting attempts to apply Lee-Yang theorem for approaching Riemann
conjecture have been pursued in (Knauf 1999) and (Julia 1994) (see also references therein quoted). Following
(Borcea \& Br\"{a}ndén 2009b, Introduction), the Lee-Yang theorem seems to have retained an aura of mystique. In
his 1988 Gibbs lecture, Ruelle proclaimed: ''I have called this beautiful result a failure because, while it has
important applications in physics, it remains at this time isolated in mathematics''. Ruelle's statement was
apparently motivated by the fact that the Lee-Yang theorem also inspired speculations about possible statistical
mechanics models underlying the zeros of Riemann or Selberg zeta functions and the Weil conjectures, but ''the
miracle has not happened''. Nevertheless, only recently Lee-Yang theorem has received new attention from
mathematician, as witnessed, for instance, by the recent works of J. Borcea and P. Br\"{a}ndén (see (Borcea \&
Br\"{a}ndén 2008; 2009a,b), (Br\"{a}ndén 2011) and references therein) whose research program makes often
reference to Laguerre-P\'{o}lya, Hermite-Bielher and P\'{o}lya-Schur classes of complex functions. Indeed,
recently Lee-Yang like problems and techniques have appeared in various mathematical contexts such as
combinatorics, complex analysis, matrix theory and probability theory. The past decade has also been marked by
important developments on other aspects of phase transitions, conformal invariance, percolation theory. However,
as A. Hinkkanen has observed, the power in the ideas behind the Lee-Yang theorem has not yet been fully
exploited: ''It seems that the theory of polynomials, linear in each variable, that do not have zeros in a given
multidisk or a more general set, has a long way to go, and has so far unnoticed connections to various other
concepts in mathematics''. Anyway, from a general overview of almost all these works concerning Riemann $\xi$
function and related applications according to the line of thought opened by Polya's works of 1926 until up the
new directions provided by Lee-Yang theorem on the wake of P\'{o}lya's work itself, it turns out that
Weierstrass-Hadamard factorizations are the key formal tools employed in these treatments, besides to be the
pivotal source from most of entire function theory sprung out, as well as to be crucial formal techniques widely
employed in the modern treatment of the theory of polynomials and their zeros (see (Gil' 2010), (Fisk 2008) and
(Rahman \& Schmeisser 2002)).

In conclusion, we may state that two main results have been at the early origins of the Lee-Yang theorems,
especially as regard the unit circle theorem, namely a 1934 paper of Aurel Wintner, from which Lee and Yang have
drawn useful hints to determine the properties of the distribution function $g(\theta)$ and related geometrical
settings of the zeros of grand partition function for the physical systems analyzed by them, and a trick used to
prove a lemma due to P\'{o}lya, namely $(\circ)$ of his Hilfssatz II (see previous section), thanks to which Lee
and Yang proved the non-trivial basis of induction corresponding to the case $n=2$ concerning the auxiliary
polynomial $\mathfrak{B}(z_1,...,z_n)$ - see (Lee \& Yang 1952, Appendix II) - used for proving, by induction, a
more general theorem than the Theorem 3 of section IV, that is to say, the unit circle theorem. Nevertheless, as
Lee and Yang themselves point out at the beginning of Section V, Point A. of (Lee \& Yang 1952), the
distribution function $g(\theta)$ has been used only to estimate the number of zeros in the unit circle, once
this last geometrical arrangement had already been determined by means of other routes (that is to say, via unit
circle theorem), and not to properly determine this last settlement. In any event, Lee and Yang have been
pioneers in opening a possible avenue in mathematical physics, even if their appreciated work dealt only with
particular physical systems, still waiting general mathematical tools which would have generalized and extended
this Lee and Yang model to a wider class of physical systems. This truly difficult task has been undertaken by
other authors (recalled above), amongst whom are T. Asano\footnote{For instance, the interesting theory of \it
polynomial contractions \rm due to Taro Asano, might have fruitful applications in algebraic geometry, and vice
versa, i.e., tools and methods of this last subject might turn out to be useful for statistical mechanics of
phase transitions and its rigorous results, always along the line outlined by Asano. In this regard, see the
exposition given in (Ruelle 2007, Chapter 17). See also (Glimm \& Jaffe 1987, Chapter 4) for other modern
treatments and extension of Lee-Yang model, above all in relation to quantum field theory context.}, D. Ruelle,
M. Suzuki, C.M. Newman, E.H. Lieb, and A.D. Sokal, with very interesting results which, nevertheless, have not
reached the expected goal. Nevertheless, just due to the great difficult to exactly determine the grand
canonical partition function of an arbitrary thermodynamical system, often the Lee-Yang model runs well when is
applied to the state equation and its possible singularities. Following (Ruelle 1988, Section 3), Lee-Yang
theorem, conjectured on a physical basis related to ferromagnetic spin system, originally took some effort to
prove. A later idea, due to T. Asano in extending Lee-Yang model to quantum case, now permits a different but
short proof (see (Ruelle 1969, Chapter 5) as well as (Ruelle 1988, Appendix)) of this theorem. Notwithstanding
its remarkable importance, Ruelle has nevertheless said of this beautiful result to be a failure because, while
it has important applications in physics, it remains at this time isolated in mathematical physics and
mathematics. In textual words of Ruelle, one might think of a connection with zeta functions (and the Weil
conjectures), the idea of such a connection being not absurd but the miracle has not happened, so that one still
does not know what to do with the circle theorem. Ruelle says too that this connection with Riemann zeta
function and related conjecture is not fully meaningless because there exist interesting applications of certain
ideas of statistical mechanics to differentiable dynamics, made possible by the introduction of Markov
partitions which transform the problems of ergodic theory for hyperbolic diffeomorphisms or flows into problems
of statistical mechanics on the ''lattice'' $\mathbb{Z}$. Among the many applications of the method, Ruelle
mentions Ya.G. Sinai's beautiful proof that hyperbolic diffeomorphisms do not necessarily have a smooth
invariant measure. Also, since the geodesic flow on manifolds of negative curvature is hyperbolic, one has the
possibility of studying zeta functions of A. Selberg's type, and, using Markov partitions, these zeta functions
are expressed as certain sorts of partition functions, which can be studied by statistical mechanics. Thus, one
obtains for instance a kind of ''prime number theorem'' for the lengths of closed geodesies on a compact
manifold of negative curvature not necessarily constant (see (Mayer 1980, Chapter IV)).\\\\\bf 3.3. On some
other applications of the theory of entire functions, and all that. \rm In reviewing the main moments of the
history of Riemann zeta function and related still unsolved conjecture known as Riemann Hypothesis (RH), as for
instance masterfully exposed in (Bombieri 2006) as well as in the various treatises, textbooks and survey papers
on the subject (see, for instance, (Whittaker \& Watson 1927), (Chandrasekharan 1958), (Ingham 1964), (Ivi\c{c}
1985), (Titchmarsh 1986), (Patterson 1988), (Karatsuba \& Voronin 1992), (Karatsuba 1994), (Edwards 2001), (Chen
2003), (Conrey 2003), (Gonek 2004) and (Borwein et al. 2008)), one realizes that a crucial point which would
have deserved major historical attention is the one concerning Hadamard factorization theorem, which is the
central point around which has revolved our attention and that has casted a precious bridge with entire function
theory, opening a new avenue in complex analysis. This point has been sufficiently treated in the above sections
which have seen involved the figures of Riemann, Weierstrass and Hadamard, so that we herein sum up, in passing,
the main points of what has been before discussed in such a manner to be a kind of preamble of what will be said
herein. As has been seen, Hadamard formulated his celebrated 1893 theorem as a continuation and completion of a
previous 1883 theorem stated by Poincaré as regard the order of an entire function factorized according to the
Weierstrass factorization theorem of 1876, applying the results so obtained to the Riemann $\xi$ function which,
in turn, had already been factorized by Riemann himself in his 1859 seminal paper. This celebrated Hadamard
result was the pivotal point through which the entire function theory entered into the realm of Riemann zeta
function. After Hadamard, it was then Pólya, in the 1920s, to achieve some further remarkable outcomes along
this research's path emphasizing the entire function theory perspective of Riemann zeta function also making use
of the 1893 Hadamard work, until to reach the recent outcomes of which we will briefly refer in this section.
First of all, according to (Davenport
1980, Chapters 8, 11 and 12), in passing we recall that\\

\it <<In his epoch-making memoir of 1860 (his only paper on the theory of numbers), Riemann showed that the key
to the deeper investigation of the distribution of the primes lies in the study of $\zeta(s)$ as a function of
the complex variable $s$. More than 30 years were to elapse, however, before any of Riemann's conjectures were
proved, or any specific results about primes were established on the lines which he had indicated. Riemann
proved two main results: (a) The function $\zeta(s)$ can be continued analytically over the whole plane and is
then meromorphic, its only pole being a simple pole at $s=1$ with residue 1. In other words,
$\zeta(s)-(s-1)^{-1}$ is an integral function. (b) $\zeta(s)$ satisfies the functional
equation\begin{equation}\pi^{-\frac{s}{2}}\Gamma(\frac{s}{2})\zeta(s)=\pi^{-\frac{1-s}{s}}\Gamma(\frac{1-s}{2})
\zeta(1-s)\end{equation}which can be expressed by saying that the function on the left is an even function of
$s-1$. The functional equation allows the properties of $\zeta(s)$ for $\sigma<0$ to be inferred from its
properties for $\sigma>1$. In particular, the only zeros of $\zeta(s)$ for $\sigma<0$ are at the poles of
$\Gamma(s/2)$, that is, at the points $s=-2,-4,-6,...$. These are called the trivial zeros. The remainder of the
plane, where $0<\sigma<1$, is called the critical strip. \rm [...] \it Riemann further made a number of
remarkable conjectures, amongst which is the follows: the entire function $\xi(s)$ defined by (entire function
because it has no pole for $\sigma\geq 1/2$ and is an even function of $s-1/2$) has the product
representation\begin{equation}\xi(s)=e^{A+Bs}\prod_{\rho}\Big(1-\frac{s}{\rho}\Big)e^{\frac{s}{\rho}}\end{equation}where
$A$ and $B$ are constants and $\rho$ runs through the zeros of $\zeta(s)$ in the critical strip. This was proved
by Hadamard in 1893. It played an important part in the proofs of the prime number theorem by Hadamard and de la
Vallée-Poussin. \rm [...] \it The next important progress in the theory of the $\zeta(s)$ function, after
Riemann's pioneering paper, was made by Hadamard, who developed the theory of entire functions of finite order
in the early 1890s and applied it to $\zeta(s)$ via $\xi(s)$. His results were used in both the proofs of the
prime number theorem, given by himself and by de la Vallée-Poussin, though later it was found that for the
particular purpose of proving the prime number theorem, they could be dispensed>>.\\\\\rm Following (Bombieri
2006), one of the main tools to study the mathematical properties of Riemann zeta function $\zeta(s)$ (hereafter
RZF), defined by
\begin{equation}\zeta(s)\doteq\sum_{n\in\mathbb{N}}\frac{1}{n^s}, \ \ \ s\in\mathbb{C}, \ \ \ \Re(s)>1,\end{equation}
is the related \it Riemann functional equation, \rm which was established in (Riemann 1858) and is defined as
follows (see also (Titchmarsh 1986, Sections 2.4 and 2.6), (Katz \& Sarnak 1999, Section 1))
\begin{equation}\pi^{-\frac{s}{2}}\Gamma(\frac{s}{2})\zeta(s)=\pi^{-\frac{1-s}{s}}\Gamma(\frac{1-s}{2})
\zeta(1-s).\end{equation}According to (Motohashi 1997, Preface), ever since Riemann's mastery use of theta
transformation formula in one of his proofs of the functional equation for the zeta-function, number-theorists
have been fascinated by various interactions between zeta-function and automorphic forms (see also (Maurin
1997)). From a proper historical viewpoint, following (Cahen 1894, Introduction) and (Torelli 1901, Chapter
VIII, Section 62), it seems have been O.X. Schl\"{o}milch, in (Schl\"{o}milch 1858), to provide a first form of
functional equation satisfied by $\zeta(s)$. Furthermore, following (Davis 1959), around 1890s, it was
discovered that first forms of the functional equation $\zeta(s)=\zeta(1-s)\Gamma(1-s)2^s\pi^{s-1}\sin(\pi s/2)$
seem to be already present in some Eulerian studies on gamma function ever since 1740s, where there is no proof
of it but a verification of its validity only for integer values and for some rational number, like 1/2 and 3/2.
Anyway, infinite products have been at the basis of the theory of Riemann zeta function since its institution:
indeed, the primary relation upon which Riemann based his 1859 paper, is the celebrated \it Euler's product
\rm(given in the 1748 Euler's \it Introductio in Analysin Infinitorum\rm)
\begin{equation}\sum_{n\in\mathbb{N}}\frac{1}{n^s}=\prod_{p\in\mathcal{P}}\frac{1}{1-p^{-s}}\end{equation}
for each $s\in\mathbb{C},\ \Re(s)>1$, where $\mathcal{P}=\{p;p\in\mathbb{N}, p\ \mbox{prime}, p>1\}$. Following
(Ingham 1964, Introduction, 6.), as has been said above, this latter Euler's identity was first used by Euler
himself only for a fixed value, namely $s=1$ (besides for some rational number), while Tchebycheff used it with
$s$ real. Subsequently, Riemann considered the left hand side of (13) as a complex function of $s$, called \it
zeta function, \rm denoted by $\zeta(s)$ and defined for $s\in\mathbb{C},\ \Re(s)>1$; afterwards, Riemann will
give the analytical continuation of such a function to the whole complex plane through the above mentioned
functional equation, obtaining a meromorphic function with only a simple pole at $s=1$, and using it to study
number theory questions through the right hand side of (4). It has been this putting into relationship number
theory with complex analysis via (4), the first revolutionary and pioneering result\footnote{From an
epistemological standpoint, this revolutionary idea's correlation, sets up by the two sides of equation (51), is
quite similar to that provided, for example, by \it Einstein's field equations \rm (1915) of General Relativity,
$R_{\mu\nu}-(1/2)g_{\mu\nu}R=8\pi GT_{\mu\nu}$ (in the natural units), which relates geometrical properties of
space-time (on the left-hand side) with physical field properties (on the right-hand-side). Besides Riemann and
Einstein, also H. Weyl was a pioneer in putting into relation conceptual areas before considered very far
between them. This type of conceptual correlation of ideas is one of the main epistemological processes with
which often scientific creativity expresses.} achieved by Riemann in his seminal paper.

As has been said above, from the symmetric form of (12) (see (Ivi\c{c} 1986, Section 1.2), it is possible, in
turn, to define the Riemann $\xi$ function (Riemann 1858) as follows (see (Whittaker \& Watson 1927, Section
13.4))
\begin{equation}\xi(t)=\Big(1/2s(s-1)\pi^{-s/2}\Gamma(s/2)\zeta(s)\Big)_{s=\frac{1}{2}+it}\end{equation}
which is an even entire function of order one with simple poles in $s=0,1$, and whose zeros
verify\footnote{Indeed, let $t=a+ib$ and $s=c+id$, so that, from $s=\frac{1}{2}+it$, it follows that
$is=\frac{1}{2}i-t$, whence $t=\frac{1}{2}i-is=i(\frac{1}{2}-s)$, that is
$t=a+ib=i(\frac{1}{2}-s)=i(\frac{1}{2}-(c+id))=i(\frac{1}{2}-c-id)=i((\frac{1}{2}-c)-id)=d+i(\frac{1}{2}-c)$
whence $a=d$ and $b=\frac{1}{2}-c$, that is to say $\Re(t)=\Im(s), \Im(t)=\frac{1}{2}-\Re(s)$, whence
$|\Im(t)|=|\frac{1}{2}-\Re(s)|\leq\frac{1}{2}$ since $c=\Re(s)\in [0,1]$ in the critical strip. In fact, if
$\frac{1}{2}\leq c\leq 1$, then $|\frac{1}{2}-c|=c-\frac{1}{2}\leq 1-\frac{1}{2}=\frac{1}{2}$ because $c\leq 1$,
whereas, if $0\leq c\leq\frac{1}{2}$, then $|\frac{1}{2}-c|=\frac{1}{2}-c\leq\frac{1}{2}$ since $0\leq c$, so
that, anyway, we have $|\frac{1}{2}-c|\leq\frac{1}{2}$.} $|\Im(t)|\leq 1/2$ (Riemann 1858). This last estimate
was then improved in $|\Im(t)|<1/2$ both by Hadamard (1896a) and by de la Vallée-Poussin (1896), but
independently of each other. The RH asserts that $\Im(t)=0$, that is to say $t\in\mathbb{R}$. Following
(Ivi\c{c} 1989) and (Gonek 2004), it is plausible to conjecture that all the zeros of RZF, along the critical
line, are simple, this assertion being supported by all the existing numerical evidences (see for example (van
de Lune et al. 1986)). Subsequently, Hadamard (1893) gave a fundamental Weierstrass infinite product expansion
of Riemann zeta function, of the following type (see, for example, (Karatsuba 1994, Chapter 1, Section 3.2) and
(Bateman \& Diamond 2004, Chapter 8, Section 8.3))
\begin{equation}\xi(t)=ae^{bt}\prod_{\rho\in Z(\zeta)}\Big(1-\frac{t}{\rho}\Big)e^{\frac{t}{\rho}}\end{equation}
where $a,b$ are constants and $Z(\zeta)$ is the set of all the complex non-trivial zeros of the Riemann zeta
function $\zeta(s)$, so that $Z(\zeta)\subseteq {t; t\in\mathbb{C}, 0<\Re(t)<1}$, with $card Z(\zeta)=\infty$
(G.H. Hardy). This Hadamard paper was considered by H. Von Mangoldt (1854-1925) as \it ''the first real progress
in the field in 34 years'' since the only number theory Riemann 1859 paper \rm (see (Von Mangoldt 1896) and
(Edwards 2001, Section 2.1)), having provided the first basic link between Riemann zeta function theory and
entire function theory. Nevertheless, in relation to the Riemann zeta function, Hadamard work didn't have that
right historical attention which it would have deserved, since a very few recalls to it have been paid in the
related literature. From above Hadamard product formula, it follows an infinite product expansion of Riemann
zeta function of the following type (see, for example, (Landau 1909, Band I, Erstes Kapitel, § 5.III-IV), (Ayoub
1963, Chapter II, Section 4), (Erdélyi 1981, Section 17.7), (Titchmarsh 1986, Section 2.12), (Narkiewicz 2000,
Chapter 5, Section 5.1, Number 2) and (Voros 2010, Chapter 4, Section 4.3))
\begin{equation}\zeta(s)=\frac{e^{(\ln 2\pi-1-\frac{\gamma}{2})s}}{2(s-1)\Gamma(1+\frac{s}{2})} \prod_{\rho\in
Z(\zeta)}(1-\frac{s}{\rho})e^{\frac{s}{\rho}}=\Theta(s)\prod_{\rho\in
Z(\zeta)}(1-\frac{s}{\rho})e^{\frac{s}{\rho}}\end{equation} where $\gamma$ is the Euler-Maclaurin constant. The
function $\Theta(s)$ is non-zero into the critical strip $0<\Re(s)<1$, so that it is quickly realized that any
question about zeros of $\zeta(s)$ might be addressed to the above infinite product factor, which is an entire
function, and, likewise, as regard the above Hadamard product formula for $\xi$. Therefore, it seems quite
obvious to account for the possible relationships existing with entire function theory, following this fruitful
perspective opened by Hadamard. Out of the best treatises on entire function theory, there are those of Boris
Ya. Levin (see (Levin 1980; 1996)). In particular, the treatise (Levin 1980) is hitherto the most complete one
on the distributions of zeros of entire functions, which deserves a considerable attention. As regard, then, the
above Hadamard product formula, in reviewing the main textbooks on Riemann zeta function, amongst which
(Chandrasekharan 1958, Lectures 4, 5 and 6), (Titchmarsh 1986, Chapter II), (Ivi\c{c} 1986, Section 1.3),
(Patterson 1988, Chapter 3), (Karatsuba \& Voronin 1992, Sections 5 and 6), (Edwards 2001, Chapter 2) and (Chen
2003, Chapter 6), it turns out that such a fundamental factorization, like the one provided by Hadamard, has
been used to study some properties of this special function, for instance in relation to its Euler infinite
product expansion or in relation to its growth order questions. But, in such treatises, it isn't exposed those
results properly related to the possible links between Riemann zeta function theory and entire function theory,
from Hadamard and P\'{o}lya works onward. Only recently, there have been various studies which have dealt with
entire function theory aspects of Riemann $\xi$ function d'aprés Pólya and Hadamrd work, and, in this regard, we
begin mentioning some valuable considerations very kindly communicated to me by Professor Jeffrey C. Lagarias
(private communication). He first says that, although there are strong circumstantial evidences for RH, no one
knows how to prove it and no promising mechanism for a proof is currently known. In particular, there are many
approaches to it, and it is not clear whether the complex variables approaches based on Laguerre-P\'{o}lya
($LP$) and Hermite-Biehler ($HB$) connections with Riemann zeta function theory via Riemann $\xi$ function (see
(Levin 1980, Chapters VII and VIII)) are going to get anywhere. He refers that, maybe, P\'{o}lya might have been
the first to have established the $LP$ connection on the basis of the previous work made by J.L.W. Jensen, and
recalled in the previous sections. The truth of Riemann hypothesis requires that $\xi(s)$ falls into the $HB$
class under suitable change of variable (see (Lagarias 2005)), even if Lagarias stresses the fact that this was
already known for a long time, by which reason it requires further historical examination. Also Louis De Branges
has made some interesting works in this direction, no matter by his attempts to prove Riemann hypothesis which
yet deserve as well a certain attention because they follow a historical method, as kindly De Branges himself
said to me (private communication). Nevertheless, Lagarias refers too that who has been the first to state this
connection to $HB$ class is historically yet not wholly clear. Further studies even along this direction have
been then made, amongst others, by G. Csordas, R.S. Varga, M.L. Patrick, W. Smith, A.M. Odlyzko, J.C. Lagarias,
D. Montague, D.A. Hejhal, D.A. Cardon, S.R. Adams and some other. Finally, Lagarias concludes stating that the
big problem is to find a mechanism that would explain why the Riemann $\xi$ function would fall into this $HB$
class of functions.

Herein, we briefly remember the main lines of some of these works. For instance, the work (Cardon \& de Gaston
2005) starts considering the Laguerre-P\'{o}lya class which, as is known, consists of the entire functions
having only real zeros with Weierstrass products of the form
\begin{equation}cz^me^{\alpha z-\beta z^2}\prod_k\Big(1-\frac{z}{\alpha_k}\Big)e^{\frac{z}{\alpha_k}}\end{equation}
where $c,\alpha,\beta,\alpha_k$ are real, $\beta\geq 0,\alpha_k\neq 0$, $m$ is a non-negative integer, and
$\sum_k(1/\alpha_k^2)<\infty$. An entire function belongs to $LP$ if and only if it is the uniform limit on
compact sets of a sequence of real polynomials having only real zeros (see (Levin 1980, Chapter VIII, Theorem
3)). One of the reasons for studying the Laguerre-P\'{o}lya class is its relationship to the Riemann zeta
function. Let $\xi(s)=(1/2)s(s-1)\pi^{-s/2}\Gamma(s/2)\zeta(s)$, where $\zeta(s)$ is the Riemann zeta function.
Then $\xi(1/2 + iz)$ is an even entire function of genus 1 that is real for real $z$. The Riemann hypothesis,
which predicts that the zeros of $\xi(s)$ have real part 1/2, can be stated as $\xi(1/2+iz)\in LP$. Furthermore,
evidence suggests that most, if not all, of the zeros of $\xi(s)$ are simple. Hence, functions in $LP$ with
simple zeros are especially interesting also in issues concerning Riemann hypothesis. Then, following (Lagarias
\& Montague 2011, Section 1.1) and references therein, there have been many studies of properties of the Riemann
$\xi$-function. This function motivated the study of functions in the $LP$ class (see (P\'{o}lya 1927) and
(Levin 1980, Chapter VIII)), to which the function $\xi(z)$ would belong if the Riemann hypothesis were true. It
motivated the study of properties of entire functions represented by Fourier integrals that are real and bounded
on the real axis (see (P\'{o}lya 1926a,b; 1927a) as well as (Titchmarsh 1980, Chapter X)) and related Fourier
transforms (see (Wintner 1936)). It led to the study of the effect of various operations on entire functions,
including differential operators and convolution integral operators, preserving the property of having zeros on
a line, as well as various necessary conditions for the $\xi$-function to have real zeros, have been verified,
amongst others by D. Craven, G. Csordas, W. Smith, P.P. Nielsen, D.A. Cardon, S.A. de Gaston, T.S.Norfolk, and
R.S. Varga. In (Newman 1976), the author introduced a one-parameter family of Fourier cosine integrals, given
for real $\lambda$ by $\displaystyle\xi_{\lambda}(z)\doteq 2\int_0^{\infty}e^{\lambda u^2}\Phi(u)\cos zu du$
with $\Phi(u)$ given by (29). Here $\xi_0(z) = \xi(z)$ as given by (28), so this family of functions
$\xi_{\lambda}$ can be viewed as deformations of the $\xi$-function. It follows from a 1950 result of N.G. de
Bruijn that the entire function $\xi_{\lambda}(z)$ has only real zeros for $\lambda\geq 1/8$. In (Newman 1976),
the author proved that there exists a real number $\lambda_0$ such that $\xi_{\lambda}(t)$ has all real zeros
for $\lambda\geq\lambda_0$, and has some non-real zeros for each $\lambda<\lambda_0$. The Riemann hypothesis
holds if and only if $\lambda_0\leq 0$, and C.M. Newman conjectured that the converse inequality $\lambda\geq 0$
holds. Newman also stated that his conjecture represents a quantitative version of the assertion that the
Riemann hypothesis, if true, is just barely true. The rescaled value $\Lambda=4\lambda_0$ was later named by
Csordas, Norfolk and Varga, the \it de Bruijn-Newman constant, \rm and they proved that $\Lambda\geq -50$.
Successive authors obtained better bounds obtaining by finding two zeros of the Riemann zeta function that were
unusually close together. Successive improvements of examples on close zeta zeros led to the lower bound
$\Lambda>-2.7\times 10^{-9}$, obtained by A.M. Odlyzko. Recently H. Ki, Y-O. Kim and J. Lee established that
$\Lambda<1/2$. The conjecture that $\Lambda = 0$ is now termed the \it de Bruijn-Newman conjecture. \rm Odlyzko
observed that the existence of very close spacings of zeta zeros, would imply the truth of the de Bruijn-Newman
conjecture. In another direction, one may consider the effects of differentiation on the location and spacing of
zeros of an entire function $F(z)$. In 1943 P\'{o}lya (see (P\'{o}lya 1943)) conjectured that an entire function
F(z) of order less than 2 that has only a finite number of zeros off the real axis, has the property that there
exists a finite $m_0\geq 0$ such that all successive derivatives $F^{(m)}(z)$ for $m_0\geq 0$ have only real
zeros. This was proved by Craven, Csordas and Smith in 1987, with a new proof given by Ki and Kim in 2000. In
2005, D.W. Farmer and R.C. Rhoades have shown (under certain hypotheses) that differentiation of an entire
function with only real zeros will yield a function having real zeros whose zero distribution on the real line
is ''smoothed''. Their results apply to the Riemann $\xi$-function, and imply that if the Riemann hypothesis
holds, then the same will be true for all derivatives $\xi^{(m)}(s) = d^m\xi(s)/ds^m, m\geq 1$. Various general
results are given (Cardon \& de Gaston 2004), while for more extensive informations about other researches on
Riemann $\xi$ function, we refer to (Lagarias \& Montague 2011) and references therein. In any way, from what
has just been said above, it turns out quite clear what fundamental role has played many part of P\'{o}lya work
on Riemann $\xi$ function in the development of entire function theory. We refer to (Korevaar 2013) and
reference therein for more historical informations in this regard.

Finally, what follows is the content of a private communication with which Enrico Bombieri who has very kindly
replied to my request to have some his comments and hints about some possible applications of entire function
theory on Riemann zeta function theory. He kindly refers that, very likely, the 1893 Hadamard work was mainly
motivated by the possible applications to Riemann zeta function, as we have above widely discussed. On the other
hand, the general theory of complex and special functions had a great growth impulse just after the middle of
18th century above all thanks to the pioneering works of Weierstrass, H.A. Schwarz, Nevanlinna brothers and
others. But Hadamard was the first to found a general theory which will receive its highest appreciation with
the next works of Nevanlinna brothers. Afterwards, the attempts to isolate entire function classes comprising
Riemann zeta function (properly modified to avoid its single poles in $s=0,1$) have been quite numerous (amongst
which those by De Branges), with interesting results but unfruitful as regard the possible applications to
Riemann hypothesis. Nevertheless, nowadays only a few mathematicians carry on along this path, amongst whom G.
Csordas and co-workers with interesting works, besides those other scholars mentioned above. For instance, along
the line of research opened by P\'{o}lya, in a recent conference, in which Bombieri was attended, Csordas
proposed to consider the class of Mellin transformation $Mf(x)$ of fast decreasing functions $f$ as
$x\rightarrow\infty$ such that each $(xd/dx)^n f(x)$ has exactly $n$ zeros for each $n\in\mathbb{N}_0$. Now, it
would seem that the Riemann zeta function may be related with this class of functions, but, at this time, there
is no exact proof of this idea to which Bombieri himself was pursuing through other ways. Many mathematicians
have besides worked on Lee-Yang theorem area hoping to meet along their routes a possible insight for the
Riemann zeta function and related conjecture, but after an initial enthusiasm, every further attempt didn't have
any sequel. As regard, then, the general context of complex function theory, this reached its apex around 1960s,
above all with the works achieved by the English school of W. Hayman on meromorphic functions and by the Russian
school of B.Ya. Levin. However, it is noteworthy to mention the recent statistical mechanics approach which
seems promising as regard zero distribution of Riemann zeta function whose behavior is however quite anomalous
with respect other complex functions, and seems to follow a Gaussian Circular Unitary Ensemble (GCUE) law (see
(Katz \& Sarnak 1999) in relation to random matrix theory). Bombieri, then, finishes mentioning some very
interesting results achieved, amongst others, by A. Beurling, B. Nyman and L. B\'{a}ez-Duarte, hence concluding
saying that, today, there still exists a little but serious group of researchers working on the relationships
between Riemann $\xi$ function, its Fourier transform and entire function theory, d'aprés Pólya work. As regard
what has just been said about GCUE law, following (Lagarias 2005), there is a great deal of evidence suggesting
that the normalized spacings between the nontrivial zeros of the Riemann zeta function have a ''random''
character described by the eigenvalue statistics of a random Hermitian matrix whose size $N\rightarrow\infty$.
The resulting statistics are the large $N$ limit of normalized eigenvalue spacings for random Hermitian matrices
drawn from the GUE distribution (''Gaussian unitary ensemble''). This limiting distribution is identical to the
large $N$ limit of normalized eigenvalue spacings for random unitary matrices drawn from the GUE distribution
(''circular unitary ensemble''), i.e., eigenvalues of matrices drawn from $U(N)$ using Haar measure, and taking
into account that the GUE and CUE spacing distributions are not the same for finite $N$. More precisely, one
compares the normalized spacings of $k$ consecutive zeros with the limiting joint probability distribution of
the normalized spacings of $k$ adjacent eigenvalues of random hermitian $N\times N$ matrices, as
$N\rightarrow\infty$. The relation of zeta zeros with random matrix theory was first suggested by the work of H.
Montgomery in 1973 which concerned the pair correlation of zeros of the zeta function. Montgomery's results
showed (conditional on the Riemnan hypothesis) that there must be some randomness in the spacings of zeros, and
were consistent with the prediction of the GUE distribution. Hence, A.M. Odlyzko, in 1987, made extensive
numerical computations with zeta zeros, now up to height $T=1022$, which show an extremely impressive fit of
zeta zero spacings with predictions of the GUE distribution. The GUE distribution of zero spacings is now
thought to hold for all automorphic $L$-functions, specifically for principal $L$-functions attached to $GL(n)$,
(see (Katz \& Sarnak 1999) and (Gonek 2004)). Further evidence for this was given by Z. Rudnick and P. Sarnak in
1996, conditionally on a suitable generalized Riemann hypothesis. They showed that the evaluation of consecutive
zero gaps against certain test functions (of limited compact support) agrees with the GUE predictions. There is
also supporting numerical evidence for certain principal $L$-functions attached to $GL(2)$. As regard, however,
Riemann zeta function theory and random matrix theory, see (Borwein et al. 2008, Chapter 4, Section 4.3).
Finally, following (Conrey \& Li 2000), the theory of Hilbert spaces of entire functions was developed by Louis
de Branges in the late 1950s and early 1960s (see (de Branges 1968)). It is a generalization of the part of
Fourier analysis involving Fourier transforms and the Plancherel formula. To be precise, the origins of Hilbert
spaces of entire functions are found in a theorem of R.E. Paley and N. Wiener that characterizes finite Fourier
transforms as entire functions of exponential type which are square integrable on the real axis. The known
examples of Hilbert spaces of entire functions belong to the theory of special functions, a subject which is
very old in relation to most of modern analysis. The foundations of the theory were laid by Euler in the century
following the discovery of the calculus whose historical approach to the subject is already so well represented
by the treatise (Whittaker \& Watson 1927). In 1986 (see (de Branges 1986)), de Branges proposed an approach to
the generalized Riemann hypothesis, that is, the hypothesis that not only the Riemann zeta function $\zeta(s)$
but also all the Dirichlet $L$-functions $L(s,\chi)$ with $\chi$ primitive, have their nontrivial zeros lying on
the critical line $\Re s = 1/2$. In his 1986 paper (see (de Branges 1986)), de Branges said that his approach to
the generalized Riemann hypothesis using Hilbert spaces of entire functions is related to the so-called
Lax-Phillips theory of scattering, exposed in (Lax \& Phillips 1976), where interesting applications to Riemann
zeta function, following the so-called \it Hilbert-P\'{o}lya approach, \rm are exposed as well, but explaining
too the difficulties of approaching the Riemann hypothesis by using the scattering theory (see also (Lax \&
Phillips 1989) and (Lapidus 2008)). However, also on the basis of what has been said above, de Branges' approach
to Riemann hypothesis formulated for Hilbert spaces of entire functions has its early historical origins above
all in a theorem on Fourier analysis due either to A. Beurling and P. Malliavin of the late 1950s (see (Beurling
\& Malliavin 1962)), and later improved by N. Levinson, as well as in other results achieved by M. Rosenblum, J.
Von Neumann and N. Wiener. Finally, we refer to (Borwein et al. 2008) for an updated and complete survey of the
most valuable approaches and attempts to solve RH, and all that, while we refer to (Lapidus 2008) and references
therein, for a comprehensive, detailed and adjourned review of almost all attempts to approach RH through
mathematical physics methods.\\\\\bf Acknowledgements. \rm This my work has benefited from many information and
remarks due to private communications kindly had with Professors Enrico Bombieri, Umberto Bottazzini, Mauro
Nacinovich, Jeff Lagarias, David Cardon, Louis de Branges, Chen Ning Yang, Barry McCoy, Paul Garrett, Iossif
Vladimirovich Ostrovskii, Sofiya Ostrovska, and Pierre-Fran\c{c}ois Rodriguez. My warm thanks to them for this.

\newpage\section*{Bibliography}\addcontentsline{toc}{section}{Bibliography}
\begin{description}\item Adams, S.R. and Cardon, D.A. (2007), Sums of entire functions having only real zeros,
\it Proceedings of the American Mathematical Society, \rm 135 (12): 3857-3866.\item Ahlfors, L.V. (1979), \it
Complex Analysis. An Introduction to the Theory of Analytic Function of One Complex Variable, \rm Third Edition,
New York, NY: McGraw-Hill, Inc.\item Amerio, L. (1982-2000), \it Analisi matematica con elementi di analisi
funzionale, \rm varie edizioni, volumi 1, 2, 3-Parte I, II, Torino, IT: UTET.\item Ananda-Rau, K. (1924), The
infinite product for $(s-1)\zeta(s)$, \it Mathematische Zeitschrift, \rm 20: 156-164.\item Ayoub, R.G. (1963),
\it An introduction to the analytic theory of numbers, \rm Providence, RI: American Mathematical Society. \item
Backlund, R. (1914), Sur les zéros de la fonction $\zeta(s)$ de Riemann, \it Les Comptes Rendus de l'Acadèmie
des Sciences de Paris, \rm 158: 1979-1981.\item Backlund, R. (1918), \"{U}ber die Nullstellen der Riemannschen
Zetafunktion, \it Acta Mathematica, \rm 41: 345-375.\item Bagnera, G. (1927), \it Lezioni sopra la Teoria delle
Funzioni Analitiche, \rm lezioni litografate raccolte dal Dott. Giovanni Ricci, Roma, IT: Stabilimento
Tipo-litografico Attilio Sampaolesi.\item Balazard, M. (2010), Un siècle et demi de recherches sur l'hypothèse
de Riemann, \it Gazette des Mathématiciens (SMF), \rm 126: 7-24.\item Baracca, A. (1980), \it Manuale critico di
meccanica statistica, \rm Catania, IT: CULC - Cooperativa Universitaria Libraria Catanese.\item Bateman, P.T.
and Diamond, H.G. (1996), A hundred years of prime numbers, \it The American Mathematical Monthly, \rm 103:
729-741.\item Bateman, P.T. and Diamond, H.G. (2004), \it Analytic Number Theory. An Introductory Course, \rm
Singapore, SG: World Scientific Publishing Company, Ltd.\item Bell, E.T. (1937), \it Men of Mathematics, \rm New
York, NY: Simon \& Schuster, Inc. (Italian Translation: (1950), \it I grandi matematici, \rm Firenze, IT:
Sansoni Editore).\item Bellacchi, G. (1894), \it Introduzione storica alla teoria delle funzioni ellittiche, \rm
Firenze, IT: Tipografia di G. Barbèra.\item Behnke, H. and Stein, K. (1948), Entwicklungen analytischer
Functionen auf Riemannschen Fl\"{a}chen, \it Mathematische Annalen, \rm 120: 430-461.\item Bergweiler, W.,
Eremenko, A. and Langley, J.K. (2003), Real entire functions with real zeros and a conjecture of Wiman, \it
Geometric and Functional Analysis, \rm 13: 975-991.\item Bergweiler, W. and Eremenko, A. (2006), Proof of a
conjecture of P\'{o}lya on the zeros of successive derivatives of real entire functions, \it Acta Mathematica,
\rm 197 (2): 145-166.\item Bernardini, C., Ragnisco, O. and Santini, P.M. (1993), \it Metodi Matematici della
Fisica, \rm Roma, IT: La Nuova Italia Scientifica.\item Berzolari, L., Vivanti, G. e Gigli, D. (a cura di)
(1930-1951), \it Enciclopedia delle matematiche elementari e complementi, con estensione alle principali teorie
analitiche, geometriche e fisiche, loro applicazioni e notizie storico-bibliografiche, \rm Volume I, Parti 1$^a$
(1930) e 2$^a$ (1932), Volume II, Parti 1$^a$ (1937) e 2$^a$ (1938), Volume III, parti 1$^a$ (1947), 2$^a$
(1950) e 3$^a$ (1951), con successive ristampe anastatiche, Milano, IT: Editore Ulrico Hoepli.\item Betti, E.
(1903-1913), \it Opere Matematiche, \rm pubblicate per cura della R. Accademia de' Lincei, in due tomi,
Milano-Roma, IT: Ulrico Hoepli Editore-Librajo della Real Casa e della R. Accademia dei Lincei.\item Beurling,
A. and P. Malliavin, P. (1962), On Fourier transforms of measures with compact support, \it Acta Mathematica,
\rm 107: 291-302.\item Biehler, C. (1879), Sur une classe d'équations algébriques dont toutes les racines sont
réeles, \it Journal für die reine und angewandte Mathematik (Crelle Journal), \rm 87: 350-352.\item Binanti, L.
(Ed.) (2001), \it Pedagogia, epistemologia e didattica dell'errore, \rm Soveria Mannelli, Catanzaro, IT:
Rubbettino Editore. \item Blaschke, W. (1915), Eine Erweiterung des Satzes von Vitali über Folgen analytischer
Funktionen, \it Berichte über die Verhandlungen der Königlich-Sächsischen Gesellschaft der Wissenschaften zu
Leipzig, Mathematisch-Physikalische Klasse, \rm 67: 194-200 (see also \it Ges. Werke, \rm 6: 187-193).\item
Blumenthal O. (1910), \it Principes de la Théorie des Fonctions Entières d'Ordre Infini, \rm Paris, FR:
Gauthier-Villars, Imprimeur-Libraires de l'Observatorie de Paris et du Bureau des Longitudes.\item Boas, R.P.
(1954), \it Entire Functions, \rm New York, NY: Academic Press, Inc. \item Bombieri, E. (2006), The Riemann
Hypothesis, in: Carlson, J., Jaffe, A. and Wiles, A. (Eds), \it The Millennium Prize Problems, \rm Published by
The Clay Mathematics Institute, Cambridge, MA, jointly with The American Mathematical Society, Providence, RI,
2006, pp. 105-124.\item Borcea, J. and Brändén, P. (2008), Lee-Yang Problems and The Geometry of Multivariate
Polynomials, \it Letters in Mathematical Physics, \rm 86: 53-61 (arXiv: math.CV/0810.1007). \item Borcea, J. and
Brändén, P. (2009a), The Lee-Yang and Pólya-Schur Programs. I. Linear operators preserving stability, \it
Inventiones Mathematicæ, \rm 177(3): 541-569 (arXiv: math.CV/0809.0401). \item Borcea, J. and Brändén, P.
(2009b), The Lee-Yang and Pólya-Schur Programs. II. Theory of stable polynomial and applications, \it
Communications on Pure and Applied Mathematics, \rm 62(12): 1595-1631 (arXiv: math.CV/0809.3087).\item Borel,
\'{E}. (1897), Sur les zéros des fonctions entières, \it Acta Mathematica, \rm 20 (1): 357-396.\item Borel,
\'{E}. (1900), \it Le\c{c}ons sur les Fonctions Entières, \rm Paris, FR: Gauthier-Villars, Imprimeur-Libraires
de l'Observatorie de Paris et du Bureau des Longitudes.\item Borel, \'{E}. (1920), \it Le\c{c}ons sur les
Fonctions Entières, \rm Deuxièmé Édition revue at augmentée d'une Note de M. G. Valiron, Paris, FR:
Gauthier-Villars, Imprimeur-Libraires de l'Observatorie de Paris et du Bureau des Longitudes, de l'\'{E}cole
Polytechnique.\item Borwein, P., Choi, S., Rooney, B. and Weirathmueller, A. (2008), \it The Riemann Hypothesis.
A Resource for the Afficionado and Virtuoso Alike, \rm CMS Books in Mathematics, New York, NY: Springer-Verlag,
Inc.\item Bottazzini, U. (1986), \it The Higher Calculus. A History of Real and Complex Analysis From Euler to
Weierstrass, \rm New York, NY: Springer-Verlag, Inc. (Italian Edition: (1981), \it Il calcolo sublime. Storia
dell'analisi matematica da Euler a Weierstrass, \rm Torino, IT: Paolo Boringhieri Editore).\item Bottazzini, U.
(1994), \it Va' pensiero. Immagini della matematica nell'Italia dell'Ottocento, \rm Bologna, IT: Società
editrice il Mulino.\item Bottazzini, U. (2003), Complex Function Theory, 1780-1900, in: Jahnke, H.N. (Ed.)
(2003), \it A History of Analysis, \rm Providence, RI and London, UK: American Mathematical Society and London
Mathematical Society, pp. 213-260.\item Bottazzini, U. and Gray, J. (2013), \it Hidden Harmony - Geometric
Fantasies. The Rise of Complex Function Theory, \rm New York, NY: Springer Science + Business Media, LLC.\item
Bourbaki, N. (1963), \it Elementi di storia della matematica, \rm Milano, IT: Giangiacomo Feltrinelli
Editore.\item Bourguet, P. (1881), Développement en séries des intégrales eulériennes, \it Annales Scientifiques
de l'\'{E}cole Normale Supérieure, 2$^{e}$ séries, \rm 10: 175-232.\item Bradley, R.E. and Sandifer, C.E.
(2009), \it Cauchy's Cours d'analyse. An Annotated Translation, \rm New York, NY: Springer Science + Business
Media, LLC.\item Brändén, P. (2011), The Lee-Yang and Pólya-Schur Programs. III. Zero-preservers on
Bargmann-Fock spaces, arXiv: math.CV/1107.1809.\item Briot, C. and Bouquet, J. (1859), \it Théorie des functions
doublement périodiques et, en particulier, des fonctions elliptiques, \rm Paris, FR: Mallet-Bachelier,
Imprimeur-Libraire.\item Burckel, R.B. (1979), \it An Introduction to Classical Complex Analysis, \rm Volume I,
New York, NY: Academic Press, Inc.\item Burkhardt, H., Wirtinger, M., Fricke, R. and Hilb, E. (Eds.)
(1899-1927), \it Encyklop\"{a}die der Mathematischen Wissenschaften mit Einschlu\ss\ ihrer Anwendungen, \rm
Zweiter Band in drei Teilen: Analysis, Leipzig, DE: Verlag und Druck von B.G. Teubner.\item Cahen, E. (1894),
\it Sur la fonction $\zeta(s)$ de Riemann et sur des fonctions analogues, \rm Thèses présentèe a la Faculté des
Sciences de Paris pour obtenir le grade de docteur en sciences mathématiques, Paris, FR: Gauthier-Villars,
Imprimeur-Libraires de Bureau des Longitudes de l'\'{E}cole Polytechnique.\item Cardon, D.A. and Nielsen, P.P.
(2000), Convolution Operators and Entire Functions with Simple Zeros, in: Bennett, M.A., Berndt, B.C., Boston,
N., Diamond, H., Hildebrand, A.J. and Phillip, W. (2002), \it Number Theory for the Millennium. I, \rm
Proceedings of The Millennial Conference on Number Theory, held on May 21-26, 2000, at the campus of the
University of Illinois at Urbana-Champaign, Natick, MA: A.K. Peters, Ltd./CRC Press, pp. 183-196.\item Cardon,
D.A. (2002), Convolution Operators and Zeros of Entire Functions, \it Proceedings of the American Mathematical
Society, \rm 130 (6): 1725-1734.\item Cardon, D.A. (2005), Fourier transforms having only real zeros, \it
Proceedings of the American Mathematical Society, \rm 133 (5): 1349-1356.\item Cardon, D.A. and de Gaston, S.A.
(2005), Differential operators and entire functions with simple real zeros, \it Journal of Mathematical Analysis
and Applications, \rm 301: 386-393.\item Carleson, L. and Gamelin, T.W. (1993), \it Complex Dynamics, \rm New
York, NY: Springer-Verlag, Inc.\item Cartier, P. (1993), Des nombres premiers \`{a} la géométrie algébrique (une
brève histoire de la fonction zeta), \it Cahiers du séminaire d'histoire des mathématiques, 2e série, \rm 3:
51-77.\item Casorati, F. (1880-82), Aggiunte a recenti lavori dei sig. Karl Weierstrass e G\"{o}sta
Mittag-Leffler sulle funzioni di una variabile complessa, \it Annali di Matematica Pura ed Applicata, \rm 10
(1): 261-278.\item Cauchy, A.L. (1827), \it Exercices de Mathématiques, \rm Volume 2, Paris, FR: Chez de Bure
Frères, Libraires du Roi et de la Bibliothèque du Roi. \item Cauchy, A.L. (1829), \it Exercices de
Mathématiques, \rm Volume 4, Paris, FR: Chez de Bure Frères, Libraires du Roi et de la Bibliothèque du Roi.\item
Cazzaniga, P. (1880-82), Espressione di funzioni intere che in posti dati arbitrariamente prendono valori
prestabiliti, \it Annali di Matematica Pura ed Applicata, \rm 10(1): 279-290.\item Chandrasekharan, K. (1958),
\it Lectures on the Riemann Zeta-Function, \rm Tata Institute of Fundamental Research, Bombay, India.\item Chen,
W.W.L. (2003), \it Distribution of Prime Numbers, \rm Lecture Notes of the course 'Elementary and Analytic
Number Theory' held at the Imperial College of London University between 1981 and 1990, Sidney, AU: Macquarie
University Publisher.\item Cheng, H. and Wu, T.T. (1967), Theory of Toeplitz Determinants and the Spin
Correlations of the Two-Dimensional Ising Model. III, \it Physical Review, \rm 164 (2): 719-735.\item Cipolla,
M. (1932), \rm Gabriele Torelli, \it Giornale di matematiche di Battaglini, \rm 70: 62-78.\item Connes, A.
(2000), Noncommutative geometry and the Riemann zeta function, in: Arnold, V., Atiyah, M., Lax, P. and Mazur, B.
(Eds.) (2000), \it Mathematics: Frontiers and Perspectives, \rm Providence, RI: American Mathematical Society.
\item Conrey, B.J. (2003), The Riemann Hypothesis, \it Notices of the American Mathematical Society, \rm 50(3):
341-353.\item Conrey, B.J. and Li, X-J. (2000), A Note on some Positivity Conditions Related to Zeta and
$L$-functions, \it International Mathematical Research Notices, \rm 18: 929-940.\item Cousin, P. (1895), Sur les
fonctions de $n$ variables complexes, \it Acta Mathematica, \rm 19: 1-62.\item Davenport, H. (1980), \it
Multiplicative Number Theory, \rm 2nd Edition revised by Hugh L. Montgomery, New York, NY: Springer-Verlag,
Inc.\item Davis, P.J. (1959), Leonhard Euler's Integral: A Historical Profile of the Gamma Function. In
Memoriam: Milton Abramowitz, \it The American Mathematical Monthly, \rm 66 (10): 849-869.\item de Branges, L.
(1968), \it Hilbert Spaces of Entire Functions, \rm Englewood Cliffs, NJ: Prentice-Hall, Inc.\item de Branges,
L. (1986), The Riemann hypothesis for Hilbert spaces of entire functions, \it Bulletin of the American
Mathematical Society, \rm 15: 1-17.\item de Bruijn, N.G. (1950), The roots of trigonometric integrals, \it Duke
Mathematical Journal, \rm 17: 197-226.\item Dedekind, R. (1876), Bernhard Riemann's Lebenslauf, in: \it 1876
Bernhard Riemann Gesammelte Mathematische Werke, Wissenschaftlicher Nachlass und Nachtr\"{a}ge - Collected
Papers, \rm Nach der Ausgabe von Heinrich Weber und Richard Dedekind neu herausgegeben von Raghavan Narasimhan,
Berlin and Heidelberg, DE: Springer-Verlag.\item de la Vallée-Poussin, C.J. (1896), Recherches analytiques sur
la théorie des nombres premiers, \it Annales de la Société Scientifique de Bruxelles, \rm 20, Part II: 183-256,
281-197.\item de la Vallée-Poussin, C.J. (1899-1900), Sur la fonction $\zeta(s)$ de Riemann et le nombre des
nombres premiers inferieurs a une limite donnee, \it Memoires couronnes de l'Acadèmie Royale des Sciences, des
Lettres et des Beaux-arts de Belgique, \rm 59: 1\item Della Sala, G., Saracco, A., Simioniuc, A. and Tomassini,
G. (2006), \it Lectures on complex analysis and analytic geometry\rm, Lecture Notes. Scuola Normale Superiore di
Pisa (New Series), Pisa, IT: Edizioni della Normale.\item Dickson, L.E. (1919-23), \it History of the Theory of
Numbers, \rm 3 Vols., Publications of The Carnegie Institution of Washington, No. 256, Washington, WA: Published
by The Carnegie Institution of Washington.\item Dieudonné, J. and Grothendieck, A. (1971), \it Éléments de
géométrie algébrique. I: Le language des schémas, \rm Berlin and Heidelberg, DE: Springer-Verlag. \item
Dieudonné, J. (1982), \it A Panorama of Pure Mathematics. A Seen by N. Bourbaki, \rm New York, NY: Academic
Press, Inc.\item Dimitrov, D.K. (2013), Lee-Yang measures and wave functions, preprint arXiv: 1311.0596v1
[math-ph] 4 Nov 2013.\item Dini, U. (1881), Alcuni teoremi sulle funzioni di una variabile complessa, in: \it In
Memoriam Dominici Chelini. Collectanea Mathematica nunc primum edita cura et studio L. Cremona et E. Beltrami,
\rm Milano, IT: Ulrico Hoepli Editore-Librajo della Real Casa, pp. 258-276.\item Dini, U. (1953-59), \it Opere,
\rm a cura dell'Unione Matematica Italiana e con il contributo del Consiglio Nazionale delle Ricerche, 5 Voll.,
Roma, IT: Edizioni Cremonese (già Casa Editrice Perrella).\item Domb, C. and Green, M.S. (Eds.) (1972), \it
Phase Transitions and Critical Phenomena, \rm Volume 1, Exact Results, London, UK: Academic Press, Ltd.\item
Dugac, P. (1973), \'{E}léments d'Analyse de Karl Weierstrass, \it Archive for History of Exact Sciences, \rm 10
(1/2): 41-176.\item Durège, H. (1864), \it Elemente der Theorie der Functionen einer complexen
ver\"{a}nderlichen Gr\"{o}sse. Mit besonderer Ber\"{u}cksichtigung der Sch\"{o}pfungen Riemann's, \rm Leipzig,
DE: Druck und Verlag Von B.G. Teubner (English Translation of the 4th Edition: (1896), \it Elements of the
theory of functions of a complex variable with especial reference to the methods of Riemann, \rm Philadelphia,
PA - Norwood, MA: G.E. Fisher \& I. J. Schwatt - Norwood Press, J.S. Cushmg \& Co. - Berwick \& Smith).\item
Ebbinghaus, H.D., Hermes, H., Hirzebruch, F., Koecher, M., Mainzer, K., Neukirch, J., Prestel, A. and Remmert,
R. (1991), \it Numbers, \rm with an Introduction by K. Lamotke, New York, NY: Springer-Verlag, Inc.\item
Edwards, H.M. (1974), \it Riemann's Zeta Function, \rm New York, NY: Academic Press, Inc.\item Eisenstein, G.
(1847), Genaue Untersuchung der unendlichen Doppelproducte, aus welchen die elliptischen Functionen als
Quotienten zusammengesetzt sind, \it Journal f\"{u}r die reine und angewandte Mathematik, \rm 35: 153-274.\item
Enriques, F. and Chisini, O. (1985), \it Lezioni sulla teoria geometrica delle equazioni e delle funzioni
algebriche, \rm 2 volumi in 4 tomi, ristampa anastatica dell'edizione originale del 1915, Bologna, IT:
Zanichelli Editore.\item Enriques, F. (1982), \it Le matematiche nella storia e nella cultura, \rm ristampa
anastatica dell'edizione originale del 1938, Bologna, IT: Nicola Zanichelli Editore.\item Erdélyi, A. (Ed.)
(1981), \it Higher Transcendental Functions, \rm Volume III, Bateman Manuscript Project, Malabar, FL: Robert E.
Krieger Publishing Company, Inc.\item Eremenko, A., Ostrovskii, I., and Sodin, M. (1998), Anatolii Asirovich
Gol'dberg, \it Complex Variables, Theory and Application: An International Journal, \rm 37 (1-4): 1-51.\item
Evgrafov, M.A. (1961), \it Asymptotic Estimates and Entire Functions, \rm New York, NY: Gordon \& Breach Science
Publishers, Inc.\item Fasano, A. and Marmi, S. (2002), \it Meccanica analitica con elementi di meccanica
statistica e dei continui, \rm nuova edizione interamente riveduta e ampliata, Torino, IT: Bollati Boringhieri
editore.\item Fine, B. and Rosenberger, G. (2007), \it Number Theory. An Introduction via the Distribution of
Prime, \rm New York, NY: Birkh\"{a}user Boston c/o Springer Science + Business Media, LLC.\item Fisk, S. (2008),
Polynomials, roots, and interlacing, arXiv: math/0612833v2 [math.CA] 11 Mar 2008.\item Forsyth, A.R. (1918), \it
Theory of Functions of a Complex Variable, \rm Third Edition, Cambridge, UK: Cambridge University Press.\item
Fou\"{e}t, L.A. (1904-1907), \it Le\c{c}ons \'{E}lémentaires dur la Théorie des Fonctions Analytiques, \rm Tome
I et II, Paris, FR: Gauthier-Villars Imprimeur-Libraires du Bureau des Longitudes de L'\'{E}cole
Polytechnique.\item Fourier, J.B.J. (1830), \it Analyse des équations déterminées, \rm Paris, FR: Chez Firmin
Didot Frères Libraires.\item Franel, J. (1896), Sur la fonction $\xi(t)$ de Riemann et son application à
l'arithmétique, \it Festschrift der Naturforschenden Gesellschaft in Z\"{u}rich-Vierteljahrsschrift, \rm 41 (2):
7-19. \item Fröhlich, J. and Rodriguez, P-F. (2012), Some applications of the Lee-Yang theorem, \it The Journal
of Mathematical Physics, \rm 53(095218): 1-15.\item Fromm, E. (1979), \it Greatness and Limitations of Freud's
Thought, \rm New York, NY: Harper \& Row Publishers.\item Genocchi, A. (1860), Formole per determinare quanti
sono i numeri primi fino ad un dato limite, \it Annali di matematica pura ed applicata, \rm tomo III:
52-59.\item Georgii, H-O. (2011), \it Gibbs Measures and Phase Transitions, \rm 2nd edition, Berlin, DE: Walter
de Gruyter.\item Gil', M. (2010), \it Localization and Perturbation of Zeros of Entire Functions, \rm Boca
Raton, FL: Chapman \& Hall/CRC Press, Taylor \& Francis Group.\item Glimm, J. and Jaffe, A. (1987), \it Quantum
Physics. A Functional Integral Point of View, \rm Second Edition, New York, NY: Springer-Verlag Inc.\item
Gol'dberg, A.A. and Ostrovski\v{i}, I.V. (2008), \it Value Distribution of Meromorphic Functions, \rm
Translations of Mathematical Monographes, Vol. 236, Providence, RI: American Mathematical Society.\item
Goldstein, C., Schappacher, N. and Schwermer, J. (Eds.) (2007), \it The Shaping of Arithmetic after C.F. Gauss's
Disquisitiones Arithmetic\ae, \rm Berlin and Heidelberg, DE: Springer-Verlag.\item Goldstein, J.T. (1973), A
History of the Prime Number Theory, \it The American Mathematical Monthly, \rm 80 (6):599-615.\item Gonchar,
A.A., Havin, V.P. and Nikolski, N.K. (Eds.) (1997), \it Complex Analysis I. Entire and Meromorphic Functions.
Polyanalytic Functions and Their Generalizations, \rm Encyclopaedia of Mathematical Sciences, Volume 85, Berlin
and Heidelberg, DE: Springer-Verlag.\item Gonek, S.M. (2004), Three Lectures on the Riemann Zeta-Function,
arXiv: math.NT/0401126v1 (2004).\item Gordan, P. (1874), Über den grössten gemeinsame Factor, \it Mathematische
Annalen, \rm 7 (1): 433-448.\item Gram, J.P. (1895), Note sur le calcul de la fonction $\zeta(s)$ de Riemann,
\it Bulletin de l'Académie Royal des sciences et des lettres de Danemark, \rm p. 303.\item Grattan-Guinness, I.
(Ed.) (2005), \it Landmark Writings in Western Mathematics 1640-1940, \rm Amsterdam, NH: Elsevier B.V.\item
Greenhill, A.G. (1892), \it The Applications of Elliptic Functions, \rm London, UK: The MacMillan Company.\item
Gruber, C., Hintermann, A. and Merlini, D. (1977), \it Group Analysis of Classical Lattice Systems, \rm Lecture
Notes in Physics, No. 60, Berlin and Heidelberg, DE: Springer-Verlag.\item Guichard, C. (1884), Sur les
fonctions entières, \it Annales scientifiques de l'École Normale Supérieure, \rm 3e Série, 1: 427-432.\item
Hadamard, J. (1892), Essai sur l'étude des fonctions données par leur développement de Taylor, Thèse de Doctorat
de la Faculté des Sciences de Paris, \it Journal de Mathématiques Pures et Appliquées, \rm 4e Série, Tome 8:
101-186.\item Hadamard, J. (1893), Étude sur les propriétés des fonctions entières et en particulier d'une
fonction considérée par Riemann, \it Journal de Mathématiques Pures et Appliquées, \rm 4e Série, Tome 9:
171-215. \item Hadamard, J. (1896a), Sur la distribution des zéros de la fonction $\xi(s)$ et ses conséquences
arithmétiques, \it Bulletin de la Société Mathématique de France, \rm 24: 199-220.\item Hadamard, J. (1896b),
Sur les fonctions entières (Extrait d'une lettre addressée à M. Picard), \it Les Comptes Rendus de l'Académie
des Sciences de Paris, \rm 122 (1): 1257-1258.\item Hadamard, J. (1896c), Sur les fonctions entières, \it
Bulletin de la Société Mathématique de France, \rm 24: 199-200.\item Hadamard, J. (1897), Sur le séries de
Dirichlet, \it Proc. Verb. Soc. Sci. Phys. Nat., Bordeaux, \rm 18 février, pp. 41-45. \item Hadamard, J. (1901),
\it Notice sur les travaux scientifiques de M. Jacques Hadamard, \rm Paris, FR: Gauthier-Villars,
Imprimeur-Libraires du Bureau des Longitudes de L'\'{E}cole Polytechnique.\item Haglund, J. (2009), Some
conjectures on the zeros of approximates to the Riemann $\Xi$-function and incomplete gamma functions, preprint
available at arXiv: 0910.5228v1 [math.NT] 27 Oct 2009. \item Hancock, H. (1910), \it Lectures on the Theory of
Elliptic Functions, \rm Volume I: Analysis, New York, NY: John Wiley \& Sons - London, UK: Chapman \& Hall,
Ltd.\item Hardy, G.H. (1914), Sur les zéros de la function $\zeta(s)$ de Riemann, \it Comptes Rendus de
l'Academie des Sciences de Paris, \rm 158: 1012-1014.\item Hardy, G.H. and Wright, E.M. (1960), \it An
Introduction to the Theory of Numbers, \rm 4th Edition, Oxford, UK: Oxford University Press.\item Heins, M.
(1968), \it Complex Function Theory, \rm New York, NY: Academic Press, Inc.\item Hejhal, D.A. (1990), On a
result of G. P\'{o}lya concerning the Riemann $\xi$-function, \it Journal d'Analyse Mathématique, \rm 55:
59-95.\item Hermite, C. (1856a), Sur les nombre des racines d'une équation algébrique comprise entre des limites
données, \it Journal für die reine und angewandte Mathematik (Crelle Journal), \rm 52: 39-51.\item Hermite, C.
(1856b), Extract d'une lettre de M. Ch. Hermite de Paris a M. Borchardt de Berlin sur le nombre des racines
d'une équation algébrique comprises entre des limites données, \it Journal für die reine und angewandte
Mathematik (Crelle Journal), \rm 52: 39-51.\item Hermite, C. (1873), \it Cours d'Analyse de L'\'{E}cole
Polytechnique, \rm Premiére Partie, Paris, FR: Gauthier-Villars et C.le, \'{E}diteurs-Libraires du Bureau des
Longitudes de L'\'{E}cole Polytechnique, Successeur De Mallet-Bachelier.\item Hermite, C. (1881), Sur quelques
points de la théorie des fonctions, \it Bulletin des sciences mathématiques et astronomiques, \rm 2e Série,
5(1): 312-320.\item Hille, E. (1959), \it Analytic Function Theory, \rm Volume I, New York, NY: Chelsea
Publishing Company.\item  Hoare, G.T.Q. and Lord, N.J. (2002), Integrale, longueur, aire - the centenary of the
Lebesgue integral, \it The Mathematical Gazette, \rm 86 (505): 3-27.\item Huang, K. (1987), \it Statistical
Mechanics, \rm 2nd Edition, New York, NY: John Wiley \& Sons, Inc.\item Ingham, A.E. (1964), \it The
Distribution of Prime Numbers, \rm reprint of the first 1932 edition, New York, NY and London, UK:
Stechert-Hafner Service Agency, Inc.\item It\^{o}, K. (Ed.) (1993), \it Encyclopedic Dictionary of Mathematics,
\rm 2 Voll., 2nd Edition, by the Mathematical Society of Japan, Cambridge, MA: The MIT Press.\item Ivi\c{c}, A.
(1985), \it The Riemann Zeta-Function. The Theory of Riemann Zeta-Function with Applications, \rm A
Wiley-Interscience Publication, New York, NY: John Wiley and Sons, Inc. \item Ivi\c{c}, A. (1987), Some recent
results on the Riemann zeta-function, in: \it Théorie des nombres, \rm Proceedings of the International
Conference on Number Theory, Laval, edited by J.M. De Koninck, Berlin, DE: Walter De Gruyter.\item Janovitz, A.
and Mercanti, F. (2008), \it Sull'apporto evolutivo dei matematici ebrei mantovani nella nascente nazione
italiana, \rm Teramo, IT: Monografie di Eiris.\item Jensen, J.L.W. (1891), \it Gammafunktionens Theorie i
element\ae r Fremstilling, \rm Nyt Tidsskriftf or Mathematik, AfdelingB, 2: pp. 33-35, 57-72, 83-84 (English
Translation: (1916), An Elementary Exposition of the Theory of the Gamma Function, \it Annals of Mathematics,
\rm 17 (2): 124-166.\item Jensen, J.L.W. (1898-99), Sur un nouvel et important théorème de la théorie des
fonctions, \it Acta Mathematica, \rm 22: 359-364.\item Jordan, C., Poincaré, H.J., Hermite, C., Darboux, G. and
Picard, E. (1892), Prix Décernés - Année 1892. Géométrie. Grand Prix des Sciences Mathématiques, \it Comptes
Rendus de l'Academie des Sciences de Paris, \rm Tome CXV, Number 25: 1120-1122.\item Julia, B. (1994),
Thermodynamic limit in number theory: Riemann-Beurling gases, \it Physica A, \rm 203: 425-436.\item Kahane, J-P.
(1991), Jacques Hadamard, \it The Mathematical Intelligencer, \rm 13 (1): 23-29.\item Karatsuba, A.A. and
Voronin, S.M. (1992), \it The Riemann Zeta-Function, \rm Berlin-New York: Walter de Gruyter.\item Karatsuba,
A.A. (1994), \it Complex Analysis in Number Theory, \rm Boca Raton-Ann Arbor-London-Tokio: CRC Press.\item Katz,
N.M and Sarnak, P. (1996), Zeros of zeta functions and symmetry, \it Bulletin of the American Mathematical
Society, \rm 36: 1-26.\item Ki, H. and Kim, Y-O. (2000), On the Number of Nonreal Zeros of Real Entire Functions
and the Fourier-P\'{o}lya Conjecture, \it Duke Mathematical Journal, \rm 104 (1): 45-73.\item Ki, H., Kim, Y-O.
and Lee, J. (2009), On the de Bruijn-Newman constant, \it Advances in Mathematics, \rm 222 (1): 281-306.\item
Ki, H. and Kim, Y-O. (2003), De Bruijn's question on the zeros of Fourier transforms, \it Journal d'Analyse
Mathématique, \rm 91: 369-387.\item Kim, Y-O. (1990), A proof of the Pólya-Wiman conjecture, \it Proceedings of
the American Mathematical Society, \rm 109: 1045-1052. \item Klein, F. (1979), \it Development of Mathematics in
the 19th century, \rm translated by M. Ackerman, Volume IX of the Series \it Lie Groups: History, Frontiers and
Applications, \rm edited by R. Hermann, Brookline, MA: MathSciPress.\item Knauf, A. (1999), Number Theory,
Dynamical Systems and Statistical Mechanics, \it Reviews in Mathematical Physics, \rm 11 (8): 1027-1060.\item
Kolmogorov, A.N. and Yushkevich, A.P. (Eds.) (1996), \it Mathematics of the 19th Century, \rm 3 Vols., Basel,
CH: Birkh\"{a}user Verlag.\item Korevaar, J. (2013), Early work of N.G. (Dick) de Bruijn in analysis and some of
my own, \it Indagationes Mathematic\ae, \rm 24: 668-678.\item Kragh, H. (1987), \it An Introduction to
Historiography of Science, \rm Cambridge, UK: Cambridge University Press (Italian Translation: (1990), \it
Introduzione alla storiografia della scienza, \rm Bologna, IT: Zanichelli Editore).\item Kudryavtseva, E.A.,
Saidak, F. and Zvengrowski, P. (2005), Riemann and his zeta function, \it Morfismos, \rm 9 (2): 1-48.\item
Lagarias, J.C. (2005), Zero Spacing Distributions for Differenced L-Functions, \it Acta Arithmetica, \rm 120(2):
159-184 (arXiv: math.NT/0601653v2).\item Lagarias, J.C. and Montague, D. (2011), The integral of the Riemann
$\xi$-function, \it Commentarii Mathematici Universitatis Sancti Pauli, \rm 60(1-2): 143-169 (arXiv:
math.NT/1106.4348v3).\item Laguerre, E.N. (1882a), Sur quelques équations transcendantes, \it Comptes Rendus de
l'Académie des Sciences de Paris, \rm XCIV: 160-163.\item Laguerre, E.N. (1882b), Sur la détermination du genre
d'une fonction transcendante entière, \it Comptes Rendus de l'Académie des Sciences de Paris, \rm XCIV:
635-638.\item Laguerre, E.N. (1882c), Sur les fonctìons du genre zèro et du genre un, \it Comptes Rendus de
l'Académie des Sciences de Paris, \rm XCV: 828-831.\item Laguerre, E.N. (1883), Sur la théorie des équations
numériques, \it Journal de Mathématiques Pures et Appliquées, 3e Série, \rm IX: 99-146;\item Laguerre, E.N.
(1884), Sur le genre de quelques fonctìons enfières, \it Comptes Rendus de l'Académie des Sciences de Paris, \rm
XCVIII: 79-81.\item Laguerre, E. (1898-1905), \it Oeuvres de Laguerre, \rm publiées sous les auspices de
l'Académie des Sciences de Paris par Ch. Hermite, H. Poincaré et E. Rouché, Tomes I, II, Paris, FR:
Gauthier-Villars et Fils, Imprimeurs-Libraires du Bureau des Longitudes de L'\'{E}cole Polytechnique.\item
Landau, E. (1908), Beitr\"{a}ge zur analytischen zahlentheorie, \it Rendiconti del Circolo Matematico di
Palermo, \rm 26 (1): 169-302.\item Landau, E. (1909a), \"{U}ber die Verteilung der Nullstellen der Riemannschen
Zetafunktion und einer Klasse verwandter Funktionen, \it Mathematischen Annalen, \rm 66 (4): 419-445.\item
Landau, E. (1909b), \it Handbuch der Lehre von der Verteilung der Primzahlen, \rm Band I, II, Leipzig and
Berlin, DE: Druck and Verlag Von B.G. Teubner.\item Landau, E. (1918), \"{U}ber die Blaschkesche
Verallgemeinerung des Vitalischen Satzes, \it Berichte über die Verhandlungen der Königlich-Sächsischen
Gesellschaft der Wissenschaften zu Leipzig, Mathematisch-Physikalische Klasse, \rm 70: 156-159 (see also \it
Collected Works, \rm 7: 138-141).\item Landau, E. (1927), \"{U}ber die Zetafunktion und die Hadamardsche Theorie
der ganzen Funktionen, \it Mathematische Zeitschrift, \rm 26: 170-175.\item Lang, S. (1974), \it Analisi Reale e
Complessa, \rm Torino, IT: Editore Boringhieri.\item Lang, S. (1987), \it Elliptic Functions, \rm Second
Edition, New York, Berlin and Heidelberg, DE: Springer-Verlag, Inc.\item Lang, S. (1999), \it Complex Analysis,
\rm Fourth Edition, New York, NY: Springer-Verlag, Inc.\item Lapidus, M.L. (2008), \it In Search of the Riemann
Zeros. Strings, Fractal Membranes and Noncommutative Spacetime, \rm Providence, RI: American Mathematical
Society.\item Laugwitz, D. (1999), \it Bernhard Riemann 1826-1866. Turning Points in the Conception of
Mathematics, \rm Boston, MA: Birkh\"{a}user.\item Lavis, D.A. and Bell, G.M. (1999), \it Statistical Mechanics
of Lattice Systems, Volume 1: Closed-Form and Exact Solutions, Volume 2: Exact, Series and Renormalization Group
Methods, \rm Berlin and Heidelberg, DE: Springer-Verlag.\item Lax, P.D. and Phillips, R.S. (1976), \it
Scattering Theory for Automorphic Forms, \rm Annals of Mathematics Studies, No. 87, Princeton, NJ: Princeton
University Press.\item Lax, P. and Phillips, R.S. (1989), \it Scattering Theory, \rm Revised Edition of the
first 1967 Edition, Series in Pure and Applied Mathematics, Volume No. 26, New York, NY: Academic Press,
Inc.\item Lebowitz, J.L., Ruelle, D. and Steer, E.R. (2012), Location of the Lee-Yang zeros and absence of phase
transitions in some Ising spin systems, \it Journal of Mathematical Physics, \rm 53: 095211/1-13.\item Lee, T.D.
and Yang, C.N. (1952), Statistical Theory of Equations of State and Phase Transitions. II. Lattice Gas and Ising
Model, \it Physical Review, \rm 87(3): 410-419. \item Levin, B. Ja. (1980), \it Distribution of Zeros of Entire
Functions, \rm Revised Edition of the 1964 first one, Providence, Rhode Island: American Mathematical Society.
\item Levin, B.Ya. (1996), \it Lectures on Entire Functions, \rm in collaboration with G.Yu. Lyubarskii, M.
Sodin and V. Tkachenko, Providence, Rhode Island: American Mathematical Society.\item Lindel\"{o}f, E. (1905),
Sur les fonctions entières d'ordre entier, \it Annales Scientifiques de l'\'{E}cole Normale Supérieure, \rm 22
(3): 369-395.\item Lindwart, E. and P\'{o}lya, G. (1914), \"{U}ber einen Zusammenhang zwischen der Konvergenz
von Polynomfolgen und der Verteilung ihrer Wurzeln, \it Rendiconti del Circolo Matematico di Palermo, \rm 37:
297-304.\item Loria, G. (1946), \it Guida allo studio della storia delle matematiche, \rm 2$^a$ edizione rifusa
ed aumentata, Milano, IT: Editore Ulrico Hoepli.\item Loria, G. (1950), \it Storia delle matematiche dall'alba
della civiltà al tramonto del secolo XIX, \rm 2$^a$ edizione riveduta e aggiornata, Milano, IT: Editore Ulrico
Hoepli.\item Lunts, G.L. (1950), The analytic work of N.I. Lobachevskii, \it Uspekhi Matematicheskikh Nauk, \rm
5 1(35): 187-195.\item Ma, S-K. (1985), \it Statistical Mechanics, \rm Singapore, SG: World Scientific
Publishing Company.\item Mackey, G.W. (1978), \it Unitary Group Representations in Physics, Probability and
Number Theory, \rm San Francisco, CA: The Benjamin/Cummings Publishing Company.\item Mandelbrojt, S. (1967),
Théorie des fonctions et théorie des nombres dans l'\OE uvre de Jacques Hadamard, \it L'Enseignement
Mathématique, \rm 13 (1): 25-34.\it Mandelbrojt, S. and Schwartz, L. (1965), Jacques Hadamard (1865-1963), \it
Bulletin of the American Mathematical Society, \rm 71 (1): 107-129.\item Manning, K.R. (1975), The Emergence of
the Weierstrassian Approach to Complex Analysis, \it Archive for History of Exact Sciences, \rm 14 (4):
297-383.\item Marcolongo, R. (1918-1919), Necrologio di Luciano Orlando, \it Rendiconti del Seminario Matematico
della Facoltà di Scienze della R. Università di Roma, \rm 5, Numero Speciale: 11-15.\item Marcolongo, R. (1931),
Gabriele Torelli, \it Rendiconti della Reale Accademia delle Scienze Fisiche e Matematiche di Napoli, \rm 4 (1):
109-118.\item Marden, M. (1949), \it The Geometry of the Zeros of a Polynomial in a Complex Variable, \rm
Mathematical Surveys Number III, New York, NY: American Mathematical Society Publications.\item Marden, M.
(1966), \it Geometry of Polynomials, \rm Providence, RI: American Mathematical Society.\item Marku\v{s}evi\v{c},
A.I. (1965-67), \it Theory of functions of a complex variable, \rm I, II and III volume, Englewood Cliffs, NJ:
Prentice-Hall, Inc.\item Marku\v{s}evi\v{c}, A.I. (1966), \it Entire Functions, \rm New York, NY: American
Elsevier Publishing Company, Inc.\item Marku\v{s}evi\v{c}, A.I. (1988), \it Teoria delle funzioni analitiche,
\rm Mosca-Roma: Edizioni Mir-Editori Riuniti.\item Maroni, A. (1924), Necrologio di Giacomo Bellacchi, \it
Periodico di Matematiche, \rm IV (4): 264-265.\item Mashreghi, J. and Fricain, E. (Eds.) (2013), \it Blaschke
Products and Their Applications, \rm New York, NY: Springer Science + Business Media, LLC.\item Maurin, K.
(1997), \it The Riemann Legacy. Riemann Ideas in Mathematics and Physics, \rm Dordrecht, NL: Kluwer Academic
Publishers.\item Mayer, D.H. (1980), \it The Ruelle-Araki Transfer Operator in Classical Statistical Mechanics,
\rm Lecture Notes in Physics, No. 123, Berlin and Heidelberg, DE: Springer-Verlag.\item Maz'ya, V. and
Shaposhnikova, T. (1998), \it Jacques Hadamard. A Universal Mathematician, \rm Providence, RI and London, UK:
American Mathematical Society and London Mathematical Society. \item McCoy, B.M. and Wu, T.T. (1966), Theory of
Toeplitz Determinants and the Spin Correlations of the Two-Dimensional Ising Model. I, \it Physical Review, \rm
149 (1): 380-401.\item McCoy, B.M. and Wu, T.T. (1967a), Theory of Toeplitz Determinants and the Spin
Correlations of the Two-Dimensional Ising Model. II, \it Physical Review, \rm 155 (2): 438-452.\item McCoy, B.M.
and Wu, T.T. (1967b), Theory of Toeplitz Determinants and the Spin Correlations of the Two-Dimensional Ising
Model. IV, \it Physical Review, \rm 162 (2): 436-475.\item McCoy, B.M. and Wu, T.T. (1973), \it The
Two-dimensional Ising Model, \rm Cambridge, MA: Harvard University Press.\item McCoy, B.M. (2010), \it Advanced
Statistical Mechanics, \rm Oxford, UK: Oxford University Press.\item McMullen, T.C. (1994), \it Complex Dynamics
and Renormalization, \rm Annals of Mathematics Studies, No. 135, Princeton, NJ: Princeton University Press.\item
Meier, P. and Steuding, J. (2009), L'hypothese de Riemann, \it Dossier Pour la Science, \rm 377: 1-17.\item
Mertens, F. (1897), \"{U}ber eine zahlentheoretische Function, \it Sitzungberichte der Kleine Akademie der
Wissenschaften in Wien Mathematik-Naturlick Classe, \rm IIa, 106: 761-830.\item Mittag-Leffler, G. (1884), Sur
la représentation analytique des fonctions monogènes uniformes d'une variable indépendante, \it Acta
Mathematica, \rm 4(1): 1-79.\item Monastyrsky, M. (1999), \it Riemann, Topology and Physics, \rm Second Edition,
Boston, MA: Birkh\"{a}user.\item Montgomery, H.L. and Vaughan, R.C. (2006), \it Multiplicative Number Theory: I.
Classical Theory, \rm Cambridge Studies in Advanced Mathematics No. 97, Cambridge, UK: Cambridge University
Press.\item Montel, P. (1932), \it Le\c{c}ons sur les Fonctions Entières ou Méromorphes, \rm Paris, FR:
Gauthier-Villars et C.le, \'{E}diteurs-Libraires du Bureau des Longitudes de L'\'{E}cole Polytechnique.\item
Morosawa, S., Nishimura, Y., Taniguchi, M. and Ueda, T. (2000), \it Holomorphic Dynamics, \rm Cambridge Studies
in Advanced Mathematics, No. 66, Cambridge, UK: Cambridge University Press.\item Motohashi, Y. (1997), \it
Spectral Theory of the Riemann Zeta-Function, \rm Cambridge, UK: Cambridge University Press.\item
M\"{u}ller-Hartmann, H. (1977), Theory of the Ising model on a Cayley tree, \it Zeitschrift f\"{u}r Physik B,
Condensed Matter, \rm 27 (2): 161-168.\item Narkiewicz, W. (2000), \it The Development of Prime Number Theorem:
From Euclid to Hardy and Littlewood, \rm Berlin and Heidelberg, DE: Springer-Verlag.\item Nathanson, M.B.
(2000), \it Elementary Methods in Number Theory, \rm New York, NY: Springer-Verlag, Inc.\item Neuenschwander, E.
(1996), \it Riemanns Einf\"{u}hrung in die Funktionentheorie. Eine quellenkritische Edition seiner Vorlesungen
mit einer Bibliographie zur Wirkungsgeschichte der Riemannschen Funktionentheorie, \rm Abhandlungen der Akademie
der Wissenschaften Zu G\"{o}ttingen. Mathematisch-Physikalische Klasse. 3. Folge. Series Orbis Biblicus Et
Orientalis, Band 44. Abhandlungen Der Akademie Der Wissenschaften. Mathematisch-Physikalische Klasse. 3. Folge,
G\"{o}ttingen, DE: Vandenhoeck \& Ruprecht.\item Nevanlinna, F. and Nevanlinna, R. (1924), \"{U}ber die
Nullstellen der Riemannschen Zetafunktion, \it Mathematische Zeitschrift, \rm 20: 253-263.\item Newman, C.M.
(1976), Fourier transform with only real zeros, \it Proceedings of the American Mathematical Society, \rm 61
(2): 245-251.\item Newman, F.W. (1848), On $\Gamma a$ especially when $a$ is negative, \it The Cambridge and
Dublin Mathematical Journal, \rm 3: 57-60.\item Newman, C.M. (1974), Zeros of the partition function for
generalized Ising systems, \it Communication in Pure and Applied Mathematics, \rm 27: 143-159. \item Newman,
C.M. (1991), The GHS inequality and the Riemann hypothesis, \it Constructive Approximation, \rm 7: 389-399.\item
Niven, I., Zuckerman, H.S. and Montgomery, H.L. (1991), \it An Introduction to the Theory of Numbers, \rm Fifth
Edition, New York, NY: John Wiley \& Sons, Inc.\item Orlando, L. (1903), Sullo sviluppo della funzione
$(1-z)e^{(z+z^2/2+...+z^{p-1}/(p-1)!)}$, \it Giornale di Matematiche di Battaglini per il progresso degli studi
nelle università italiane, \rm Volume XLI, $40^o$ della $2^{a}$ Serie: 377-378.\item Ostrovski\v{i}, I.V.
(1994), M.G. Krein's investigations in the theory of entire and meromorphic functions and their further
development, \it Ukrainian Mathematical Journal, \rm 46 (1-2): 87-100.\item Ostrovski\v{i}, I.V. and Sodin, M.
(1998), Anatolii Asirovich Gol'dberg, \it Complex Variable, Theory and Applications. An International Journal,
\rm 37 (1-4): 1-51.\item Ostrovski\v{i}, I.V. and Sodin, M. (2003), The Scientific School of B.Ya. Levin (in
Russian), \it Zhurnal Matematicheskoi Fiziki, Analiza, Geometrii, \rm 10 (2): 228-242.\item Painlèvé, P.
(1898a), Sur la représentation des fonctions analytiques uniformes, \it Comptes Rendus de l'Acadèmie des
Sciences de Paris, \rm CXXVI: 200-202.\item Painlèvé, P. (1898b), Sur le développement des fonctions  uniformes
ou holomorphes dans un domain e quelconque, \it Comptes Rendus de l'Acadèmie des Sciences de Paris, \rm CXXVI:
318-321.\item Patterson, S.J. (1988), \it An introduction to the theory of the Riemann zeta function, \rm
Cambridge, UK: Cambridge University Press.\item Pepe, L. (2012), Mascheroni and Gamma, in: Burenkov, V.I.,
Goldman, M.L., Laneev, E.B. and Stepanov V.D. (Eds.) (2012), \it Peoples' Friendship University of Russia.
Progress in Analysis. \rm Proceedings of the 8th Congress of the International Society for Analysis, its
applications and computation, 22-27 August 2011, 2 Vols., Moscow, RU: Peoples' Friendship University of Russia
Publishers, pp. 23-36.\item Petersen, J. (1899), Quelques remarques sur les fonctions entières, \it Acta
Mathematica, \rm 23: 85-90.\item Picard, É. (1881), Sur la décomposition en facteurs primaires des fonctions
uniformes ayant une ligne de points singuliers essentiels, \it Comptes Rendus de l'Académie des Sciences de
Paris, \rm 92: 690-692.\item Picard, \'{E}. (1897), Karl Weierstrass, Nécrologie, \it Revue générale des
sciences pures et appliquées, \rm 8 (5): 173-174.\item Pierpont, J. (1914), \it Function of a complex variable,
\rm Boston, MA: Ginn \& Company Publisher.\item Pincherle, S. (1899-1900), \it Lezioni sulla teoria delle
funzioni analitiche tenute nella R. Università di Bologna, \rm Bologna, IT: Litografia Sauer e Barigazzi.\item
Pincherle, S. (1922), \it Gli elementi della teoria delle funzioni analitiche, \rm Parte prima, Bologna, IT:
Nicola Zanichelli editore.\item Pizzarello, D. (1900), \it Sulle funzioni trascendenti intere, \rm Memoria
presentata come tesi di laurea nella R. Università di Roma il 14 Novembre 1899, Messina, IT: Tipografia
Dell'epoca.\item Poincaré, H.J. (1882), Sur les transcendantes entières, \it Comptes Rendus de l'Acadèmie des
Sciences de Paris, \rm XCV: 23-26.\item Poincaré, H.J. (1883), Sur les fonctions entières, \it Bulletin de la
Société Mathématique de France, \rm XI: 136-144.\item Poincaré, H.J. (1899), L'oeuvre mathématique de
Weierstrass, \it Acta Mathematica, \rm 22(1): 1-18.\item P\'{o}lya, G. (1913), \"{U}ber Ann\"{a}herung durch
Polynome mit lauter reellen Wurzeln, \it Rendiconti del Circolo Matematico di Palermo, \rm 36: 279-295.\item
Pólya, G. (1926a), Bemerkung Über die Integraldarstellung der Riemannsche $\xi$-Funktion, \it Acta Mathematica,
\rm 48 (3-4): 305-317, reprinted in: Boas, R.P. (Ed.) (1974), \it George Pólya Collected Papers, \rm Vol. II,
Locations of Zeros, Mathematics of Our Time, Cambridge, MA and London, UK: The MIT Press.\item P\'{o}lya, G.
(1926a), On the zeros of certain trigonometric integrals, \it Journal of the London Mathematical Society, \rm 1:
98-99.\item P\'{o}lya, G. (1927a), \"{U}ber trigonometrische Integrale mit nur reelen Nullstellen, \it Crelle
Journal f\"{u}r die Reine und Angewandte Mathematik, \rm 158, 6-18.\item P\'{o}lya, G. (1927b), \"{U}ber die
algebraisch-funktionentheoretischen Untersuchungen von J.L.W.V. Jensen, \it Det Kongelige Danske Videnskabernes
Selskab Mathematisk-fysiske Meddelelser (K\o benhavn), \rm 7 (17): 1-34.\item P\'{o}lya, G. (1930), Some
problems connected with Fourier's work on transcendental equations, \it The Quarterly Journal of Mathematics,
\rm 1 (1): 21-34.\item P\'{o}lya, G. (1943), On the zeros of the derivative of a function and its analytic
character, \it Bulletin of the American Mathematical Society, \rm 49: 178-191.\item P\'{o}lya, G. Szeg\H{o}, Z.
(1998a), \it Problems and Theorems in Analysis I. Series. Integral Calculus. Theory of Functions. \rm Reprint of
the 1978 Edition, Berlin and Heidelberg, DE: Springer-Verlag.\item P\'{o}lya, G. Szeg\H{o}, Z. (1998b), \it
Problems and Theorems in Analysis II. Theory of Functions. Zeros. Polynomials. Determinants. Number Theory.
Geometry, \em Reprint of the 1976 Edition, Berlin and Heidelberg, DE: Springer-Verlag.\item Pradisi, G. (2012),
\it Lezioni di metodi matematici della fisica, \rm Pisa, IT: Edizioni della Normale - Scuola Normale Superiore
Pisa.\item Pringsheim, A. (1889), \"{U}ber die Convergenz unendlicher Produkte, \it Mathematische Annalen, \rm
33: 119-154.\item Rademacher, H. (1973), \it Topics in Analytic Number Theory, \rm Berlin and Heidelberg, DE:
Springer-Verlag.\item Rahnman, Q.I. and Schmeisser, G. (2002), \it Analytic Theory of Polynomials, \rm Oxford,
UK: Oxford University Press.\item Regge, T. (1958), Analytic Properties of Scattering Matrix, \it Il Nuovo
Cimento, \rm VIII (5): 671-679.\item Remmert, R. (1998), \it Classical Topics in Complex Function Theory, \rm
New York, NY: Springer-Verlag New York, Inc.\item Riemann, B. (1858-1859), \"{U}ber die Anzahl der Primzahlen
unter Einer Gegebene Grösse, \it Monatsberichte der K\"{o}niglich Preu\ss ischen Berliner Akademie der
Wissenschaften\rm: 671-680.\item Riera Matute, A. (1970), Cultura y naturaleza, \it Anuario filos\'{o}fico, \rm
3 (1): 287-315. \item Rouse Ball, W.W. (1937), \it Le matematiche moderne sino ad oggi, \rm versione del Dott.
Dionisio Gambioli, riveduta e corretta dal Prof. Gino Loria, Bologna, IT: Nicola Zanichelli Editore.\item
Ruelle, D. (1969), \it Statistical Mechanics. Rigorous Results, \rm New York, NY: W.A. Benjamin, Inc.\item
Ruelle, D. (1988), Is our mathematics natural? The case of equilibrium statistical mechanics, \it Bulletin of
the American Mathematical Society, \rm 19(1): 259-268.\item Ruelle, D. (1994), \it Dynamical Zeta Functions for
Piecewise Monotone Maps of the Interval, \rm CRM Monograph Series, Volume 4, Providence, RI: American
Mathematical Society.\item Ruelle, D. (2000), Grace-like Polynomials, in: Cucker, F. and Rojas, J.M. (Eds.)
(2000), \it Foundations of Computational Mathematics, \rm Proceedings of the Smilefest, Hong Kong and Singapore,
SG: World Scientific Publishing Company, Ltd., pp. 405-421 (arXiv: math-ph/0009030). \item Ruelle, D. (2007),
\it The Mathematician's Brain, \rm Princeton, NJ: Princeton University Press.\item Ruelle, D. (2010),
Characterization of Lee-Yang Polynomials, \it Annals of Mathematics, \rm 171: 589-603 (arXiv:
math-ph/0811.1327v1).\item Runge, C. (1885), Zur Theorie der eidentigen analytischen Functionen, \it Acta
Mathematica, \rm 6: 229-244.\item Saks, R. and Zygmund, A. (1952), \it Analytic Functions, \rm Monografje
Matematyczne, Vol. 28, Warsawa, PL: Pa\'{n}stwowe Wydawnietwo Naukowe. \item Sansone, G. (1972), \it Lezioni
sulla teoria delle funzioni di una variabile complessa, \rm Volume I, quarta edizione, Padova, IT: CEDAM - Casa
Editrice Dott. Antonio Milani.\item Scharlau, W. and Opolka, H. (1985), \it From Fermat to Minkowski. Lectures
on the Theory of Numbers and Its Historical Development, \rm New York, NY: Springer-Verlag, Inc.\item Schering,
E.C.G. (1881), Das Anschliessen einer Function an algebraische Functionen in unendlich vielen Stellen, \it
Abhandlungen der Königlichen Gesellschaft der Wissenschaften zu Göttingen, \rm Band 27: 3-64.\item
Schl\"{o}milch, O.X (1844), Einiges \"{u}ber die Eulerschen Integrale der zweiten, \it Art. Grunert Archiv, \rm
4: 167-174.\item Schl\"{o}milch, O.X. (1848), \it Analytische Studien, 1. Theorie der Gammafunktionen, \rm
Leipzig, DE: Vertag Von Wilhelm Engelmann.\item Schl\"{o}milch, O.X. (1858), \"{U}ber eine Eigenschaft gewisser
Reihen, \it Zeitschrift f\"{u}r Mathematik und Physik, \rm 3: 130-132.\item Siegel, C.L. (1932), \"{U}ber
Riemanns Nachlass zur analytischen Zahlentheorie, \it Quellen und Studien zur Geschichte der Mathematik,
Astronomie und Physik. Abteilung B: Studien, \rm 2: 45-80.\item Simon, B. (1974), \it $P(\phi_2)$ Euclidean
Quantum Field Theory, \rm Princeton, NJ: Princeton University Press.\item Stieltjes, J.T. (1885), Sur une
fonction uniforme, \it Les Comptes Rendus de l'Acadèmie des Sciences de Paris, \rm 101: 153-154. \item
Stillwell, J. (2002), \it Mathematics and Its History, \rm Second Edition, New York, NY: Springer-Verlag,
Inc.\item Stopple, J. (2003), \it A Primer of Analytic Number Theory. From Pythagoras to Riemann, \rm Cambridge,
UK: Cambridge University Press.\item Tannery, J. and Molk, J. (1893), \it Éléments de la théorie des fonctions
elliptiques, \rm Tome I, Paris, FR: Gauthier-Villars et Fils, Imprimeurs-Libraires de l'École Polytechnique, du
Bureau des Longitudes.\item Tenenbaum, G. and Mendès France, M. (2000), \it The Prime Numbers and Their
Distribution, \rm Providence, RI: American Mathematical Society.\item Titchmarsh, E.C. (1986), \it The Theory of
the Riemann Zeta-Function, \rm Oxford, UK: At Clarendon Press.\item Torelli, G. (1901), \it Sulla totalità dei
numeri primi fino ad un limite assegnato, \rm Memoria estratta dagli Atti della R. Accademia delle Scienze
Fisiche e Matematiche di Napoli, Volume XI, Serie 2$^a$, N. 1, Napoli, IT: Tipografia della Reale Accademia
delle Scienze Fisiche e Matematiche, diretta da E. De Rubertis fu Michele.\item Tricomi, F. (1968), \it Funzioni
Analitiche, \rm II$^a$ Edizione, Bologna, IT: Nicola Zanichelli Editore.\item Tchebycheff, P.L. (1895), \it
Teoria delle congruenze, \rm traduzione italiana con aggiunte e note della Dott.ssa Iginia Massarini, Roma, IT:
Ermanno Loescher \& C.\item Ullrich, P. (1989), Weierstraß Vorlesung zur ''Einleitung in die Theorie der
analytischen Funktionen'', \it Archive for History of Exact Sciences, \rm 40(2): 143-172.\item Ullrich, P
(2003), Review of ''Riemanns Einf\"{u}hrung in die Funktionentheorie. Eine quellenkritische Edition seiner
Vorlesungen mit einer Bibliographie zur Wirkungsgeschichte der Riemannschen Funktionentheorie by Erwin
Neuenschwander. Abhandlungen der Akademie der Wissenschaften in G\"{o}ttingen. G\"{o}ttingen (Vandenhoeck \&
Ruprecht), 1996, pp. 232'', \it Historia Mathematica, \rm 30(2): 221-223.\item Valiron, G. (1949), \it Lectures
on the General Theory of Integral Functions, \rm New York, NY: Chelsea Publishing Company.\item van de Lune, J.,
te Riele, H.J.J. and Winter, D.T. (1986), On the zeros of the Riemann zeta-function in the critical strip, IV,
\it Mathematics of Computation, \rm 46: 667-681.\item Vivanti, G. (1901), \it Teoria delle funzioni analitiche,
\rm Milano, IT: Ulrico Hoepli Editore-Librajo della Real Casa (with a 1906 German translation entitled \it
Theorie der Eindeutigen Analytischen Funktionen Umarbeitung Üntee Mitwirkung des Verfassers Deutsch
Herausgegeben Von A. Gutzmer, \rm published in Leipzig, DE, by Dr\"{u}ck und Verlag Von B.G. Teubner, and with a
1928 revised and enlarged second edition entitled \it Elementi della teoria delle funzioni analitiche e delle
funzioni trascendenti intere, \rm published by the same publishing house of the first Italian edition).\item Von
Mangoldt, H. (1896), Zu Riemanns Abhandlung <<\"{U}ber di Anzahl der Primzahlen unter einer gegebenen Grenze>>,
\it Crelle Journal f\"{u}r die Reine und Angewandte Mathematik, \rm 114 (4): 255-305.\item Von Mangoldt, H.
(1898), \"{U}ber eine Anwendung der Riemannschen Formel f\"{u}r die Anzahl der Primzahlen unter einer gegebenen
Grenze, \it Crelle Journal f\"{u}r die Reine und Angewandte Mathematik, \rm 119 (1): 65-71.\item Von Mangoldt,
H. (1905), Zur Verteilung der Nullstellen der Riemannschen Funktion $\xi(t)$, \it Mathematische Annalen, \rm 60:
1-19.\item Voros, A. (2010), \it Zeta Functions over Zeros of Zeta Functions, \rm Lecture Notes of the Unione
Matematica Italiana, Berlin and Heidelberg, DE: Springer-Verlag.\item Watkins, J.J. (2014), \it Number Theory.
An Historical Approach, \rm Princeton, NJ: Princeton University Press.\item Weierstrass, K. (1856a), \"{U}ber
die Theorie der analytischen Facultäten, \it Journal für die reine und angewandte Mathematik, \rm 51: 1-60.\item
Weierstrass, K. (1856b), Theorie der Abel'schen Functionen, \it Journal für die reine und angewandte Mathematik
(Crelle's Journal), \rm 52: 285-380.\item Weierstrass, K. (1876), Zur Theorie der eindeutigen analytischen
Funktionen, \it Mathematische Abhandlungen der Akademie der Wissenschaften zu Berlin, \rm S. 11-60 (French
Translation: (1879), Mémoire sur les fonctions analytiques uniformes. Traduit par E. Picard, \it Annales
Scientifiques de l'\'{E}cole Normale Supérieure, \rm 2$^e$ Série, 8: 111-150).\item Weierstrass, K. (1894-1915),
\it Mathematische Werke von Karl Weierstrass, \rm Bands 1-6, Berlin, DE: Mayer \& M\"{u}ller.\item Weil, A.
(1975), \it Essais historiques sur la théorie des nombres, \rm Monographie N$^0$ 22 de L'Enseignement
Mathématique, Université de Genève, Genève, CH: Imprimerie Kundig.\item Weil, A. (1976), \it Elliptic Functions
according to Eisenstein and Kronecker, \rm Berlin and Heidelberg, DE: Springer-Verlag.\item Weil, A. (1984), \it
Number Theory. An Approach Through History from Hammurapi to Legendre, \rm Boston, MA: Birkh\"{a}user, Inc.\item
Weil, A. (1989), Prehistory of the Zeta-Function, in: Aubert, K.E., Bombieri, E. and Goldfeld, D. (Eds.) (1989),
\it Number Theory, Trace Formulas and Discrete Groups, \rm Symposium in Honor of Atle Selberg, Oslo, Norway,
July 14-21, 1987, San Diego, CA: Academic Press, Inc., pp. 1-9.\item Weyl, H. and Weyl, F.J. (1965), \it
Meromorphic Functions and Analytic Curves, \rm Reprint of the 1th 1943 edition published in Annals of
Mathematics Studies (No. 12) by Princeton University Press, Princeton, NJ, New York, NY: Kraus Reprint
Corporation.\item Whittaker, E.T. and Watson, G.N. (1927), \it A Course in Modern Analysis. An Introduction to
the General Theory of Infinite Processes and of Analytic Functions; with an Account of the Principal
Transcendental Functions, \rm Fourth Edition, Cambridge, UK: Cambridge University Press.\item Wintner, A.
(1934), \"{U}ber die asymptotische Verteilungsfunktion rekurrenter Winkelvariablen, \it Monatshefte f\"{u}r
Mathematik und Physik, \rm 41 (1), 1-6.\item Wintner, A. (1935), A Note on the Riemann $\xi$-function, \it
Journal of the London Mathematical Society, \rm 10: 82-83.\item Wintner, A. (1936), On a class of Fourier
transforms, \it American Journal of Mathematics, \rm 58: 45-90.\item Wintner, A. (1947), On an oscillatory
property of the Riemann $\Xi$-function, \it Mathematical Not\ae, \rm 7: 177-178.\item Yang, C.N. and Lee, T.D.
(1952), Statistical Theory of Equations of State and Phase Transitions. I. Theory of Condensation, \it Physical
Review, \rm 87(3): 404-409.\item Yang, C.N. (1961), \it Elementary Particles. A Short History of Some
Discoveries in Atomic Physics - 1959 Vanuxem Lectures, \rm Princeton, NJ: Princeton University Press (Italian
Translation: (1969), \it La scoperta delle particelle elementari, \rm Torino, IT: Paolo Boringhieri Editore).
\item Yang, C.N. (2005), \it Selected Papers (1945-1980) With Commentary, \rm  World Scientific Series in 20-th
Century Physics, Vol. 36, Singapore, SG: World Scientific Publishing Company, Ltd.\item Zhang, G-H. (1993), \it
Theory of Entire and Meromorphic Functions. Deficient and Asymptotic Value and Singular Directions, \rm
Translations of Mathematical Monographs, Volume 122, Providence, RI: American Mathematical
Society.\newpage\subsection*{Digital libraries, online databases and electronic archives
consulted}http://eudml.org/\\http://gallica.bnf.fr/\\http://gdz.sub.uni-goettingen.de/
\\http://www.emis.de/MATH/JFM/JFM.html/\\https://archive.org/\\http://ebooks.library.cornell.edu/m/math/
\\http://quod.lib.umich.edu/u/umhistmath/\\https://openlibrary.org/\\http://matematica.unibocconi.it/i-matematici/
\\http://www-history.mcs.st-and.ac.uk/history/\\http://mathematica.sns.it/\\http://www.emis.de/projects/JFM/
\\http://www.sism.unito.it/\\http://www.mathematik.uni-bielefeld.de/~rehmann/DML/dml links.html/\\
http://mathematics.library.cornell.edu/additional/Digital-Projects/\\http://www.euro-math-soc.eu/digital-libraries/\\
http://www.ams.org/dmr/\\http://www.numdam.org/\\https://projecteuclid.org/\\
http://empslocal.ex.ac.uk/people/staff/mrwatkin//zeta/physics.htm\\http://arxiv.org/

\end{description}

\end{document}